\documentclass[preprint]{elsarticle}
\usepackage[bookmarks,bookmarksopen,bookmarksdepth=2]{hyperref}
\usepackage{amsmath}
\usepackage{amssymb}
\usepackage{hyperref}
\usepackage{textcomp}
\usepackage{color}
\usepackage{graphicx}
\usepackage{slashed}
\usepackage{epstopdf}
\usepackage{upgreek}
\usepackage{soul}
\usepackage[makeroom]{cancel}

\newcommand{\dbar}{\bar{d}}

\title{Review of the theoretical and experimental status of dark matter identification with cosmic-ray antideuterons}

\begin{document}
\clubpenalty = 10000 \widowpenalty = 10000 \displaywidowpenalty = 10000

\author[cu,slac]{T.~Aramaki}
\address[cu]{Columbia Astrophysics Laboratory, Columbia University, New York, NY 10027, USA}
\address[slac]{SLAC National Accelerator Laboratory, Menlo Park, CA, USA}

\author[ucb]{S.~Boggs}
\address[ucb]{Space Sciences Laboratory, University of California at Berkeley, Berkeley, CA 94709, USA}
 
\author[infnto]{S.~Bufalino}
\address[infnto]{INFN Sezione di Torino, 10125 Torino, Italy}

\author[oslo]{L.~Dal}
\address[oslo]{Department of Physics, University of Oslo, 0316 Oslo, Norway}

\author[uhm]{P.~von~Doetinchem\corref{cor1}}
\ead{philipvd@hawaii.edu}
\address[uhm]{Department of Physics and Astronomy, University of Hawaii at Manoa, Honolulu, HI 96822, USA}
\cortext[cor1]{Corresponding author}

\author[infnto,uto]{F.~Donato}
\address[uto]{Department of Physics, University of Torino, 10125 Torino, Italy}

\author[infnto,uto]{N.~Fornengo}

\author[isas]{H.~Fuke}
\address[isas]{Institute of Space and Astronautical Science, Japan Aerospace Exploration Agency (ISAS/JAXA), Sagamihara, Kanagawa 252-5210, Japan}

\author[uham]{M.~Grefe}
\address[uham]{II.\ Institut f\"ur Theoretische Physik, Universit\"at Hamburg, 22761 Hamburg, Germany}

\author[cu]{C.~Hailey}

\author[uma]{B. Hamilton}
\address[uma]{Department of Physics, University of Maryland, College Park, MD 20742, USA}

\author[tum]{A.~Ibarra}
\address[tum]{Physik-Department T30d, Technische Universit\"at M\"unchen, 85748 Garching, Germany}
 
\author[gsfc]{J.~Mitchell}
\address[gsfc]{NASA/Goddard Space Flight Center, Greenbelt, MD 20771, USA}

\author[ucla]{I.~Mognet}
\address[ucla]{Department of Physics and Astronomy, University of California, Los Angeles, CA 90095, USA}
 
\author[ucla]{R.A.~Ong}

\author[uhm]{R.~Pereira}

\author[hav]{K.~Perez}
\address[hav]{Haverford College, Haverford, PA 19041, USA}

\author[lapth,lapp]{A.~Putze}
\address[lapth]{LAPTh, Universite Savoie Mont Blanc, CNRS, 74941 Annecy-le-Vieux, France}
\address[lapp]{LAPP, Universite Savoie Mont Blanc, CNRS/IN2P3, 74941 Annecy-le-Vieux, France}

\author[oslo]{A.~Raklev}

\author[lapth]{P.~Salati}

\author[gsfc]{M.~Sasaki}

\author[umi]{G.~Tarle}
\address[umi]{Department of Physics, University of Michigan, Ann Arbor, MI 48109, USA}

\author[sis]{A.~Urbano}
\address[sis]{SISSA - International School for Advanced Studies, 34136, Trieste, Italy}
 
\author[infnto,uto]{A.~Vittino}

\author[tum]{S.~Wild}

\author[mit]{W.~Xue}
\address[mit]{Center for Theoretical Physics, Massachusetts Institute of Technology, Cambridge, MA 02139, USA}

\author[kek]{K.~Yoshimura}
\address[kek]{High Energy Accelerator Research Organization (KEK), Tsukuba, Ibaraki 305-0801, Japan}

\begin{abstract}

Recent years have seen increased theoretical and experimental effort towards the first-ever detection of cosmic-ray antideuterons, in particular as an indirect signature of dark matter annihilation or decay. 
In contrast to indirect dark matter searches using positrons, antiprotons, or $\upgamma$-rays, which suffer from relatively high and uncertain astrophysical backgrounds, searches with antideuterons benefit from very suppressed conventional backgrounds, offering a potential breakthrough in unexplored phase space for dark matter. This article is based on the first dedicated cosmic-ray antideuteron workshop, which was held at UCLA in June 2014. It reviews broad classes of dark matter candidates that result in detectable cosmic-ray antideuteron fluxes, as well as the status and prospects of current experimental searches. The coalescence model of antideuteron production and the influence of antideuteron measurements at particle colliders are discussed. This is followed by a review of the modeling of antideuteron propagation through the magnetic fields, plasma currents, and molecular material of our Galaxy, the solar system, the Earth's geomagnetic field, and the atmosphere. Finally, the three ongoing or planned experiments that are sensitive to cosmic-ray antideuterons, BESS, AMS-02, and GAPS, are detailed. As cosmic-ray antideuteron detection is a rare event search, multiple experiments with orthogonal techniques and backgrounds are essential. Therefore, the combination of AMS-02 and GAPS antideuteron searches is highly desirable. Many theoretical and experimental groups have contributed to these studies over the last decade, this review aims to provide the first coherent discussion of the relevant dark matter theories that antideuterons probe, the challenges to predictions and interpretations of antideuteron signals, and the experimental efforts toward cosmic antideuteron detection.

\end{abstract}


\maketitle

\tableofcontents

\section{Introduction\label{s-1}}

The fields of cosmic-ray, antimatter, and dark matter physics have developed together over the last fifty years. Antideuterons, which are nuclei composed of one antiproton and one antineutron, were identified in the early 1960s among the secondaries from nuclear targets exposed to the proton beams of the CERN Proton Synchrotron~\cite{CERNdbar} and Brookhaven Alternating Gradient Synchrotron~\cite{PhysRevLett.14.1003}. By this time deuterons, which are nuclei composed of one proton and one neutron, had already been identified in nuclear emulsions exposed to cosmic rays~\cite{1932PhRv...39..164U}. The production of deuterons and antideuterons in collider experiments, however, allowed for tests of models describing their formation~\cite{PhysRevLett.7.69,Butler196277,PhysRev.129.836,PhysRev.129.854}. From this, the picture of a two-stage formation process emerged, with the production of a nuclear cascade within which several nucleons interact to form a nucleus. 

By the 1970s, the existence of dark matter had been established on galactic scales by precision galaxy rotation measurements~\cite{1976AJ.....81..687R,1976AJ.....81..719R}. The presence of enough hidden, heavy, normal objects to account for this dark matter has now been ruled out by astronomical surveys and cosmological simulations~\cite{darkmatter}, which confirm that dark matter is roughly five times more abundant in the universe than baryonic matter~\cite{planck}. Current experiments instead focus on searching for entirely new particles that will make up this mass. Among the most theoretically well-motivated candidates for such a particle is the weakly interacting massive particle, or WIMP.

It is generally acknowledged that the identification of dark matter will require the interplay of direct searches (which probe scattering cross sections through the recoil of dark matter off of target nuclei), indirect searches (which probe annihilation cross sections and decay lifetimes through the detection of annihilation or decay products), and collider searches (which probe production cross sections through the products of high-energy particle collisions).
There are currently about twenty operating or planned direct search experiments \cite{porter,baudis,panci}. 
The current best exclusion limits come from the LUX and CDMS experiments~\cite{akerib14, Agnese:2014aze,2015arXiv151203506L}. However, these limits conflict with earlier CDMS-II~\cite{agnese13}, DAMA/LIBRA~\cite{dama1,dama2,dama3,dama4,dama5}, and CoGeNT \cite{cogent1,cogent2,cogent3,cogent4,cogent5} results, which are consistent with light ($\approx10$~GeV) dark matter. The results of CDMS-II and CoGeNT can be made compatible with the limits of LUX \cite{akerib14}, CDMSlite \cite{agnese13} and XENON \cite{aprile12} if dark matter is not isospin-invariant \cite{nobile131,nobile132,fox}. 
In any case, the high backgrounds and low signal amplitudes of direct detection experiments make analysis, particularly at low WIMP mass, challenging, and direct experiments optimized for low-energy recoils, such as PICO \cite{pico}, are still orders of magnitude less sensitive for the spin-independent case than LUX or CDMS.

If dark matter was in thermal equilibrium in the early universe, and froze out when the temperature dropped due to expansion, it is a natural assumption that dark matter particles are able to interact with each other and produce Standard Model particles. Indirect searches exploit possible kinematic differences between the production of cosmic rays through dark matter and standard astrophysical processes to identify dark matter signals. Cosmic-ray antiparticles are ideal candidates for indirect searches, but are often challenging due to high or uncertain levels of astrophysical background. 
Recently, results from PAMELA \cite{pamela}, Fermi \cite{fermi}, and AMS-02 \cite{ams02first,ams02elposflux2014,ams4} show a structure in the positron data that might be interpreted as originating from dark matter with TeV-scale mass.
Antiproton data do not show such a significant feature \cite{besspbar,2010PhRvL.105l1101A}. These null antiproton results instead have been used to constrain a variety of dark matter models \cite{pbarci,Fornengo:2013osa, Delahaye:2013yqa}, though their statistical accuracy at low energies is limited. 

Antideuterons may be generated in dark matter annihilations or decays, offering a potential breakthrough in unexplored phase space for dark matter. In this process, annihilation or decay proceeds via particle physics, producing particles that then hadronize to an antiproton, $\bar{p}$, and antineutron, $\bar{n}$, pair that coalesce to form an antideuteron, $\bar{d}$. The unique strength of a search for low-energy antideuterons lies in the ultra-low astrophysical background for this channel and was pointed out for the first time by the authors of~\cite{Donato:1999gy}. By now, the topic has been further discussed in more depth for more than a decade, e.g., in ~\cite{Fornengo:2013osa,Baer:2005tw,Donato:2008yx,Duperray:2005si,Ibarra:2012cc}. The dominant conventional sources for secondary (background) antideuteron production are cosmic-ray protons or antiprotons interacting with the interstellar medium~\cite{Duperray:2005si}. However, the high threshold energy for antideuteron production and the steep energy spectrum of cosmic rays mean there are fewer particles with sufficient energy to produce secondary antideuterons, and those that are produced have relatively large kinetic energy. In addition, the low binding energy of antideuterons makes energy loss through collisions difficult. 

\begin{figure}
\centering
\includegraphics[width=0.65\linewidth]{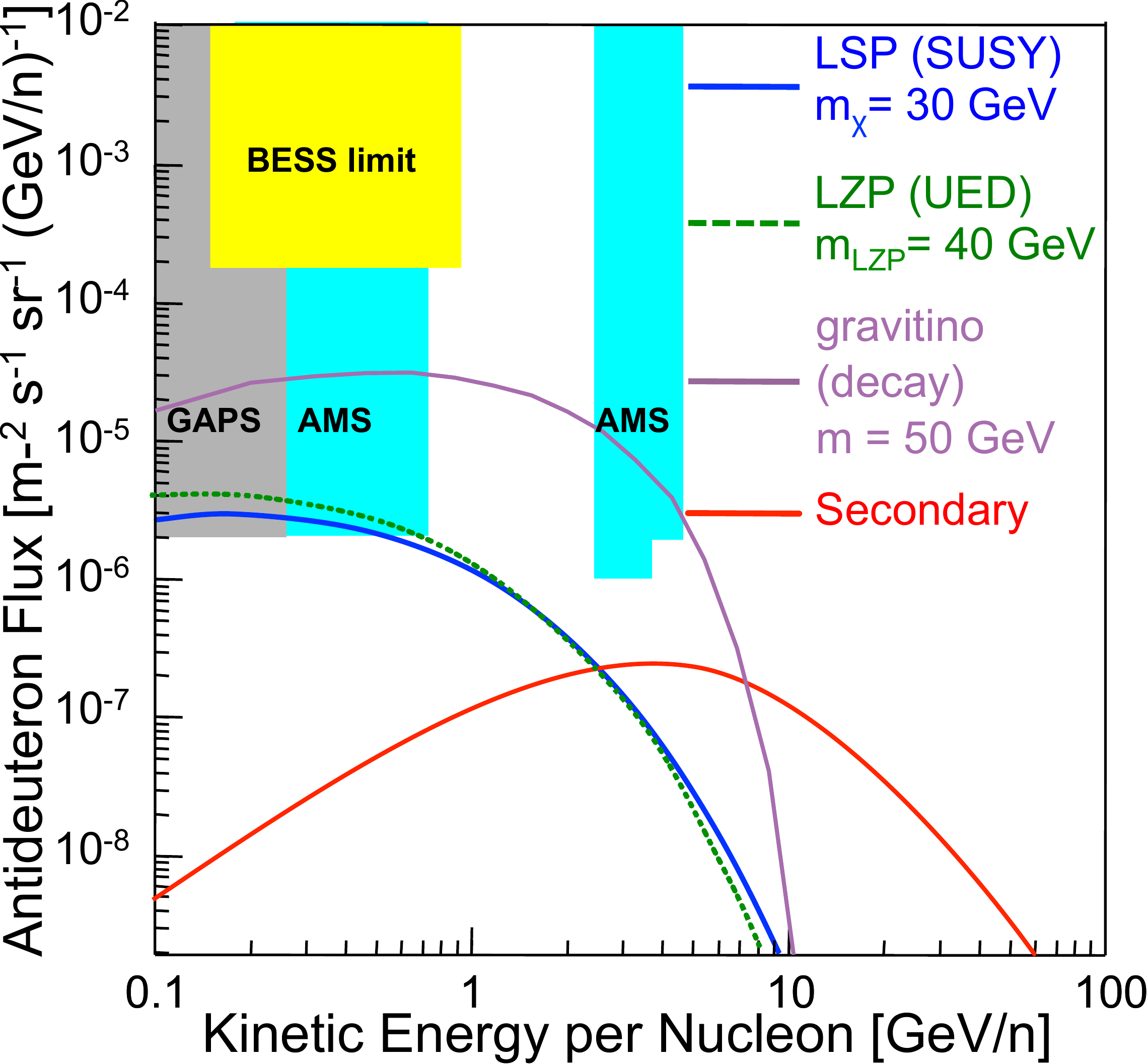}
\caption{\label{f-dmdbar}
Predicted antideuteron flux as a function of kinetic energy per nucleon for a 30~GeV neutralino, a 40~GeV extra-dimensional Kaluza-Klein neutrino, and a 50~GeV gravitino~\cite{Baer:2005tw,Donato:2008yx,Dal:2014nda,Ibarra2013a}. The antideuteron limits from BESS are shown~\cite{Fuke:2005it}, along with the projected sensitivities of AMS-02 for the superconducting-magnet configuration~\cite{ams02dbaricrc2007} after 5 years of operation and GAPS after three 35-day flights~\cite{gaps,Aramaki2015}. The MED Galactic propagation scenario is assumed (Sec.~\ref{subsec:transport_ISM}). These predictions use a coalescence momentum that is set to 195\,MeV (Sec.~\ref{s-3}) and the Einasto dark matter density profile (Sec.~\ref{subsec:DMdensity}). For the solar modulation parameters see Sec.~\ref{subsec:transport_HELIO}.}
\end{figure}

Fig.~\ref{f-dmdbar} shows the antideuteron flux expected from three benchmark dark matter scenarios. These dark matter candidates include a lightest supersymmetric particle (LSP) neutralino from the minimal supersymmetric model (MSSM), a 5D warped GUT Dirac neutrino (LZP), and an LSP gravitino. The expected secondary/tertiary background~\cite{Ibarra2013a} is also shown. This figure reveals why low-energy antideuterons are such an important approach: the flux from a wide range of viable dark matter models exceeds the background flux by more than two orders of magnitude in the energy range below 0.25\,GeV/$n$, and by more than an order of magnitude up to 1\,GeV/$n$. 

The large signal-to-background ratio for low-energy antideuterons does not rely on any boosting mechanisms due to, e.g., dark matter clumpiness, Sommerfeld enhancement, or large Galactic Halo size. Introducing such effects would further increase the signal strength. This is in contrast to dark matter signal predictions for positrons, antiprotons, and $\upgamma$-rays, which typically constitute a small component on top of the astrophysical background. Sec.~\ref{s-2} reviews various dark matter and other beyond the Standard Model candidates that produce an antideuteron signal evading current experimental bounds, but within the sensitivity of ongoing or planned cosmic-ray antideuteron experiments.

The coalescence model describes the process through which an antideuteron is formed through the merging of the antiproton and antineutron. As outlined in Sec.~\ref{s-3}, several implementations of this model have been developed. The first analyses \cite{Donato:1999gy,Baer:2005tw,Donato:2008yx,Chardonnet:1997dv,Brauninger:2009pe,Ibarra:2009tn} were performed within a purely analytical framework, while more recent ones \cite{Fornengo:2013osa,Ibarra:2012cc,Ibarra2013a,Dal:2014nda,Kadastik:2009ts,Cui:2010ud,Grefe:2011dp,Dal:2012my,Grefe:2015jva} exploit an event-by-event Monte Carlo approach. These coalescence models are then tuned to collider measurements of deuteron and antideuteron production, as described in Sec.~\ref{s-coll}.
 
The predicted cosmic antideuteron flux arriving at the Earth relies on the modeling of charged-particle transport both in the Galactic medium and in the heliosphere. These transport models are discussed in Sec.~\ref{s-4}. The modeling of Galactic transport is the main source of theoretical uncertainty, while solar modulation is relevant to shaping the low-energy tail of the predicted flux.

Despite the large signal-to-background ratio, any search for cosmic-ray antideuterons is a rare event search, and multiple experiments with complementary techniques and backgrounds will be necessary to build confidence in any detection. Cosmic antideuterons have not yet been detected, with the current best flux upper limits provided by the BESS experiment \cite{Fuke:2005it}. 
More sensitive limits will be provided by the AMS-02 experiment \cite{Battiston:2008zza,Incagli:2010zz,Bertucci:2011zz,Kounine:2012zz}, which is currently taking data onboard of the International Space Station.
The General Antiparticle Spectrometer (GAPS) experiment \cite{Mori:2001dv,2009AIPC.1166..163H,Hailey2013290,2014NIMPA.735...24M,2014AdSpR..53.1432F}, which is proposed for several Antarctic balloon campaigns, will provide essential complementary sensitivity to both antideuterons and antiprotons in an unprecedented low-energy range. Sec.~\ref{s-5} presents the experimental sensitivities and status of the BESS, AMS-02, and GAPS antideuteron searches.

A number of reviews on the status of dark matter searches have been published in recent years, with discussions related to antideuteron searches appearing in, e.g., the following~\cite{Bertone:2004pz,2010ARA&A..48..495F,2011PPN....42..650R,2011ARA&A..49..155P}.

\section{Prospects for dark matter detection with antideuterons \label{s-2}}

Many dark matter models are capable of producing an antideuteron flux that is within the reach of currently operating or planned experiments. Sec.~\ref{s-anni} discusses antideuterons resulting from dark matter self-annihilations, while Sec.~\ref{s-decay} discusses antideuterons from decays of long-lived dark matter. The purpose of the discussion is not to detail every theoretical model that can result in a detectable flux of antideuterons; instead, the following review demonstrates that measurements of cosmic-ray antideuterons are sensitive to a wide range of theoretical models, probing dark matter masses from $\mathcal{O}(1\text{\,GeV})$ to $\mathcal{O}(1\text{\,TeV})$. Antihelium nuclei can be produced by many of the same beyond the Standard Model processes as antideuterons, providing an even higher signal-to-background ratio at low energies, but also significantly lower predicted flux levels. The prospects and challenges of future dark matter searches using antihelium signatures are discussed in Sec.~\ref{s-antihe}.

Before starting the discussion, it is important to point out the dominant uncertainties on the antideuteron flux predictions presented here, which will be more deeply addressed in Sec.~\ref{s-3} and \ref{s-4}. These uncertainties are due to the hadronization and coalescence models used to describe antideuteron formation, as well as the propagation models used to describe antideuteron transport in the Galactic and solar environments. The antideuteron fluxes presented in this section assume a conservative boost factor, due to dark matter clumps in the Galactic halo, of $f=1$. However, a boost factor of $f=2$--3, as is consistent with current theoretical expectations, would increase dark matter fluxes by a factor $f$ over those discussed below~\cite{Baer:2005tw,Lavalle:1900wn,PhysRevD.81.043532}. Such a boost factor is only relevant for dark matter annihilation, as dark matter decay depends linearly on the dark matter density.

It is also vital to note that every process capable of producing antideuterons will also produce a much larger flux of antiprotons. Thus, the strength of any prospective antideuteron signature from dark matter is constrained by the (current) non-observation of a signature in the cosmic antiproton spectrum. However, the ultra-low background for low-energy antideuteron searches still opens new phase space for dark matter observations. Furthermore, detecting deviations from the astrophysical antiproton flux requires very high statistics. Antideuterons provide an additional search channel with very strongly suppressed astrophysical backgrounds compared to antiprotons, and thus can act as an essential probe to confirm or rule out potential deviations in the antiproton spectrum due to processes like dark matter annihilation or decay. On the other hand, a non-detection of a signal above background in the antiproton channel might also just be a consequence of experimental limitations and can serve as additional motivation to pursue antideuteron searches.

\subsection{Dark matter annihilation\label{s-anni}}

\begin{figure}
\centering
\includegraphics[width=0.65\linewidth]{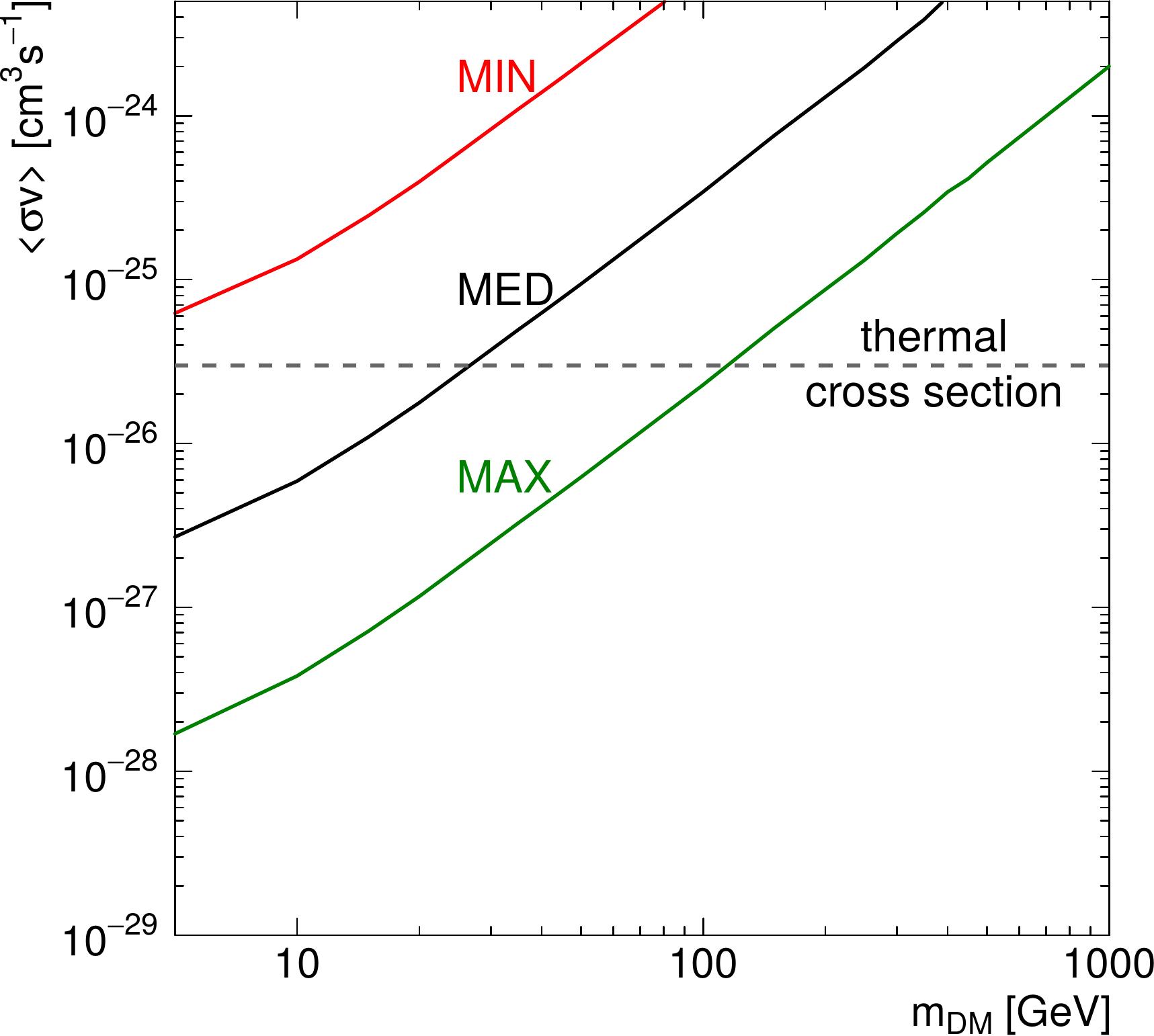}
\caption{The sensitivity of GAPS to a three standard deviation detection in the antideuteron channel of a dark matter WIMP annihilating into $u\bar{u}$, shown as a function of annihilation cross section, $\langle \sigma v \rangle$, and dark matter mass, $m_{\text{DM}}$. The red, black, and green lines indicate the GAPS sensitivity in the MIN, MED, and MAX Galactic propagation scenarios (Sec.~\ref{subsec:transport_ISM}). The dark matter density profile, solar modulation parameters, and coalescence momentum are as defined in Fig.~\ref{f-dmdbar}. \textit{Source:} Adapted from \cite{Fornengo:2013osa}.}
\label{fig:reachabilities}
\end{figure}

It is common to illustrate experimental sensitivity to annihilating dark matter by first assuming an annihilation cross section into a particular Standard Model channel, without considering the details of the underlying dark matter model. Analytical models or Monte Carlo simulations can then be used to calculate the resulting flux level of different cosmic-ray species. Since larger dark matter particle mass implies lower dark matter particle density, the lowest annihilation cross section that a measurement is sensitive to then scales quadratically with the dark matter mass.

The prospects for a three standard deviation detection with the GAPS experiment of a dark matter WIMP annihilating into $u\bar{u}$, as a function of annihilation cross section and dark matter mass, are shown in Fig.~\ref{fig:reachabilities}. The three solid lines refer to the GAPS sensitivity assuming the MIN, MED and MAX Galactic propagation models (Sec.~\ref{subsec:transport_ISM}), while the dashed line represents the thermally averaged cross section for a thermal-relic WIMP. The theoretical and experimental uncertainty for the propagation model and coalescence scheme are both on the order of $\mathcal{O}(10)$. The experimental sensitivity reaches well below the thermal-relic cross section for dark matter masses below $m_{\text{DM}} =$ 20 (100) GeV for the MED (MAX) propagation scenario. It is important to stress that recent measurements disfavor the MIN model, as will be further discussed in Sec.~\ref{subsec:twozone_constraints}. Antideuteron searches thus both provide complementary sensitivity to direct search experiments, which are most sensitive to intermediate-mass dark matter, and provide increased sensitivity to light dark matter models, where direct search techniques are challenging. For example, light candidates in the next-to-minimal supersymmetric model (NMSSM) still evade direct, collider, and Fermi dwarf galaxy bounds~\cite{Cerdeno:2014cda}, but would be probed by ongoing and proposed antideuteron experiments. An example of the sensitivity to standard supersymmetric particles is shown in Fig.~\ref{f-dmdbar}, which presents the signal predicted for a 30~GeV supersymmetric neutralino annihilating into $b\bar b$. 

This signal is within the detectable range of both GAPS and AMS-02, and is motivated by the diffuse $\upgamma$-ray excess observed surrounding the Galactic Center by Fermi~\cite{Daylan:2014rsa}. The excess shows a high significance and the interpretation is ongoing~\cite{2015PhRvL.114u1303K,2015arXiv150205703A,2015arXiv150402477O,2015arXiv150605119C,2015arXiv150902164C,TheFermi-LAT:2015kwa}.

In addition, the authors of \cite{Cui:2010ud} state that the gluon-gluon decay channel of dark matter could produce an antideuteron flux in the detectable range for masses below $\mathcal{O}(100\,\text{GeV})$. Also shown in Fig.~\ref{f-dmdbar} is the case of a right-handed Kaluza--Klein neutrino of warped 5-dimensional grand unified theories with a conserved $Z_3$ parity (LZP)~\cite{Baer:2005tw, Agashe:2004ci, Agashe:2004bm, Hooper:2005fj}, which for a mass of 40~GeV has a major annihilation channel into $Z$ bosons. The authors of \cite{Baer:2005tw} interpret especially the LZP mass range close to the $Z$ pole of 40--50\,GeV as an interesting benchmark case for WIMP pair annihilation into $Z$ that still produces the correct thermal relic density.

\begin{figure}
\centering
\includegraphics[width=0.65\linewidth]{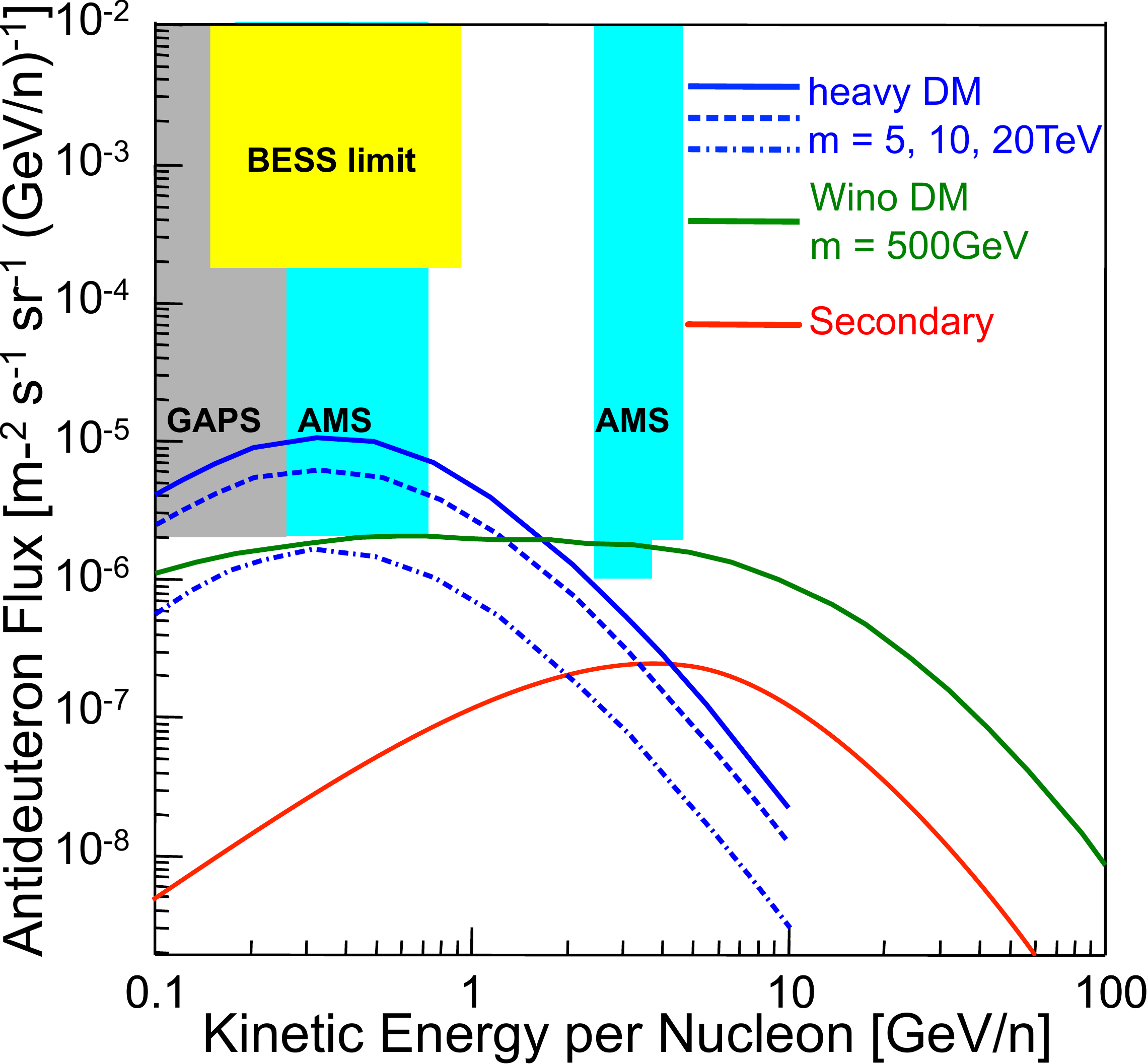}
\caption{\label{f-heavy}
Predicted antideuteron flux for annihilation of dark matter with $m_{\text{DM}} =$~5, 10, 20~TeV~\cite{Brauninger:2009pe} (blue lines, top to bottom) into $b\bar b$ with enhanced annihilation cross sections ($\langle\sigma v\rangle_{5\,\text{TeV}}=3\cdot10^{-22}$\,cm$^3$/s, $\langle\sigma v\rangle_{10\,\text{TeV}}=7\cdot10^{-22}$\,cm$^3$/s, $\langle\sigma v\rangle_{20\,\text{TeV}}=20\cdot10^{-22}$\,cm$^3$/s).  The predicted antideuteron flux from pure-Wino dark matter~\cite{Hryczuk:2014hpa} (solid green line) with $m_{\text{DM}}= 0.5$\,TeV, $\langle\sigma v\rangle=4.82\cdot10^{-25}$\,cm$^3$/s with Sommerfeld enhanced annihilations into $W^+W^-$. The MAX propagation model is used for all predictions.}
\end{figure}

Antideuteron signals from annihilation of heavy ($m_{\text{DM}} =$ 0.5--20\,TeV) dark matter may also be detectable by AMS-02 and GAPS, for the MAX propagation scenario~\cite{Brauninger:2009pe}. These multi-TeV mass dark matter particles are motivated by the possible high-energy cosmic positron excess observed by AMS-02~\cite{ams02first,ams02elposflux2014,ams4}. Accounting for the lack of a similar excess in the available antiproton data, it requires both high dark matter mass and an enhanced annihilation cross section, such as provided by the Sommerfeld mechanism~\cite{ArkaniHamed:2008qn}. The prediction for such multi-TeV mass particles annihilating into $b\bar b$ is shown in Fig.~\ref{f-heavy}. Annihilation into $W^+W^-$ would be disfavored in this scenario. Another model of heavy supersymmetric dark matter, also shown in Fig.~\ref{f-heavy}, is pure-Wino dark matter~\cite{Hryczuk:2014hpa}. The thermal relic dark matter density is reached for Wino dark matter at masses in the TeV range, which is close to the detectable range in the low-energy region of GAPS and AMS-02, and in the detectable range for the high-energy region of AMS-02. It is interesting to note that the antideuteron spectra from quark and heavy gauge boson annihilation channels are qualitatively different. Under the assumption that dark matter is moving slowly and that the dark matter halo and the Earth's rest frame are roughly the same, the hadronization in gauge boson channels is happening in a boosted frame while the hadronization in quark channels is essentially happening in the Earth's rest frame. As a result, the shapes of the antideuteron spectra originating from quark annihilation channels only vary slightly with mass and the gauge boson channels show a stronger dependence.

Recent Higgs portal dark matter models discuss the possibility of dark matter annihilating via Higgs exchange processes~\cite{Silveira:1985rk,McDonald:1993ex,Burgess:2000yq,Patt:2006fw}.
In these models, the scalar field $S$ plays the role of cold dark matter in the universe with a scalar gauge singlet mass, $m_{\rm S}$. 
There are three mass regions where it is possible to reproduce  the correct dark matter relic abundance. 
The low-mass region  ($m_{\rm S} \lesssim 50$\,GeV), in which  the dominant annihilation channel is into $b\bar{b}$ pairs, is completely ruled out by direct detection experiments and the bound on the invisible  decay width of the Higgs~\cite{Cline:2013gha}.  
The resonant region  ($m_{\rm S} \approx 50 - 70$\,GeV), in which  dark matter annihilation proceeds via the on-shell exchange of the Higgs boson, is severely constrained by cosmic antiproton~\cite{Urbano:2014hda} and $\upgamma$-ray~\cite{Feng:2014vea} measurements.
The high-mass region ($m_{\rm S} \gtrsim 70$\,GeV), in which  the dominant annihilation channels are into electroweak gauge bosons, two Higgses (if $m_{\rm S} > 2m_{h}\approx 252$\,GeV),  and $t\bar{t}$ pairs (if $m_{\rm S} > 2m_{t}\approx 347$\,GeV), are still only marginally disfavored by direct and indirect detection experiments~\cite{Urbano:2014hda}, and for $m_{\rm S}\gtrsim 100$\,GeV one can obtain a good dark matter candidate. This region extends up to the TeV scale.  Dark matter candidates of this type with intermediate dark matter masses are able to produce an antideuteron signal well above the astrophysical background, but not yet within the reach of the current experiments.

\subsection{Decaying dark matter\label{s-decay}}

Dark matter may also exist in the form of very long-lived, but unstable particles whose decays can produce antideuterons~\cite{Ibarra:2013cra}. A particularly well-motivated and well-studied case of decaying dark matter is the gravitino in supergravity models with $R$-parity violation~\cite{Takayama:2000uz,Buchmuller:2007ui,Lola:2008bk}. In the most general case, $R$-parity violation adds one bilinear and three trilinear renormalizable operators to the superpotential~\cite{Barbier:2004ez}:
\begin{equation}
W_{\slashed{R}} \sim \mu_iH_uL_i+\lambda_{ijk}L_iL_j\bar{E}_k+\lambda'_{ijk}L_iQ_j\bar{D}_k+\lambda_{ijk}^{\prime\prime}\bar{U}_i\bar{D}_j\bar{D}_k\,,
\end{equation} 
where $H_u$, $L_i$ and $Q_i$ are the up-type Higgs, lepton and quark SU(2) doublet superfields, respectively, and $E_i$, $U_i$ and $D_i$ are the lepton, up-type quark and down-type quark SU(2) singlet superfields, respectively. Summation over generation indices $i,j,k$ and suppressed gauge indices is assumed. The first three operators violate lepton number and the last operator violates baryon number. Moreover, there could be additional bilinear and trilinear $R$-parity violating operators in the soft supersymmetry breaking terms. Proton stability is not violated as long as only specific subsets of those operators are present in the theory~\cite{Ibanez:1991pr}. A theory with bilinear $R$-parity violation only violates lepton number conservation, while theories with trilinear $R$-parity violation usually assume the presence of either only lepton number violating operators or only baryon number violation. Since the gravitino decay rate is quadratically suppressed by the Planck mass, $1/M_{\text{Pl}}^2$, moderately small $R$-parity violating parameters are sufficient to make the gravitino long-lived on cosmological time scales ($\tau \gg 4\times 10^{17}\,$s).

Taking into account constraints on the gravitino lifetime derived from the PAMELA antiproton observations~\cite{Adriani:2012paa}, i.e.\ $\tau > 10^{26}$--$10^{28}$\,s~\cite{Delahaye:2013yqa}, antideuteron signals from gravitino decays into the channels $\gamma\nu_i$, $Z\nu_i$, $W\ell_i$, and $h\nu_i$ via bilinear R-parity violation~\cite{Ishiwata:2008cu,Covi:2008jy} exceed the astrophysical background at low energies, but are not within the detectable region of AMS-02 or GAPS~\cite{Grefe:2011dp,Grefe:2015jva,Grefe:2011kh}. Gravitino decays via trilinear $R$-parity violating operators, however, may be detectable~\cite{Dal:2014nda,Monteux:2014tia}. In these models, the gravitino typically decays into three final state particles: $\nu_id_j\bar{d}_k$, $\ell_iu_j\bar{d}_k$, $u_id_jd_k$~\cite{Moreau:2001sr}. The purely leptonic operator $L_iL_j\bar{E}_k$ leads to final states $\nu_i\ell_j\ell_k$ and thus does not produce any antideuterons. A particularly relevant case is the baryon number violating operator $\bar{U}_i\bar{D}_j\bar{D}_k$ that triggers gravitino decay into three quarks. 
The production of three quarks in the hard process leads to a larger number of antiprotons and antineutrons in the final state. While this leads to stronger constraints from antiproton observations, the increased probability for antideuteron formation leads to an enhancement of the expected antideuteron flux compared to the antiproton flux. Therefore, this particular trilinear $R$-parity violating gravitino decay channel allows for a flux that is within the reach of AMS-02 and GAPS while obeying the antiproton constraints~\cite{Bomark:2009zm}. This is demonstrated for the case of a 50\,GeV gravitino in Fig.~\ref{f-dmdbar}.

Note that since the antideuteron flux only drops linearly with increasing gravitino mass (as opposed to $1/m_{\text{DM}}^2$ for annihilating dark matter), the gravitino decay signal may dominate over the background of secondary antideuterons at high energies as well. This opens the possibility of using high-energy antideuterons (i.e.\ $T > 100$\,GeV) to look for signals from heavier decaying dark matter. 

\subsection{Prospects for dark matter detection with antihelium}
\label{s-antihe}

\begin{figure}
\centering
\includegraphics[width=0.65\linewidth]{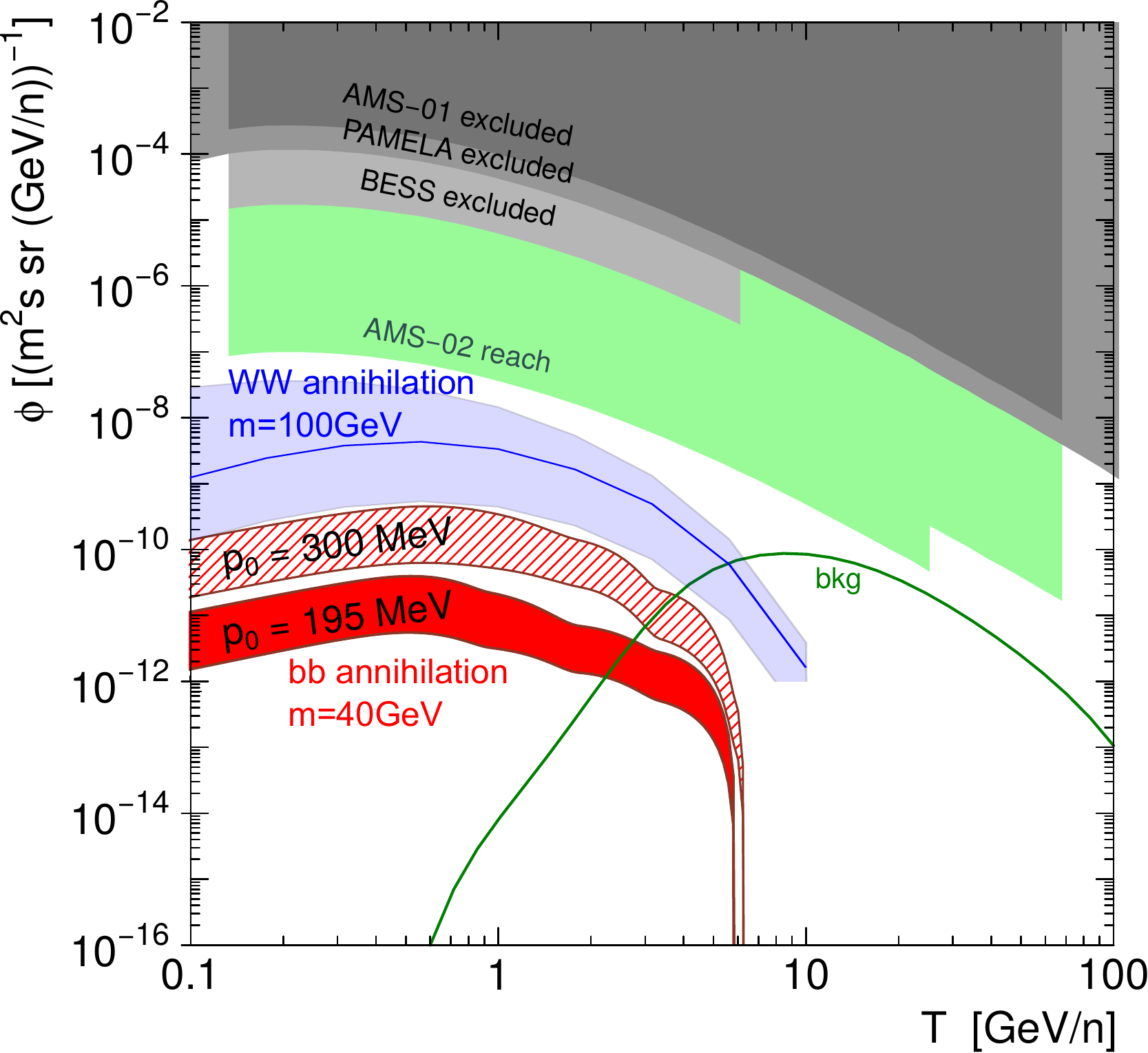}
\caption{Predicted antihelium flux from various dark matter models. The red regions show the flux from of a 40\,GeV dark matter particle annihilating into $b\bar{b}$~\cite{Cirelli:2014qia} with $\langle \sigma v \rangle= 3\cdot10^{-26}$\,cm$^3$\,s$^{-1}$. 
The solid red band corresponds to coalescence momentum $p_0 =195$\,MeV and the hatched red band to $p_0 = 300$\,MeV (Sec.~\ref{s-3}). Only those configurations that are compatible with PAMELA antiproton constraints are considered. 
The predicted antihelium flux of a 100\,GeV dark matter particle annihilating into $W^+W^-$ with $\langle \sigma v \rangle= 3\cdot10^{-26}$\,cm$^3$\,s$^{-1}$ is shown in blue~\cite{Carlson:2014ssa}. For both cases, the band width reflects variations due to different propagation model parameters.
The astrophysical background is represented by the green solid line. The projected sensitivity of AMS-02 after 5 years of operation is shown by the light green region. The gray regions correspond to the bounds imposed by previous BESS~\cite{2012PhRvL.108m1301A}, PAMELA~\cite{Mayorov:2011zz}, and AMS-01~\cite{Alcaraz:2000ss} measurements (the helium flux from PAMELA was used to translate from the $\overline{{\rm He}}/{\rm He}$ results given in these references). The dark matter density profile is Einasto.}
\label{fig:antihelium_prospects}
\end{figure}

Due to the larger energy threshold necessary to produce secondary antihelium nuclei ($31\,m_p = 29.09\,$GeV, instead of $17\,m_p = 15.95\,$GeV to produce antideuterons), antihelium signatures of dark matter annihilation or decay have an even higher signal-to-background ratio of $10^3-10^5$ (Fig.~\ref{fig:antihelium_prospects}) at low energies than antideuteron signatures (about $10^2$) ~\cite{Cirelli:2014qia,Carlson:2014ssa}.  
However, in these models also the  antihelium formation from dark matter is suppressed compared to antideuteron formation. 

Fig.~\ref{fig:antihelium_prospects} shows the predicted fluxes for a 40\,GeV dark matter particle annihilating into $b\bar{b}$ and a 100\,GeV dark matter particle annihilating into $W^+W^-$. The thermal-relic annihilation cross section is used for both predictions. 
The width of the shaded bands is defined by the variation in flux predicted using a range of Galactic propagation model parameters that produce an antiproton flux compatible with PAMELA measurements. As discussed in Sec.~\ref{subsec:antihelium_coalescence}, there is a large uncertainty due to the choice of the coalescence momentum, $p_0$. This is illustrated by the two red bands corresponding to $p_0 = 195$\,MeV and $p_0 = 300$\,MeV for the $b\bar b$ annihilation channel. The classical coalescence model~\cite{PhysRev.129.836,PhysRev.129.854} predicts a flux dependence proportional to $p_0^6$ for antihelium, as opposed to $p_0^3$ for antideuterons. Thus constraining the value of the coalescence momentum is critical to constraining antihelium flux predictions. The sensitivity of AMS-02 with 5 years of integration time~\cite{2010arXiv1009.5349K} is indicated by the light green region. These antihelium signatures are outside the current sensitivity of both AMS-02 and GAPS, but motivate future, more sensitive experiments. 

\section{Antideuteron production\label{s-3}}

Although antideuterons were discovered for the first time in 1965~\cite{PhysRevLett.14.1003,CERNdbar}, their formation is not well understood. However, understanding antideuteron production is crucial for the interpretation of the cosmic-ray data, which impacts both the antideuteron background expectation from interactions of primary/secondary cosmic rays with the interstellar medium as well as the formation in the aftermath of dark matter annihilations or decays. In recent years event-by-event Monte Carlo techniques have been used to show that the antideuteron flux prediction carry an uncertainty of about one order of magnitude for the formation mechanism. It is currently an open question if the antideuteron production depends on the exact underlying process and on the available center-of-mass energy or if  Monte Carlo generators need further refinement. This section will explain the theoretical status as well as future outlook of the understanding of the formation of antideuterons (Sec.~\ref{s-coa}). This discussion is followed by a description of the recent experimental results of antideuteron production in hadronic decays of the $\Upsilon$ resonance from $B\!{\scriptstyle A}\!B\!{\scriptstyle A\! R}$ (Sec.~\ref{s-babar}) and of $p$-$p$, $p$-Pb, and Pb-Pb collisions in the ALICE experiments (Sec.~\ref{s-alice}). In addition, the outlooks for the operational fixed target experiment NA61/SHINE (Sec.~\ref{s-na61}) and the future $\bar{\text{P}}$ANDA experiment (Sec.~\ref{s-panda}) are discussed. The section concludes by outlining the procedure for the antihelium coalescence model (Sec.~\ref{subsec:antihelium_coalescence}).

\subsection{The coalescence model\label{s-coa}}

\subsubsection{Basic concepts}
The fusion of an antiproton and an antineutron into an antideuteron is described by the coalescence model. The coalescence model is based on the simplifying assumption that any (anti)nucleons within a sphere of radius $p_0$ in momentum space will coalesce to produce an (anti)nucleus. For the case of deuterons, the condition means that any $pn$-pair with $\Delta p < p_0$ will coalesce to produce a deuteron. The coalescence momentum $p_0$ is a phenomenological quantity, and has to be determined through fits to experimental data. The model was first proposed by Schwarzschild and Zupan\v{c}i\v{c}~\cite{PhysRev.129.854} for the production of deuterons, tritons and helium nuclei in fixed target scattering experiments.
This has since been adopted for the description of antideuteron production in collider experiments and in the context of indirect dark matter detection.

One of the strengths of this simple model is that under the assumption of isotropic and uncorrelated proton and neutron momentum distributions, analytic expressions can be found for the spectra of different nuclei, given in terms of the nucleon spectra and the coalescence parameter $p_0$.
Before the onset of Monte Carlo event generators, such analytic expressions were also entirely necessary for application of the coalescence prescription. The antideuteron spectrum can in this approach be given in terms of the antiproton and antineutron spectra through
\begin{equation} \label{eq:iso_coalescence}
 \frac{\text dN_{\bar{d}}}{\text dT_{\bar{d}}} =
 \frac{p_0^3}{6}\,\frac{m_{\bar{d}}}{m_{\bar{n}} m_{\bar{p}}}\,
 \frac{1}{\sqrt{T_{\bar{d}}^{2} + 2 m_{\bar{d}} T_{\bar{d}}}}\,
 \frac{\text dN_{\bar{n}}}{\text dT_{\bar{n}}}\,\frac{\text dN_{\bar{p}}}{\text dT_{\bar{p}}} \,,
\end{equation}
where $m_i$, $T_i$ and $dN_i/dT_i$ are, respectively, the mass, kinetic energy and differential yield per event of particle $i$.

In the literature, some variant of this expression will be found in most works before 2009, but often with a differing constant prefactor -- typically a factor of 8 -- which can be absorbed into a redefinition of $p_0$.
While the assumptions of isotropic and uncorrelated nucleon spectra might be a good approximation in low-energy or minimum bias nuclear interactions for which the model was made, 
these assumptions have been found not to hold in relevant elementary particle interactions~\cite{Kadastik:2009ts}, such as dark matter annihilations or decay, and $p$-$p$ collisions at low center-of-mass energies. 
This is due to the nucleons typically being produced in geometrically restricted QCD jets, leading to strong correlations between nucleons on a per-event basis.

In order to take these correlations into account, the state of the art has since 2009 been to apply the coalescence condition to $\bar{p}\bar{n}$-pairs on a per-event basis in Monte Carlo events.
Unfortunately, this requires some four orders of magnitude more events than the isotropic approximation in order to achieve the same level of statistical error on predictions. 
As extensively discussed below, different event generators yield different values of $p_0$ when compared to a particular experiment, indicating a substantial systematic uncertainty in the coalescence prediction. 
Even with the same event generator, experiments differing in energy and interaction type will typically give inconsistent values of $p_0$.

Since momentum is not a Lorentz invariant quantity, the frame in which the coalescence condition is applied could also potentially be of importance.
In the original coalescence model, the coalescence condition was applied in the rest frame of one of the coalescing particles, 
while in recent works, employing per-event coalescence, the condition is either applied in the center-of-mass frame of the antinucleons, or applied as a condition on the invariant momentum,
\begin{equation}
 -\Delta^2 < p_0^2
\end{equation}
where $\Delta^{\mu} = k_{\bar{p}}^{\mu} - k_{\bar{n}}^{\mu}$.
These two approaches turn out to be equivalent in the limit $m_n = m_p$, and are in practice easily interchangeable.

As pointed out in \cite{Ibarra:2012cc}, spatial separation should also be taken into account in the formation of antideuterons.
Nuclear interactions typically take place on scales of a few femtometers. Antinucleons originating from weakly decaying particles, whose long lifetimes lead to macroscopic decay lengths, 
will hence be produced too far from the primary interaction vertex to have any chance of interacting with antinucleons from the primary collision. 
To take this into account, one can consider weakly decaying particles stable in this context. 
This considerably reduces the number of antinucleons available, and correspondingly increases the value of $p_0$ required to reproduce experimental data. 
The reader should be aware that not all recent work takes spatial separation into account, and the cut on lifetime varies when used, e.g., Ref.~\cite{Ibarra:2012cc} uses 1\,mm/$c$  while \cite{Dal:2014nda} uses 1\,\AA/$c$. 
While the differences that originate from differing values of the lifetime cut are believed to be fairly small, one should keep this in mind when comparing values of $p_0$ from different sources.

\subsubsection{Determination of the coalescence momentum $p_0$}

The predicted antideuteron fluxes from dark matter annihilations or decays, as well as from cosmic-ray spallations depend strongly on the only free parameter of the coalescence model: the coalescence momentum $p_0$. Currently, this parameter cannot be calculated from first principles, and should therefore be determined from fitting the predictions of the event-by-event coalescence model described in the previous section to the experimental results on antideuteron production. 

\begin{figure}
\centering
\includegraphics[width=1\linewidth]{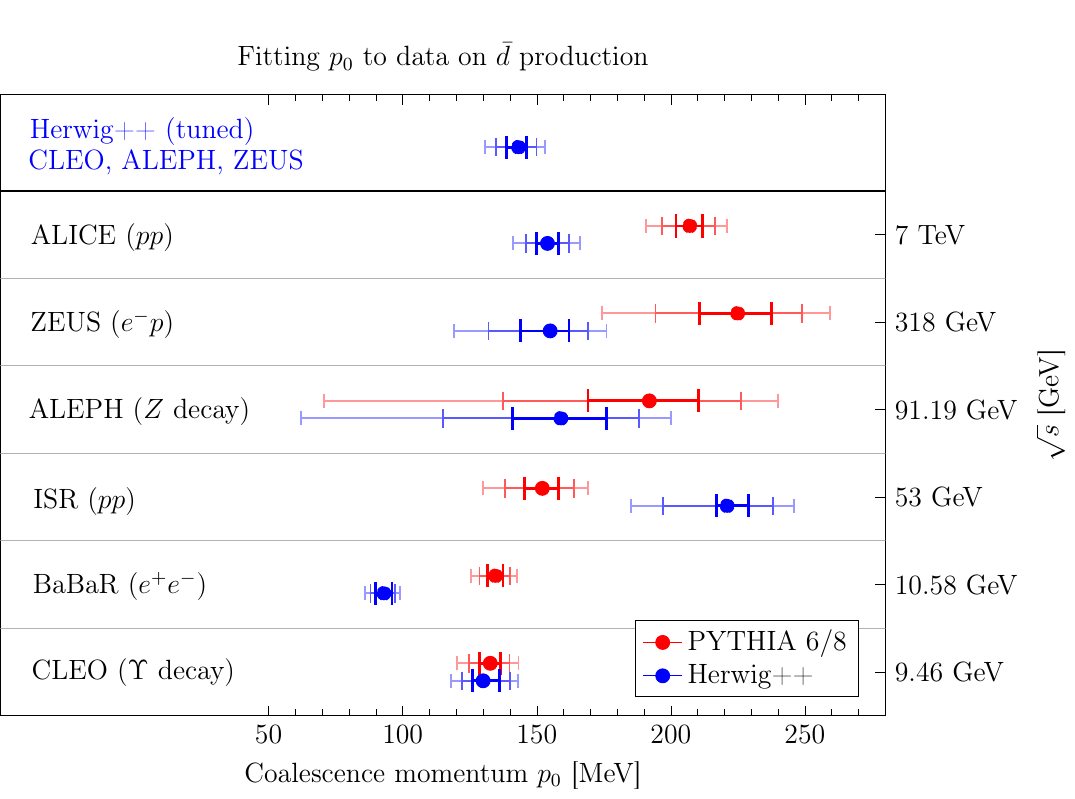}
\caption{Results of fitting the coalescence momentum $p_0$ to different datasets on antideuteron production, based on \cite{Ibarra:2012cc} and \cite{Dal:2014nda}.}
\label{fig:p0_determination}
\end{figure}

The ALEPH~\cite{Schael:2006fd}, CLEO~\cite{Asner:2006pw}, CERN ISR~\cite{Alper:1973my,Henning:1977mt}, ZEUS~\cite{Chekanov2007}, ALICE~\cite{Sharma:2012zz}, and $B\!{\scriptstyle A}\!B\!{\scriptstyle A\! R}$~\cite{Lees:2014iub} experiments have measured antideuteron production, where the recent results of the latter two are discussed in more detail in Sec.~\ref{s-babar} and \ref{s-alice}. The best-fit values of the coalescence momentum inferred from each experiment using {\tt PYTHIA}~\cite{Sjostrand:2006za} or {\tt HERWIG++}~\cite{Bahr:2008pv}, including the 1, 2 and 3 standard deviations error bands, are summarized in Fig.~\ref{fig:p0_determination}. As apparent from the plot, when inferring the coalescence momentum using {\tt PYTHIA} with default settings, there is no value of $p_0$ that can simultaneously fit all the data~\cite{Ibarra:2012cc}. On the contrary, the best-fit value of the coalescence momentum shows a dependence on the underlying process and on the center-of-mass energy involved. Interestingly when using {\tt HERWIG++ }and simultaneously tuning $p_0$ and three hadronization parameters to antideuteron data from ALEPH, CLEO and ZEUS as well as antiproton data, these three datasets can be brought in agreement with a single value of $p_0$~\cite{Dal:2014nda}. However, the data from CERN ISR and $B\!{\scriptstyle A}\!B\!{\scriptstyle A\! R}$ still give rise to a significantly different $p_0$. The hadronization model dependence is discussed more extensively in Sec.~\ref{s-had}.

As a consequence, it is currently inconclusive which value of $p_0$ and which Monte Carlo generator is most suitable for calculating the antideuteron flux from dark matter annihilations or decays as well as for the production in cosmic-ray spallation. This uncertainty has dramatic implications for the search for cosmic antideuterons, due to the strong dependence of the antideuteron yield on the coalescence momentum, $N_{\bar{d}} \propto p_0^3$. This proportionality is exact in the isotropic coalescence model, as can be seen from Eq.~\eqref{eq:iso_coalescence}, but also holds well for per-event coalescence for kinetic energies above $p_0$. 

\subsubsection{Hadronization model dependence}
\label{s-had}

Calibration against experimental data typically yields best-fit $p_0$-values that are smaller than or comparable to $\Lambda_{\rm QCD}$ -- 
the scale at which perturbative QCD breaks down.
The coalescence model is therefore sensitive to non-perturbative effects in the hadronization models of Monte Carlo event generators. This naturally includes (anti)baryon production rates from partons. However, due to the small value of the coalescence momentum $p_0$, coalescence is also highly sensitive to two-particle correlations between the (anti)nucleons.

\begin{figure}
\centering
  \includegraphics[width=0.65\linewidth]{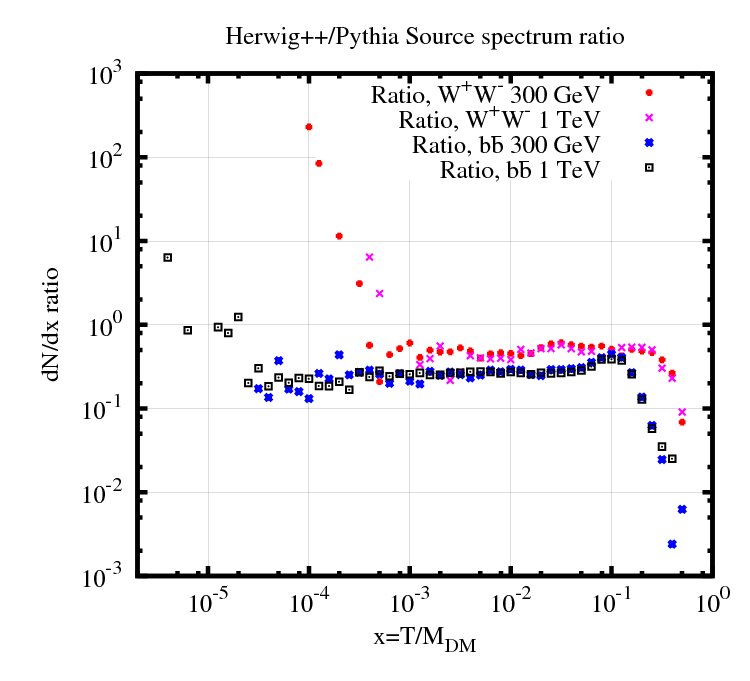}
  \caption{Ratio between the antideuteron spectra from {\tt HERWIG++} and {\tt PYTHIA 6.4} as function of the scaled kinetic energy $x\equiv T/m_{\text{DM}}$. \textit{Source:} Figure taken from \cite{Dal:2012my}.}
  \label{fig:dbar_ratio}
\end{figure}

This issue was first raised in \cite{Dal:2012my}, where the uncertainty from hadronization dependence was estimated by comparing antideuteron spectra produced in the Monte Carlo event generators {\tt HERWIG++} and {\tt PYTHIA 6.4}. 
While {\tt PYTHIA} employs the Lund string hadronization model, {\tt HERWIG++} uses cluster hadronization. Neither {\tt PYTHIA} nor {\tt HERWIG++} are properly tuned for $p$-$p$ collisions at $\sqrt{s} \approx 10$\,GeV, which is the dominant production channel of secondary antideuterons. For that purpose, the better suited Monte Carlo generator DPMJET-III~\cite{Roesler:2000he} has been employed to calculate the antideuteron background flux using the event-by-event coalescence model~\cite{Ibarra2013a}.
In the comparison, the two Monte Carlos were calibrated separately against ALEPH data on antideuteron production in hadronic $Z$-decays~\cite{Schael:2006fd}, yielding different best-fit values for the coalescence momentum $p_0$.
The resulting antideuteron spectra from dark matter annihilations obtained with {\tt HERWIG++} and {\tt PYTHIA} were found to differ by a factor of about 2--4 for most energies, depending on the process in question, but rapidly increasing at high and low energies. The ratio between the {\tt HERWIG++} and {\tt PYTHIA} antideuteron spectra as a function of the scaled kinetic energy $x\equiv T/m_{\text{DM}}$, where $m_{\text{DM}}$ is the dark matter mass, can be seen in Fig.~\ref{fig:dbar_ratio}. For the $\chi\chi\rightarrow b\bar{b}$ annihilation channel, the point at low energies below which the uncertainty rises steeply was found to lie at a kinetic energy of $T\approx10^{-2}$\,GeV, independent of the dark matter mass.  This is below the experimentally relevant energies. For the $\chi\chi\rightarrow W^+W^-$ channel, however, the point was observed to lie at a constant value of the scaled kinetic energy, $x \approx 10^{-3}$, which means that experimentally relevant kinetic energies can lie in the high uncertainty region for dark matter masses above a few TeV. The underlying reason for this fundamental difference is that while the quarks, being colored (and color-connected) objects, radiate off increasing amounts of QCD radiation with increasing energies, the gauge bosons are treated as on-shell. As long as the two $W$s are sufficiently boosted to disallow interactions between their decay products, the antideuteron spectrum equals the sum of their spectra when decaying at rest, boosted according to the speed of the $W$s.

It is important to note that most of the uncertainties affecting the determination of the antideuteron signal from dark matter annihilations or decays are strongly correlated with the corresponding uncertainties in the antiproton flux, which has been measured rather precisely by PAMELA~\cite{Adriani:2012paa}. 
For example, uncertainties stemming from unknown propagation parameters do not significantly affect the antideuteron-to-antiproton ratio expected from dark matter annihilations or decays. 
However, the uncertainty stemming from the two-particle correlations in the different hadronization models is not correlated with the antiproton flux, and hence deserves special attention when evaluating prospects for antideuteron detection. 

As the underlying mechanisms of the hadronization models are very different, one should not expect the two-particle correlations from the two main hadronization models to agree.
There is also no a priori reason to expect the two-particle correlations from one Monte Carlo to be more reliable than the other. However, as pointed out by \cite{Dal:2014nda}, better agreement with data might be achieved for both hadronization models by tuning the  phenomenological parameters in the models specifically for antideuteron production.

Each model has a large number of free parameters, which have been tuned against various experimental data, typically with a focus on the ability to reproduce particle multiplicities and jet physics results. These general tunes do not take two-particle correlations into account to any significant degree, and (anti)proton spectra have no higher weights than other particles in the tuning.
For the purpose of antideuteron production, the spectra of particles that do not contribute to the (anti)nucleon or antideuteron spectra are of no importance.
A tuning of the hadronization parameters that only takes the relevant spectra into account is more likely to be able to reliably reproduce experimentally observed antideuteron spectra.
This has previously been shown to improve the physics description when done with other processes, such as Higgs production~\cite{Richardson:2012bn}.
A further advantage of this procedure is that a combined fit of $p_0$ together with a set of hadronization parameters also allows for an estimation of the uncertainty on the antideuteron spectrum from the hadronization and coalescence tuning.

In \cite{Dal:2014nda}, the coalescence momentum $p_0$ was tuned along with three of the most relevant hadronization parameters in {\tt HERWIG++} against antideuteron data from ALEPH, CLEO and ZEUS, as well as proton/antiproton data from ALEPH and OPAL.
The inclusion of (anti)proton data in the tuning is necessary for a consistent result, as one cannot expect a reliable reproduction of the antideuteron spectrum if the spectra of its constituents are wrong.
The best-fit point of the tune can be found in Table~\ref{tab:fitresults}, and is in reasonably good agreement with the default hadronization parameters for {\tt HERWIG++}. For a short description of the role of the various parameters, please see the original article \cite{Dal:2014nda}. Note that the parameter errors are strongly correlated, and that the given numerical values therefore cannot be naively used to estimate the goodness-of-fit of other parameter points. 

\begin{table}
\centering
\caption{Results from a hadronization parameter fit compared to default values in {\tt HERWIG++}. $p_0$-values are in units of MeV.}
\begin{tabular}{ l c c c }
\hline
\hline
Parameter 		& Default value	& Value at $\chi^2_{\rm min}$ 	& Uncertainty\\[1pt] 
\hline
$p_0$ 	       		&  -- 			&	143.2						& $ ^{+6.2}_{-5.5}$					\\[1ex]
{\tt ClMaxLight}   	& 3.25 			& 	3.03						& $ ^{+0.18}_{-0.15}$				\\[1ex]
{\tt PSplitLight}   & 1.20 			& 	1.31						& $ ^{+0.19}_{-0.32}$				\\[1ex]
{\tt PwtDIquark}  	& 0.49 			& 	0.48						& $ ^{+0.15}_{-0.04}$  				\\\hline
\end{tabular}
\label{tab:fitresults}
\end{table}

The $p_0$ parameter was intended to describe the maximum difference in momenta for which two (anti)nuclei will form ions. The definition in Eq.~\ref{eq:iso_coalescence} allows $p_0$ to be determined from the measurement of (anti)nuclei spectra. However, $p_0$ not only describes the required difference in momenta of the coalescence partners, but also parameterizes a number of other effects. For instance, the available phase space for ion production depends on the available energy in the formation interaction. Comparing the deuteron and antideuteron production in $p$-$p$ interactions at the same energy it is clear that a $pn$-pair can already be produced by $pp\rightarrow pn\pi^+$ while a $\bar p\bar n$-pair requires at least six final-state nucleons ($pp\rightarrow \bar p\bar nnppp$). Therefore, if the coalescence approach is applied to describe the total deuteron and antideuteron production cross sections, $p_0$ describes not only the probability for merging an (anti)proton and an (anti)neutron, but also the probability for (anti)protons and (anti)neutrons to be produced at all. This is especially important for (anti)deuteron production close to the production threshold energy. In a similar way, the production close to the threshold also favors an anti-correlation of (anti)protons and (anti)neutrons for kinematical reasons and causes an additional phase-space suppression. The astrophysical background production of antideuterons in interactions of cosmic rays with the interstellar medium is most likely dominated by the production at the threshold, and thus strongly phase-space suppressed. However, the energy available in dark matter annihilations or decays can be much higher, and therefore the antideuteron production can be much less suppressed. Hence, using different coalescence momentum values for different dark matter masses as well as for different contributing astrophysical background processes is likely to be the right approach. Furthermore, any inconsistencies in the hadronic generators to describe the (anti)proton and (anti)nucleon spectra automatically result in a shift of the $p_0$ parameter. For instance, an underproduction of antiprotons with a generator compared to available data results in a higher $p_0$ value when trying to describe measured (anti)deuteron data. Moreover, (anti)neutron spectra are very challenging to access in typical particle physics detectors. Therefore, the typical approach is to assume that the antiproton and antineutron production cross-sections are equal, and hence it is even more challenging to tune hadronic generators to (anti)neutrons than to antiprotons. Any possible possible isospin asymmetry is an additional component of the $p_0$ value. On the quantum-theoretical level, it is also important to realize that the formation probability in the per-event simulation approach is taken to be exactly 100\% if the $p_0$ condition is met. This is unlikely to be true when considering, e.g., that (anti)proton and (anti)neutron need to have aligned spins to form a stable (anti)deuteron with spin 1. Therefore, trying to separate the actual coalescence process from other conditions that have to be met will be an important challenge for the understanding of (anti)deuteron production.

\subsubsection{Alternatives to the coalescence model\label{s-alt}}

Few alternatives to the coalescence model have been suggested. Recently,the authors of~\cite{Dal:2015sha} discussed a model in which the combination of an antinucleon pair with center-of-mass frame momentum difference $k=|\vec p_{\bar p}-\vec p_{\bar n}|$ into an antideuteron, $\bar p\bar n\to \bar d X$, is a random event with a probability given by the ratio of the cross section of the corresponding scattering process as a function of $k$, to a normalization cross section, $\sigma_{\bar p\bar n\to \bar d X}(k)/\sigma_0$. The normalization $\sigma_0$ is a free parameter to be fixed through calibration against experimental data, analogous to $p_0$ in the coalescence model. The main consequence of this approach is that it shifts the typical value of $k$ involved in the antideuteron formation: where the coalescence model restricts the phase space to $k<p_0$, with $p_0$ typically of the order of $100-200$\,MeV, this model prefers values of $\approx$1\,GeV where resonant production through the delta-resonance can occur. In turn, this affects the antideuteron spectrum, and the authors observe significantly improved fits to antideuteron production in recent $p$-$p$-data from ALICE~\cite{Serradilla:2013yda}.

\subsection{Collider measurements of antideuteron production}
\label{s-coll}

In order to make progress in the understanding of antideuteron formation and the prediction of the primary and secondary antideuteron fluxes, more experimental data and a better understanding of the physics and modeling of antideuteron formation are needed. Recently, new experimental data that can be used for the determination of the coalescence momentum have become available, and further determinations of the coalescence momentum will likely also be possible in the near future.

The $B\!{\scriptstyle A}\!B\!{\scriptstyle A\! R}$ experiment measured the production rate of antideuterons in $\Upsilon (nS)$ decays (with $n=1,2,3$) (Sec.~\ref{s-babar}) as well as in continuum electron--positron annihilations to quarks at $\sqrt{s}=10.58$\,GeV, which can be used to tune $p_0$ in these processes. This determination, especially the one from $e^+ e^- \rightarrow q \bar{q}$, is particularly relevant for the calculation of the antideuteron flux from annihilations of light dark matter particles into quark-antiquark pairs. 

The ALICE experiment at the LHC measured the production rate of antideuterons in $p$-$p$ collisions at center-of-mass energies of $\sqrt{s}=7$, 8\,TeV and will measure at 13\,TeV and 14\,TeV in the future (Sec.~\ref{s-alice}). Although at a different center-of-mass energy, this is precisely the main process participating in the antideuteron production by cosmic-ray spallations, hence the relevance of this process in calculating the secondary antideuteron flux. Collisions involving heavy ions, such as $p$-Pb and Pb-Pb collisions, are less relevant for studies of cosmic antideuterons, due to the different dynamics involved in these processes compared to dark matter annihilations, decays or cosmic-ray collisions.

To reduce the systematic uncertainties in the calculation of the secondary antideuteron flux and to correctly interpret a hypothetical future detection of cosmic antideuterons, an experiment measuring antideuteron production in $p$-$p$ collisions at low center-of-mass energies, i.e.\ $\sqrt{s} \approx 10$\,GeV, would be of utmost importance. This is precisely the main process and the most relevant energy for the production of antideuterons by cosmic-ray spallations, constituting the background for dark matter searches. In addition, studies of antideuteron production in processes like $p$-C are important for constraining instrumental backgrounds and reducing systematic effects. The operational fixed target experiment NA61/SHINE is ideally suited for these tasks (Sec.~\ref{s-na61}).

In addition, the upcoming $\bar{\text{P}}$ANDA experiment at the FAIR collider in Darmstadt, Germany will probe the production of antideuterons in interactions of antiprotons with  different fixed targets (Sec.~\ref{s-panda}). This is also an important process for the prediction of the astrophysical antideuteron background and will be discussed more in Sec.~\ref{subsec:transport}. Using different beam energies and target materials the cross section can be studied as a function of increasing energy starting slightly below the antideuteron production threshold. 

In the longer term, a high luminosity electron--positron linear collider with center-of-mass energies $\sqrt{s} \approx 10$--500\,GeV would be of great interest to investigate the antideuteron production in a single channel at different center-of-mass energies. 
Even reanalyses of LEP data could be helpful here. In particular, such study will allow to compare and tune different Monte Carlo generators and to elucidate whether $p_0$ is a constant or on the contrary scales with the center-of-mass energy. 
Moreover, the final state $q \bar q$ corresponds to the final state in the annihilation or decay in many well-motivated dark matter scenarios. Besides, the proposed super-B factory would deliver much more precise data on antideuteron production in $\Upsilon$ decays, thanks to the much larger luminosity compared to $B\!{\scriptstyle A}\!B\!{\scriptstyle A\! R}$ and BELLE. In particular, in combination with data on $\bar{d}$ production in the $e^+ e^-$ continuum, it could help to test the universality of the coalescence momentum with respect to the underlying hard process.

The following subsections will elaborate more on the different experimental aspects.

\subsubsection{$B\!{\scriptstyle A}\!B\!{\scriptstyle A\! R}$\label{s-babar}}

Recently, the $B\!{\scriptstyle A}\!B\!{\scriptstyle A\! R}$ collaboration reported precise measurements of antideuteron production in hadronic decays of $\Upsilon$ resonances ($\Upsilon(nS)$, $(n=1,2,3)$), as well as the first measurement of antideuteron production in continuum (non-resonant) $e^{+}e^{-}$ annihilation at $\sqrt{s}\approx10.58$\,GeV~\cite{Lees:2014iub}. A brief description of the measurement is provided below as an illustration of the techniques employed to measure antideuteron production rates at $e^{+}e^{-}$ colliders.

The measurements are performed using data collected with the $B\!{\scriptstyle A}\!B\!{\scriptstyle A\! R}$ detector at the asymmetric-energy PEP-II electron--positron collider at SLAC. The data used are collected running at $e^{+}e^{-}$ center-of-mass (CM) energies corresponding to $m_{\Upsilon(2S)}$, $m_{\Upsilon(3S)}$ and $m_{\Upsilon(4S)}$, along with data collected 40\,MeV below each resonance. The asymmetric energy of the collisions means that the center-of-mass (CM) frame is boosted in the laboratory (detector) frame by a factor $\beta\gamma\approx 0.56$.

Deuteron and antideuteron candidates are selected from well-reconstructed charged particle candidates with lab-frame momentum between 0.5 and 1.5\,GeV which are well-contained within the acceptance of the detector's 40-layer multicell drift chamber (DCH) (the $B\!{\scriptstyle A}\!B\!{\scriptstyle A\! R}$ detector is described in detail in \cite{babar-nim}). Thanks to the boost of the collision frame, this corresponds to a wide range of deuteron CM momenta. This momentum range is below the Cherenkov threshold for antideuterons in the quartz radiator bars of the Cherenkov particle identification system (DIRC) used at higher momenta, thus signals in the DIRC are used to veto lighter species. Finally, candidates are required to leave a well-reconstructed pattern of ionization in the DCH, with at least 24 good measurements of deposited charge sampled by the system.

The resulting candidates are binned by momentum in the center-of-mass frame of the colliding $e^{+}e^{-}$ beams. The candidates are weighted by the inverse efficiency for triggering, detection and reconstruction, as well as by a factor which corrects for the fiducial acceptance of the detector at a given deuteron CM momentum. For each candidate, specific ionization ($\text dE/\text dx$) measurements performed in the DCH and in the 5-layer Silicon Vertex Tracker (SVT) are corrected for the differing materials and combined in a weighted average that also accounts for the expected resolution of the two independent measurements. The variable of interest is the distribution of normalized residuals of these averaged measurements with respect to the most probable value given the deuteron mass hypothesis. For real deuterons and antideuterons, this distribution is nearly Gaussian and peaks at zero. The distribution for other  species lies mostly at large negative values and falls rapidly near the signal region.

A weighted unbinned simultaneous maximum-likelihood fit is used to extract the yield of antideuterons in each bin. Matter deuterons are produced at a much higher rate via interactions of primary particles with detector and beampipe materials. These provide a pure high statistics sample which allows the parameters of the signal distribution to be determined in the simultaneous fit by data rather than simulation. This fit is performed separately for $B\!{\scriptstyle A}\!B\!{\scriptstyle A\! R}$'s $\Upsilon(nS)$, $(n=2,3)$ datasets, a subset of the $\Upsilon(2S)$ dataset with well-reconstructed $\Upsilon(2S)\to \Upsilon(1S)\pi^{+}\pi^{-}$ candidates, and in the large $\sqrt{s}=m_{\Upsilon(4S)}$ dataset. The latter is used to measure production in continuum, as direct $\Upsilon(4S)$ or $B$ meson decay to antideuterons is expected to be negligibly small. This assumption is validated by cross-checks against data collected at a CM energy 40\,MeV below $m_{\Upsilon(4S)}$.

As a result of the event-by-event weighting, the fits provide directly the differential cross section (or differential decay rate) for antideuteron production. For $\Upsilon(2,3S)$ decay, below-resonance datasets are fitted simultaneously to subtract any continuum contribution. For $\Upsilon(1S)$, the background subtraction is performed by using sideband regions in the variable \begin{equation}m_{\rm recoil}=\sqrt{(E_{\rm beam}-E_{\pi\pi})^2 - (\vec{p}_{\rm beam}-\vec{p}_{\pi\pi})^2}.\end{equation} For the continuum measurement, the below-resonance yields are added to the yields in the on-resonance dataset.

\begin{figure}
\centering
\includegraphics[width=0.65\linewidth]{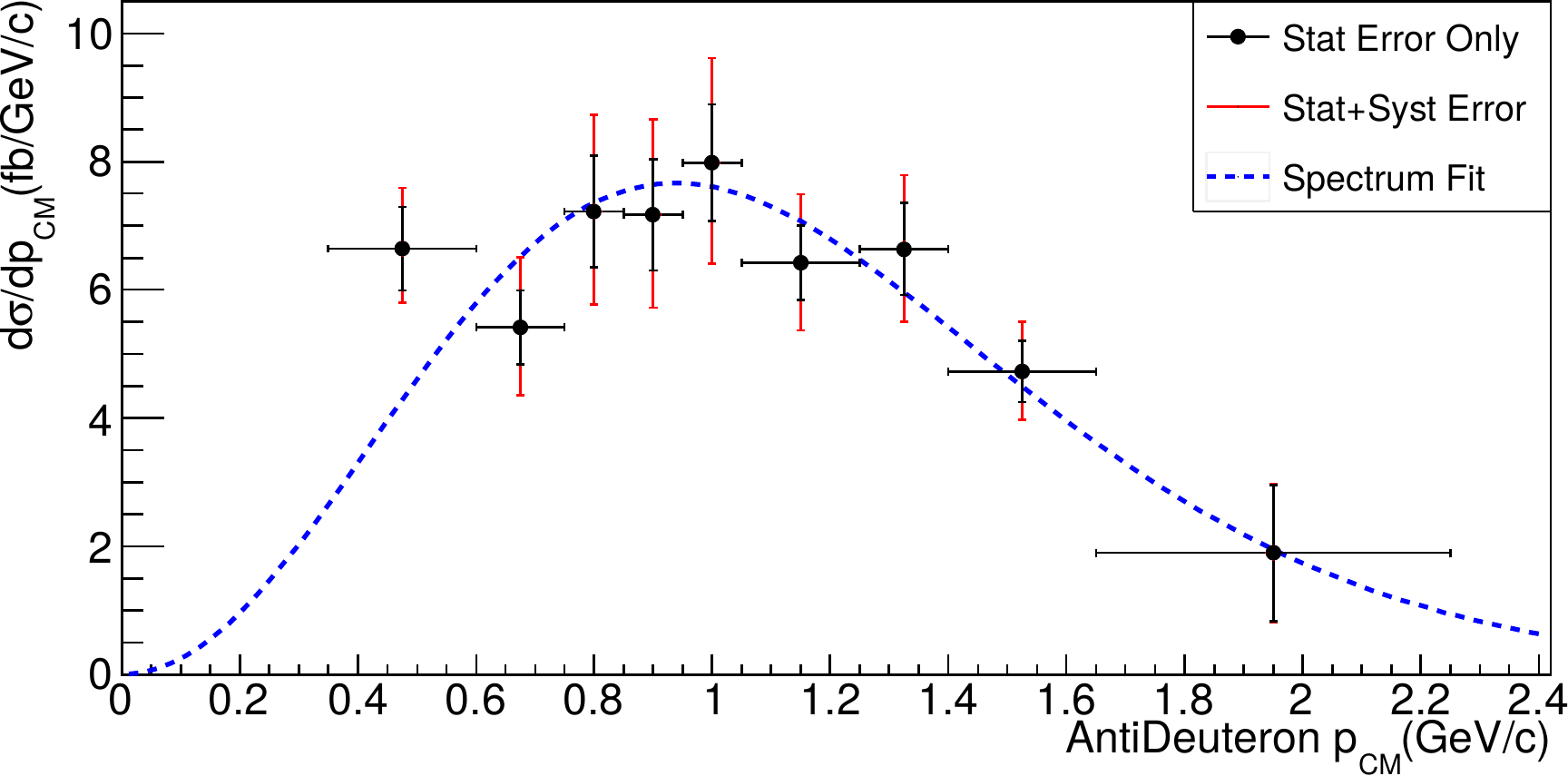}
\caption{Measured antideuteron differential spectra in $e^{+}e^{-} \to q\bar{q}$ at a CM energy of $\sqrt{s}\approx10.58$\,GeV. The points with inner (black) error bars give the measurements and their associated statistical uncertainties, the outer (red) error bars give the quadratic sum of the statistical and systematic uncertainties, and the dashed (blue) curve shows a fit to determine the total rate, from \cite{Lees:2014iub}.}
\label{fig:babarresult}
\vspace{-0.2cm}
\end{figure}

The results for continuum $e^{+}e^{-}$ annihilation are shown in Fig.~\ref{fig:babarresult}. The results are in good agreement with previous measurements of antideuteron production in $\Upsilon(1S)$ and $\Upsilon(2S)$ decay, and represent the first measurement of production in continuum $e^{+}e^{-}$ annihilation with $\sigma (e^{+}e^{-} \to \bar{d}X)=(9.63 \pm 0.41 {}^{+1.17}_{-1.01})$\,fb at $\sqrt{s}\approx10.58$\,GeV. Normalizing the continuum cross section to $\sigma_{e^{+}e^{-} \to {\rm hadrons}}$, one finds a suppression relative to production in hadronic $\Upsilon$ decay of $(3.01 \pm 0.13{}^{+0.37}_{-0.31})\cdot 10^{-6}$, as expected by the higher baryon production observed in gluon fragmentation relative to quark fragmentation.

\subsubsection{ALICE\label{s-alice}}

The results presented in this section are obtained from A Large Ion Collider Experiment (ALICE) at the LHC. Its performance and the description of its various subsystems are discussed in~\cite{alice}. Due to its unique particle identification capabilities ALICE is ideally suited to measure rarely produced nuclei. The collision energy reached at the LHC provides the opportunity to measure nuclei and the corresponding antiparticles in unprecedented abundances, although the measurement is challenging as the production probability decreases with increasing particle mass. The results presented here are obtained with the data of $p$-$p$ collisions at $\sqrt{s}$ = 7\,TeV recorded in 2011, Pb-Pb collisions at $\sqrt{s_{NN}}$ = 2.76\,TeV  recorded in 2010 and $p$-Pb collisions at $\sqrt{s_{NN}}$ = 5.02\,TeV recorded at the beginning of 2013. The following data analysis and discussion focuses on deuteron production. Symmetry arguments can be used to transfer this discussion to the production mechanism of antideuterons.

\begin{figure}
\centering
\includegraphics[width=0.65\linewidth]{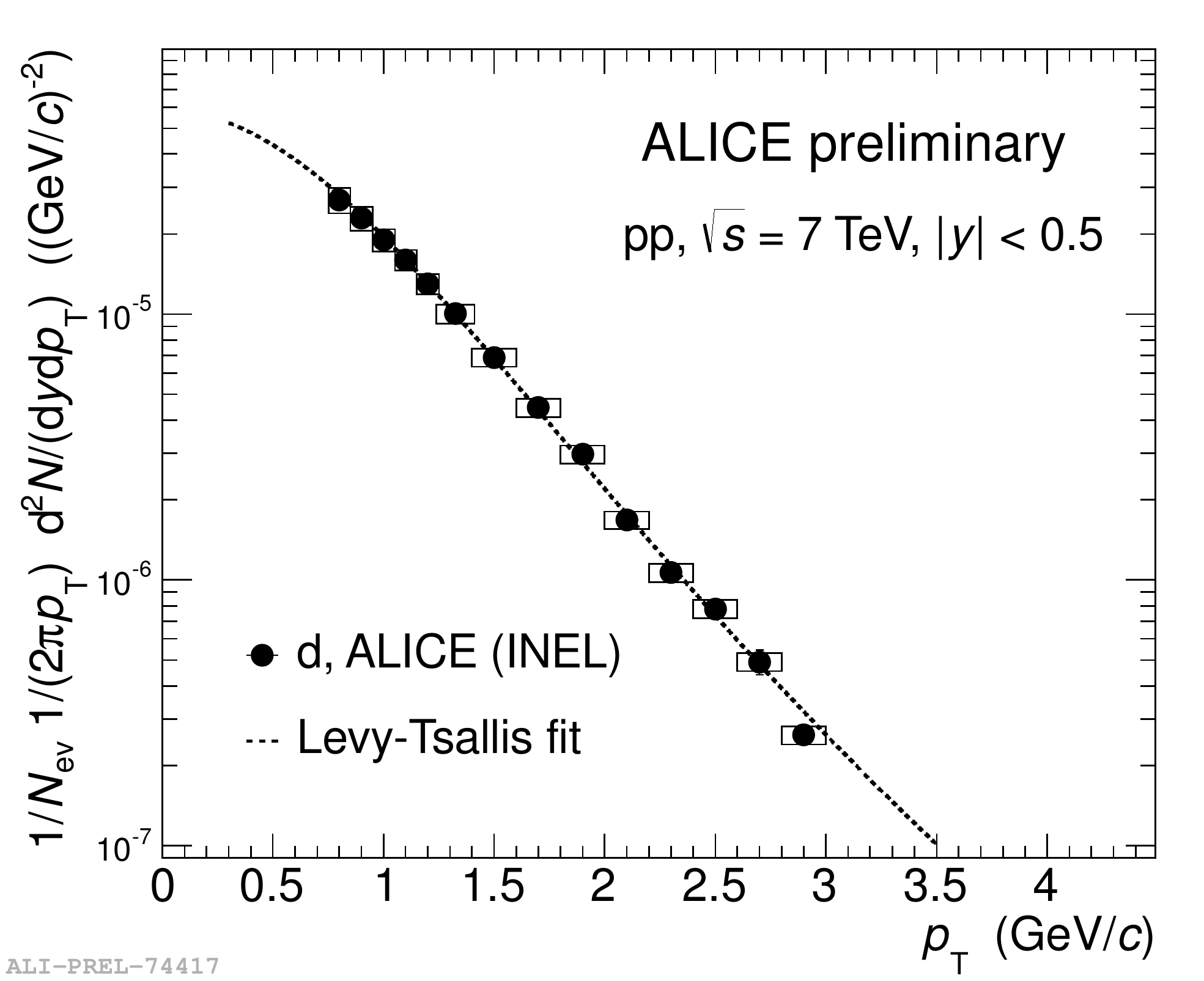}
\caption{Deuteron spectrum for $p$-$p$ collisions at $\sqrt{s}$ = 7\,TeV. The dotted line represents the fit with the Levy--Tsallis function.}
\label{fig:deut_p}
\end{figure}

Nuclei and antinuclei are identified over a wide transverse momentum ($p_{T}$) range using the combined information of the specific energy loss ($\text d E/\text d x$) measurement in the Time Projection Chamber (TPC)~\cite{alice} and the velocity measured by the Time Of Flight detector (TOF)~\cite{tof}. The measured energy loss signal of a track in the TPC is required to be within a three standard deviation region around the expected value for a given mass hypothesis described by the Bethe--Bloch formula~\cite{bethe}. Since these curves start to overlap at high momenta, the velocity measurement with the TOF is used in addition to allow for clear identification at $p_{T} > 0.8$\,GeV/c. Fig.~\ref{fig:deut_p} shows the efficiency and acceptance corrected deuteron spectra for $p$-$p$ collisions at  $\sqrt{s}$ = 7\,TeV. The measurement of deuteron transverse momentum spectra has been also performed using the data collected during the $p$-Pb run. The deuteron spectra show a hardening with increasing multiplicity/centrality that is qualitatively similar to the proton spectra. In heavy ion collisions this hardening is commonly attributed to radial flow~\cite{flow}.

\begin{figure}
\centering
\includegraphics[width=0.65\linewidth]{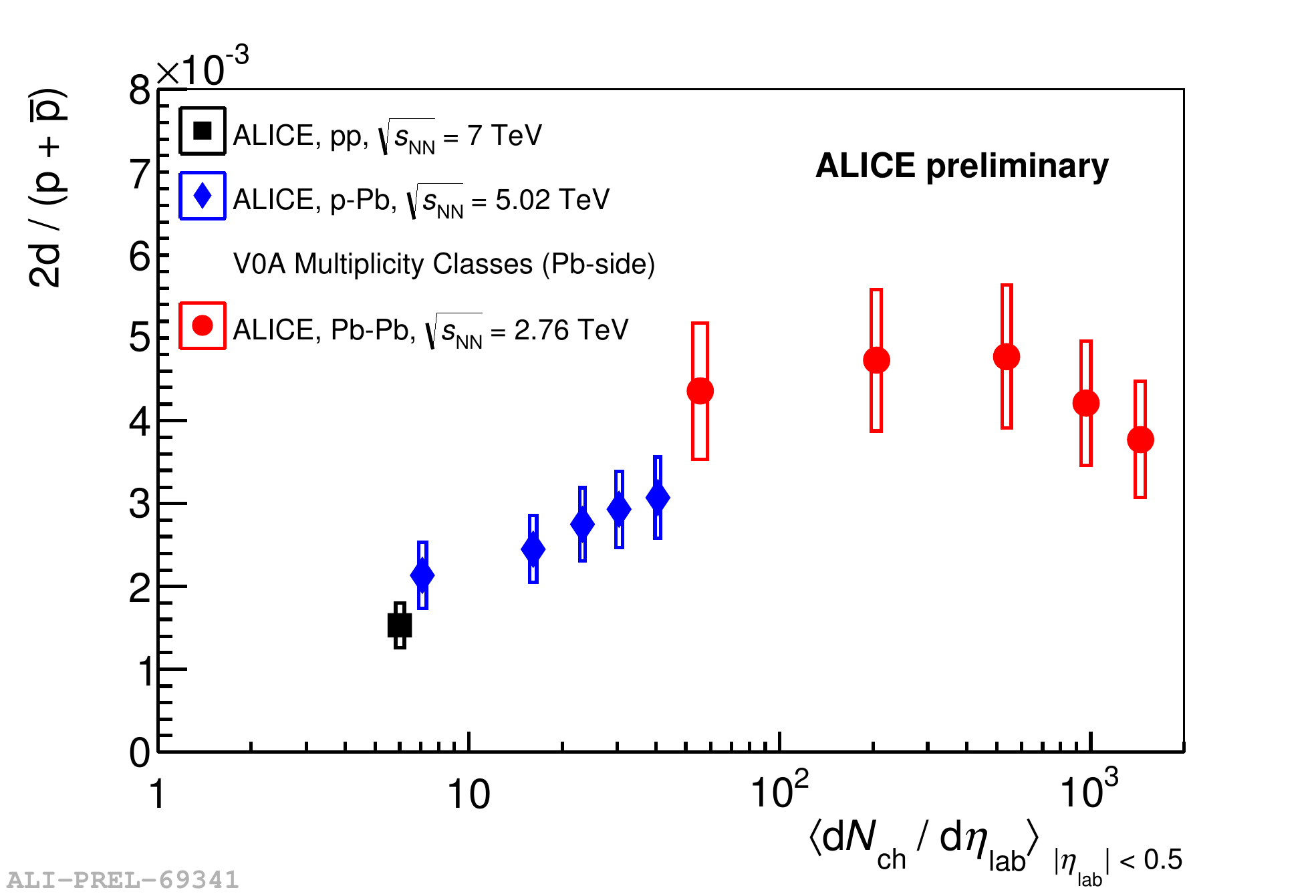}
 \caption{Deuteron-to-proton ratio as a function of charged-particle multiplicity at midrapidity for different colliding systems at different energies.}
\label{fig:deut_antideut_ratio}
\end{figure}

Fig.~\ref{fig:deut_antideut_ratio} shows the deuteron-to-proton ratio for minimum bias $p$-$p$ collisions and as a function of the multiplicity for $p$-Pb and Pb-Pb. The ratio rises with multiplicity until a saturation within errors in Pb-Pb collisions is reached. The transition between the different collision systems suggests that the $d/p$ ratio is in part determined by the event multiplicity, at least for smaller systems ($N<100$).

The production spectra of light nuclei can be understood based on the coalescence approach assuming that deuterons (and other light nuclei) are produced by protons and neutrons that are close in phase space. In the most naive picture, this would lead to an increased deuteron production for higher nucleon multiplicities. The increase of the deuteron-to-proton ratio with the charged particle multiplicity in Fig.~\ref{fig:deut_antideut_ratio} is consistent with this picture for small colliding systems like $p$-$p$ and $p$-Pb.  However, the $d/p$ ratio for Pb-Pb collisions is constant with increasing centrality, although the nucleon multiplicity increases.

A possible explanation of the Pb-Pb results can be that the increasing nucleon multiplicity is counterbalanced by the increasing source volume, leading to a constant nucleon density. This is also consistent with the rising $d/p$ ratio with multiplicity in $p$-Pb collisions, if the effect of the increasing nucleon multiplicity dominates over the effect of the increasing source volume. 

In addition to the coalescence model presented in Sec.~\ref{s-coa}, the production mechanism of antinuclei in heavy ion collision can also be discussed in a thermal model~\cite{andronic,cleymans}. The production yields depend exponentially on the chemical freeze-out temperature $T_{\text{chem}}$ and the mass $m$: 
\begin{equation}\text dN/\text dy \approx \exp\left(\displaystyle-\frac{m}{T_{\text{chem}}}\right).\end{equation}
Due to their large masses the abundance of nuclei is very sensitive to $T_{\text{chem}}$. An important question is whether the nuclei are produced  at the chemical freeze-out or at a later stage via coalescence. Besides the constant $d/p$ ratio a key observation is that in Pb-Pb collisions the absolute production yields ($\text d N/\text d y$) of light nuclei are in good agreement with thermal model calculations, as shown in \cite{floris}. On the other hand the highest $d/p$ ratio obtained in $p$-$p$ collisions is about half the value predicted by a thermal model in central AA collisions, disfavoring the statistical description based on thermal equilibrium in $p$-$p$ collisions at the LHC energies. Further studies are needed to establish whether the fast expansion conserves the particle ratios and which additional conditions in the coalescence model are required to describe the constant particle ratio in Pb-Pb.

\subsubsection{NA61/SHINE\label{s-na61}}

The NA61/SHINE (SPS Heavy Ion and Neutrino Experiment) experimental facility at the Super Proton Synchrotron (SPS) at CERN is used to study collisions of protons and heavy nuclei with fixed targets in a wide incident beam momentum range (e.g., 13--158\,GeV/$c$ protons on a hydrogen target)~\cite{na61}. Antideuterons are also produced in these interactions and NA61/SHINE is currently the best operational experiment to study antideuteron production from the threshold on. Different subdetectors are needed for the particle identification. Numerous of these components were inherited from its predecessor, the NA49 experiment~\cite{na49}. The main upgrades from NA49 to NA61/SHINE include a forward time-of-flight detector wall, improved time projection chamber readout, and five times better energy resolution with the projectile spectator detector.

A set of scintillation and Cherenkov counters as well as beam position detectors upstream of the spectrometer provide timing reference, identification and position measurements of incoming beam particles. A trigger scintillator counter placed downstream of the target is used to select events with collisions in the target area. The main tracking devices of the spectrometer are large volume time projection chambers (TPC). Two of them are located in the magnetic fields of two super-conducting dipole magnets with a maximum combined bending power of 9\,Tm, which corresponds to about 1.5\,T in the first and 1.1\,T in the second magnet. Two large TPCs are positioned downstream of the magnets symmetrically to the beam line. The fifth small TPC is placed between the two vertex time projection chambers on the beam line. It closes the gap between the beam axis and the sensitive volumes of the other TPCs. The particle identification capability of the TPCs is based on measurements of the specific energy loss, $\text d E/\text d x$, and is augmented by velocity measurements using time-of-flight detectors ($\sigma_t=110$\,ps). The high resolution forward calorimeter, the projectile spectator detector, measures energy flow around the beam direction, which in nucleus-nucleus collisions is primarily given by the projectile spectators. NA61/SHINE uses various solid nuclear targets and a liquid hydrogen target. The targets are positioned about 80\,cm upstream of the sensitive volume of the first vertex time projection chamber.

NA61/SHINE can be used to improve the parameters of the underlying (anti)deuteron coalescence model. Proton-proton interactions with incident momentum between 13 and 158\,GeV/$c$ were already recorded in 2009 and 2011 and data with higher energies will be taken in 2015. The low-energy cosmic-ray antideuteron background flux ($<1$\,GeV/$n$) is dominated by products of proton interactions with interstellar medium hydrogen at somewhat higher kinetic energies (2--5\,GeV/$n$), which is then shifted towards lower energies due to propagation through the Galaxy~\cite{Dal:2014nda}.  Calculations predict that the most relevant cosmic-ray production happens between 40 and 400\,GeV/$c$. Therefore, the SPS energies from 9 to 400\,GeV/$c$ are ideally suited for a study relevant to the antideuteron search~\cite{Ibarra2013a}. In addition, NA61/SHINE scans different ion collisions and different energies like $p$-C, $\pi$-C, Be-Be, Ar-Sc. Especially the production of antideuterons in $p$-C interactions is important to understand systematic experimental effects due to antideuteron production inside experiments.

The NA61/SHINE (anti)deuteron production can be studied as a function of rapidity and transverse momentum and used to calibrate the coalescence model. NA61/SHINE can also be used to test the reasonable assumption that the  merging of $pn$ and $\bar p \bar n$ pairs  should  proceed similarly for nucleons and antinucleons. For this purpose, it will be important to also measure deuterons  with NA61/SHINE at the same time. However, as discussed at the end of Sec.~\ref{s-had}, only specific deuteron channels allow to draw direct conclusions for antideuterons. For instance, the antideuteron production cross section in $\bar{p}p\rightarrow \bar{d}np$ should be the same as for deuteron production in $pp\rightarrow d\bar{n}p$. The advantage of measuring $pp\rightarrow d\bar{n}p$ in comparison to $pp\rightarrow\bar{d}X$ is that the cross-section is larger, because only four instead of six (anti)nucleons have to be produced. Therefore, using the $pp\rightarrow d\bar{n}p$ channel can already answer the question of how (anti)deuteron coalescence depends on the available energy. Another important cross-check for the Monte Carlo generators will be the measurement of the yield of antiprotons with the same data. The Pb-Pb data of the predecessor of NA61/SHINE, NA49, were already analyzed for antideuterons and successfully demonstrated that antideuterons can be identified in the experimental setup~\cite{dbarna49}. However, it is theoretically not clear how to apply the event-by-event coalescence model to proton-nucleus or nucleus-nucleus collisions because the physics of the creation of secondary nucleons is very different in this case.

\subsubsection{The $\bar{\text{P}}$ANDA fixed target experiment\label{s-panda}}

The upcoming $\bar{\text{P}}$ANDA experiment at the FAIR collider in Darmstadt, Germany will study the interactions of a 1.5--15\,$\text{GeV}/c$ antiproton beam with different fixed targets (e.g., hydrogen, deuterium, or even gold), and thus will also probe antideuteron physics~\cite{pandaphys,fair}. It will reach luminosities of up to $2\cdot10^{32}$\,cm$^{-2}$\,s$^{-1}$ corresponding to a $\bar{p}p$-annihilation rate of $2\cdot10^7$\,s$^{-1}$. The $\bar{\text{P}}$ANDA detector is a multi-purpose particle physics detector with $4\pi$ acceptance. The detector is divided into two main components, namely the target and the forward spectrometer.

The target spectrometer surrounds the interaction point and is split by the target injection pipe into two half shells. The typical target will be gaseous hydrogen or other heavier gases as well as solid targets for hypernuclei studies. The detector concept follows a layer structure. The interaction point is surrounded by a microvertex detector including radiation hard silicon pixels. The next layer is the central tracker and will  consist of a 24-layer straw tube tracker. The forward direction will be covered by three layers of GEM tracking stations. Particle identification of slow particles ($<1\,\text{GeV}/c$) for the barrel region will be done by a time of flight detector with a timing resolution of 50--100\,ps and the specific energy loss in the central tracker. Faster particles are identified by Cherenkov detectors for the barrel and forward endcap. The next layer is formed by the electromagnetic calorimeter made out of lead tungstate with a short radiation length and Moli\`ere radius to assure a compact design. All the former mentioned subdetectors are contained in a 2\,T magnetic field provided by a superconducting solenoid coil. Outside of the magnetic field as part of the magnetic return yoke a fine segmentation with aluminum drift tubes serves as a range tracking system for muons and pions. In the forward direction particles are deflected by a dipole magnet. The particle identification of the forward spectrometer is similar to the target spectrometer. The forward tracker measures the deflection in the magnetic field with three pairs of tracking detectors based on straw tubes. Particle identification will be done by a ring image Cherenkov detector and a time of flight system followed by a lead-scintillator sandwich electromagnetic calorimeter. Forward muons are again detected with a range tracking system laid out for higher momenta. At the very end of the line, a luminosity detector measures elastically scattered antiprotons with four silicon detector layers.

Using different beam energies and target materials the cross section can be studied as a function of increasing energy starting slightly below the antideuteron production threshold with hydrogen. Furthermore, by analyzing antiproton--deuterium interactions the inelastic scattering cross section of antideuterons with matter can be studied assuming $CPT$-symmetry. 

\subsection{The coalescence model for antihelium\label{s-coaantihe}}
\label{subsec:antihelium_coalescence}

The yield of antinuclei  produced by dark matter annihilation or decay processes decreases rapidly with the antinucleus mass number. As a rough estimate, one can assume that the probability to have coalescence is reduced by a factor ${\mathcal O}(10^{-4})$ for each additional antinucleon that participates in the process. This is why only antihelium formed by three antinucleons ($^3\overline{{\rm He}}$), will be taken into account. The production rate of $^4\overline{{\rm He}}$ is even more suppressed and will not be discussed further. Artificially-produced antihelium-3 was discovered in 1971~\cite{Antipov1971235} and antihelium-4 in 2011~\cite{2011Natur.473..353S}.

The coalescence process that leads to the production of $^3\overline{{\rm He}}$ can involve either two antiprotons and one antineutron (in which case the antihelium is produced directly) or two antineutrons and one antiproton (in which case the antihelium comes from the decay of the antitritium, a phenomenon that occurs on a time scale that is much shorter than those characterizing Galactic propagation and can therefore be considered as instantaneous for our purposes). 
Coalescence in the $(\bar{p},\bar{p},\bar{n})$ channel is expected to be suppressed by Coulomb repulsion due to the presence of two antiprotons~\cite{Chardonnet:1997dv}. If this suppression is not taken into account, the two different coalescence channels instead have practically the same yield~\cite{Carlson:2014ssa, Cirelli:2014qia}.

Two Monte Carlo coalescence models for antihelium have been proposed.
The first requires that the relative momenta of all sets of the three nucleons are smaller than the coalescence momentum $p_0$~\cite{Cirelli:2014qia},
while the second requires that each of the relative momenta lie within a ``minimal bounding momentum sphere" with diameter $p_0$~\cite{Carlson:2014ssa}. 
The flux predicted using each model can vary by, at most, 15\%~\cite{Carlson:2014ssa}.

Laboratory measurements of the antihelium production rate are extremely scarce and related to processes that are quite different from dark matter annihilation, such as proton--nucleus~\cite{Vishnevsky:1974ks,Bozzoli:1978ud,Bussiere:1980yq} or nucleus--nucleus collisions~\cite{Lemaire:1979aa}. Therefore, the value of the coalescence momentum used for dark matter signal predictions is highly uncertain. Ref.~\cite{Cirelli:2014qia} adopted the same fiducial coalescence momentum determined for the antideuteron case, i.e.\ 195\,MeV; Ref.~\cite{Carlson:2014ssa} instead rescaled the coalescence momentum derived for the antideuteron case by a suitable factor. This factor can either be determined by assuming that the coalescence momentum is proportional to the square root of the binding energy $B$ of the bound state: 

\begin{equation}
p_0^{^3\overline{\rm He}}=\sqrt{\frac{B_{\bar d}}{B_{^3\overline{\rm He}}}} \; p_0^{\bar d}=(0.357\pm0.059)\,{\rm GeV}
\end{equation} 

or by assuming a proportionality between $p_0^{^3\overline{\rm He}}$ and $p_0^{\bar{d}}$ inferred by comparing theoretical expectations with the experimental measurements of \cite{Lemaire:1979aa}: 

\begin{equation}
p_0^{^3\overline{\rm He}} = 1.28\, p_0^{\bar{d}} = (0.246\pm0.038)\,{\rm GeV}\,.
\end{equation} 

To ensure that the $\bar p$ and $\bar n$ originate from the same production vertex, as has been discussed for the antideuteron case, all decays of long-lived particles in the Monte Carlo event generator are turned off.

\section{Antideuteron propagation\label{s-4}}

After coalescence, antideuterons must propagate through the Galactic magnetic fields, plasma currents, and the interstellar medium to reach our solar system. They can then be deflected by the solar magnetic field or suffer adiabatic energy losses in the solar wind. Finally, antideuterons can be deflected away from balloon-borne or satellite detectors by the Earth's geomagnetic field and/or interact with Earth's atmosphere. Each of these processes impacts the predicted antideuteron flux seen by experiments, typically calculated at the top of atmosphere (TOA). Galactic propagation is an important uncertainty in predicting antideuteron fluxes at Earth, with well-motivated choices of transport model parameters predicting antideuteron fluxes that can differ by an order of magnitude. Recent positron data exclude the MIN Galactic propagation model, which predicts the lowest antideuteron flux levels at Earth, supporting higher antideuteron flux predictions that could be detected by GAPS or AMS-02.

The propagation of antideuterons in the Galactic environment is detailed in Sec.~\ref{subsec:transport_ISM}. Transport in the solar environment, geomagnetic field, and atmosphere is discussed in Sec.~\ref{subsec:transport_HELIO}-\ref{s-atmo} and followed by a brief introduction of antideuteron interactions with detectors (Sec.~\ref{s-geant}). The section concludes by elaborating on a possible extragalactic origin of antideuterons (Sec.~\ref{s-extrag}).

\subsection{Transport in the Galactic environment}
\label{subsec:transport_ISM}

\subsubsection{Cosmic-ray transport equation}
\label{subsec:transport}

Once produced, antinuclei propagate inside the halo of the Milky Way, where they can be deflected by the Galactic magnetic fields and local plasma currents, lose energy through interactions with the interstellar medium (ISM) and magnetic fields, and be destroyed by fission or annihilation. Although the magnetic fields are significantly turbulent, cosmic-ray transport can still be modeled by spatial diffusion, which plays a major role above about 10\,GeV, and convection, which contributes primarily at lower energies. Cosmic-ray fluxes are thus determined by a transport equation as given, e.g., in~\cite{1990acr..book.....B}.

\begin{equation}
\vec{\nabla} \! \cdot \! \left\{ - K \, \vec{\nabla} N + \vec{V_c} \, N \right\} +
\frac{\partial}{\partial E} \! \left\{
f_{\rm o} N - s_{\rm o} \frac{\partial N}{\partial E} \right\} =
q_{\rm src}(\vec{r} , E) - \Gamma_{\rm \! dst} N \,.
\label{eq:transport_CR_1}
\end{equation}

The left-hand side describes spatial diffusion ($K$) and convection ($V_c$), and the first and second order energy transport terms ($f_{\rm o}$ and $s_{\rm o}$). The right-hand side corresponds to the source term ($q_{\rm src}(\vec{r} , E)$), and the sink term describing destruction in the interstellar medium (ISM) ($\Gamma_{\rm \! dst}$). 

The diffusion coefficient $K$ depends a priori on the location $\vec{r}$ and energy $E$ of the particles. It is related to the power spectrum of the magnetic inhomogeneities, which is poorly known. Although many forms have been proposed, it is often assumed to depend only on energy as~\cite{1997A&A...321..434P,don13}:

\begin{equation}
K(\vec{r} , E) = \beta \, K_0 \left( \! \frac{\mathcal{R}}{1\,{\rm GV}} \! \right)^{\delta} \,,
\label{eq:diff_K}
\end{equation}

where the normalization $K_0$ and the exponent $\delta$ are assumed to be constant parameters, and $\beta$ and ${\mathcal R} = {pc}/{(Ze)}$ denote the cosmic-ray velocity and rigidity, respectively.

At energies below about 10\,GeV, additional processes come into play. Antinuclei are swept by the convection of the local plasma, which results from stellar winds, and drift with a velocity $\vec{V_c}$ upwards and downwards with respect to the Galactic Disk. The conservation of the cosmic-ray current in energy space yields the first and second order energy transport terms. The first order term, $f_{\rm o}(\vec{r} , E)$, corresponds to the sum of four processes: ionization, Coulomb, and adiabatic losses, and first-order reacceleration. Ionization losses take place in the neutral ISM, while Coulomb losses are dominated by scattering off thermal electrons~\citep{1994A&A...286..983M,1998ApJ...493..694M} in the completely ionized plasma.  The spatial dependence of these two terms is thus encoded in the distribution of the neutral and ionized gas.  Adiabatic losses are due to the expanding stellar winds, and their spatial dependence is thus related to the gradient of $\vec{V_c}$. 

Both the contribution to $f_{\rm o}(\vec{r} , E)$ from reacceleration and the second-order term $s_{\rm o}(\vec{r} , E)$ originate from the motion of knots in the turbulent Galactic magnetic fields. In addition to being responsible for spatial diffusion, these Alfv\'enic waves also cause energy drift and reacceleration.  A minimal reacceleration scheme is well-motivated~\citep{1994ApJ...431..705S} and allows us to calculate the $f_{\rm o}$ and $s_{\rm o}$ coefficients. Similar, albeit more empirical, forms have also been used~\citep{1995ApJ...441..209H,Maurin:2001sj}. In all these models, the strength of the reacceleration is mediated via the Alfv\'enic speed $V_a$ of the scatterers.

On the right-hand side of Eq.~(\ref{eq:transport_CR_1}), $q_{\rm src}(\vec{r} , E)$ denotes the rate at which antinuclei are produced. Primary antinuclei refer to those produced by the annihilation of dark matter species in the Milky Way halo. For antideuterons, this production rate can be written as~\cite{Donato:1999gy,Ibarra:2012cc}:

\begin{align}
q_{\dbar}^{\rm pri}(\vec{r} , E_{\dbar}) &= \frac{1}{2} \, \langle \sigma v \rangle \,
\frac{\text dN_{\dbar}}{\text dE_{\dbar}}
\left( \frac{\rho_{\text{DM}}(\vec{r})}{m_{\text{DM}}} \right)^2 & & \text{for dark matter annihilation,} 
\label{q_primary} \\
q_{\dbar}^{\rm pri}(\vec{r} , E_{\dbar}) &= \frac{1}{\tau_{\text{DM}}} \,
\frac{\text dN_{\dbar}}{\text dE_{\dbar}} \,
\frac{\rho_{\text{DM}}(\vec{r})}{m_{\text{DM}}} & & \text{for dark matter decay.} 
\label{q_primary_dec} 
\end{align}

Here, $\langle \sigma v \rangle$ is the thermally-averaged dark matter pair-annihilation cross section, $\tau_{\text{DM}}$ is the dark matter lifetime, ${\text dN_{\dbar}}/{\text dE_{\dbar}}$ is the injection spectrum of antideuterons produced by individual annihilations or decays, $\rho_{\text{DM}}(\vec{r})$ is the dark matter density at location $\vec{r}$, and $m_{\text{DM}}$ is the dark matter particle mass. Secondary antinuclei refer to those produced via interactions of high-energy cosmic-ray protons and helium nuclei on the ISM. These are the irreducible background for any dark matter annihilation signature. In the case of antideuterons produced by cosmic-ray protons impinging on hydrogen atoms at rest, the secondary production rate can be written as~\cite{Duperray:2005si,Chardonnet:1997dv}:

\begin{equation}
q_{\dbar}^{\rm sec}(\vec{r} , E_{\dbar}) = {\displaystyle \int_{E^{0}_{p}}^{+ \infty}} \,
{\displaystyle \frac{\text d\sigma_{p {\rm H} \to {\dbar} X}}{\text dE_{\dbar}}}(E_{p} \! \to \! E_{\dbar}) \, n_{\rm H}(\vec{r}) \,
v_{p} \, N_{p}(\vec{r} , E_{p}) \, \text dE_{p} \,,
\label{q_secondary}
\end{equation}

where $n_{\rm H}$ is the hydrogen density in the ISM, $v_{p}$ is the velocity of the protons, and $N_{p}$ is the differential proton density, $N_{p}(\vec{r} , E) \equiv {\text dn_{p}}/{\text dE_{p}}$. The differential production cross section, $d\sigma_{p {\rm H} \to {\dbar} X} / dE_{\dbar}$, is computed using the coalescence scheme described in Sec.~\ref{s-3} and can be adapted to account for production via helium in the ISM and in the cosmic radiation field.

Also cosmic-ray antiproton collisions with the ISM can produce antineutrons, which can then merge with cosmic-ray antiprotons into antideuterons. Production from proton and antiproton collisions with the interstellar medium have different kinematics. For the proton production enough energy for at least six nucleons must be present while for antiprotons only four are needed because the incident antiproton can coalesce with an antineutron collision product. For the case of impinging protons and antiprotons on interstellar medium hydrogen at rest, the corresponding kinetic energy production thresholds are 17\,GeV and 7\,GeV, respectively. Therefore, the two dominant production processes are $pp\rightarrow\bar dX$ and, although the antiproton flux is strongly suppressed compared to protons, $\bar pp\rightarrow \bar dX$. The antiprotonic production is especially for the lower energy part of the spectrum non-negligible \cite{Ibarra2013a}.

As antinuclei propagate inside the Galactic Disk, they can be destroyed through fission or annihilation with the ISM. In the case of antideuterons, the annihilation rate $\Gamma_{\rm \! dst}$ is given by~\cite{Duperray:2005si}:

\begin{equation}
\Gamma_{\rm \! dst}^{\dbar} = (n_{\rm H} + 4^{2/3}n_{\rm He}) \, v_{\dbar} \,
\sigma_{\rm ine}(\dbar p \rightarrow X) \, ,
\label{eq:spallation}
\end{equation}

where the hydrogen and helium densities in the ISM are assumed homogeneous with $n_{\rm H}$ = 0.9\,cm$^{-3}$ and $n_{\rm He}$ = 0.1\,cm$^{-3}$, $v_{\dbar}$ is the velocity of the antideuteron, and the cross section $\sigma_{\rm ine}$ accounts for inelastic interactions of antideuterons on the ISM. To calculate this cross section, which has not yet been measured, one can follow the assumptions outlined in \cite{Brauninger:2009pe}. See also the discussion in~\cite{Grefe:2015jva}.

Antideuterons rarely survive inelastic collisions. However, they can sometimes undergo inelastic, but non-annihilating collisions, and transfer enough energy to excite the target proton as a $\Delta$ resonance. As they typically lose a substantial fraction of their initial energy, these antideuterons are effectively re-injected into the total flux with a degraded momentum. This mechanism redistributes antideuterons towards lower energies and flattens their spectrum, as first remarked by~\cite{Duperray:2005si,Bergstrom:1999jc}. The antideuterons that result from these inelastic collisions are referred to as tertiary antideuterons, with a production rate given by~\cite{Duperray:2005si}:

\begin{eqnarray}
q_{\dbar}^{\rm ter}(\vec{r} , E_{\dbar}) & = & {\displaystyle \int_{E_{\dbar}}^{+ \infty}} \,
{\displaystyle \frac{\text d\sigma_{\dbar p \to \dbar X}}{\text dE_{\dbar}}}(E'_{\dbar} \! \to \! E_{\dbar}) \,
n_{\rm H}(\vec{r}) \, v'_{\dbar} \, N(\vec{r} , E'_{\dbar}) \, \text dE'_{\dbar} \nonumber \\
& &- \sigma_{\dbar p \to \dbar X}(E_{\dbar}) \, n_{\rm H}(\vec{r}) \, v_{\dbar} \, N(\vec{r} , E_{\dbar}) \,.
\label{q_tertiary}
\end{eqnarray}

In this expression, the differential cross section for inelastic and non-annihilating interactions is often approximated
by~\cite{Bergstrom:1999jc}:

\begin{equation}
{\displaystyle \frac{\text d\sigma_{\dbar p \to \dbar X}}{\text dE_{\dbar}}}(E'_{\dbar} \! \to \! E_{\dbar}) =
{\displaystyle \frac{\sigma_{\dbar p \to \dbar X}(E'_{\dbar})}{T'_{\dbar}}} \,,
\end{equation}

where ${T'_{\dbar}}$ denotes the initial antideuteron kinetic energy. To account for the tertiary production rate Eq.~(\ref{q_tertiary}) in the transport equation Eq.~(\ref{eq:transport_CR_1}), one needs to subtract the cross section for inelastic, non-annihilating interactions from $\sigma_{\rm ine}$ in the calculation of the annihilation rate $\Gamma_{\rm \! dst}$.

Once the density $N$ is derived from Eq.~(\ref{eq:transport_CR_1}), the interstellar flux at the solar system is given by~\cite{Fornengo:2013osa}:
\begin{align}
\phi_{\bar{d}}(E) &= \frac{\beta_{\bar{d}}}{4\pi}\,n_{\bar{d}}(r=r_{\odot},z=0,E) \\ 
&=\frac{\beta_{\bar{d}}}{4\pi}\left(\frac{\rho_{\odot}}{m_{\text{DM}}}\right)^{2}R_{\bar{d}}^{\text{ann}}(E)\,\frac{1}{2}\,
\langle \sigma v \rangle\,\frac{\text dN_{\bar{d}}}{\text dE_{\dbar}} & & \text{for dark matter annihilation}, \\
\label{eq:antid_flux}
&=\frac{\beta_{\bar{d}}}{4\pi}\,\frac{\rho_{\odot}}{m_{\text{DM}}}\,R_{\bar{d}}^{\text{dec}}(E)\,\frac{1}{\tau_{\text{DM}}}\,
\frac{\text dN_{\bar{d}}}{\text dE_{\dbar}} & & \text{for dark matter decay},
\end{align}

where the propagation function $R_{\bar{d}}(E)$ encodes all the dependence on astrophysical parameters and is determined numerically. The specific form of $R_{\bar{d}}(E)$ can be found in \cite{Fornengo:2013osa}.

\subsubsection{Two-zone diffusion model}
\label{subsec:twozone}

A full numerical treatment is generally required to solve the transport equation Eq.~(\ref{eq:transport_CR_1}), as described, e.g., in~\cite{1998ApJ...509..212S}. However, analytical (or semi-analytical) solutions may be derived assuming a simplified description of the spatial dependence of some parameters. The two-zone diffusion model~\cite{1969ocr..book.....G,1980Ap&SS..68..295G}, based on the description of the Galaxy as a thin gaseous disk embedded in a thick diffusive halo, has proven to be successful in reproducing the nuclear~\cite{1992ApJ...390...96W,Maurin:2001sj,2002A&A...381..539D}, antiproton~\cite{2001ApJ...563..172D}, and radioactive isotopes~\cite{2002A&A...381..539D} data. It also allows us to treat contributions from dark matter (or other exotic) sources located in the magnetic halo~\cite{Donato:2003xg,2005PhRvD..72f3507B,2005PhRvD..71f3512S}. This model has been extensively detailed in~\cite{Maurin:2001sj,2002A&A...394.1039M,2002astro.ph.12111M}.

In the two-zone model, the Galaxy is defined as a cylinder with a magnetic halo of half-height $z=L$ and radius $r=R$. The ISM and the nuclei accelerators are contained in a thin-disk of half-height $h \ll L$. The halo thickness $L$ is a free parameter of the model, and $h$ and $R$ are set to 100\,pc and 20\,kpc, respectively. The diffusion coefficient Eq.~(\ref{eq:diff_K}) is assumed to be the same throughout the Galactic magnetic halo, and the convective wind is assumed to be of constant magnitude directed outwards perpendicular to the Galactic Plane with $\vec{V_c}=V_c \vec{e}_z$. The reacceleration strength, mediated by the Alfv\'en velocity $V_a$,  is confined to the thin disk. The first and second order terms $f_{\rm o}$ and $s_{\rm o}$ in Eq.~(\ref{eq:transport_CR_1}) follow the formulation given in~\cite{Maurin:2001sj}.

In the framework of the two-zone diffusion model, the details of Galactic propagation depend entirely on the values that are assigned to the parameters ($L$, $V_{c}$, $K_{0}$, $\delta$, $V_a$). These parameters are usually constrained by measurements of cosmic-ray observables such as the primary fluxes and secondary-to-primary ratios (in particular B/C). However, fitting to existing B/C data does not lift the strong degeneracy between these five transport parameters. In particular, many sets of these parameters lead to the same B/C ratio and the same secondary (standard) antiproton flux~\cite{2001ApJ...563..172D}. This degeneracy is broken for sources located in the diffusive halo, leading to large astrophysical uncertainties for the relevant fluxes~\cite{Donato:2003xg,2002A&A...388..676B,2003A&A...398..403B}. In particular, a large uncertainty for dark matter searches stems from the degeneracy between the normalization of the diffusion coefficient $K_0$ and the halo size $L$. Indeed, stable secondary-to-primary ratios mostly constrain the cosmic-ray escape time $\propto L/K_0$. However, $L$ has a strong impact on dark matter signal predictions since it is proportional to the amount of cosmic rays induced by dark matter annihilation or decay within the diffusive Galactic Halo that contribute to the local cosmic-ray flux. 

\begin{figure}
\centering
\includegraphics[width=0.65\linewidth]{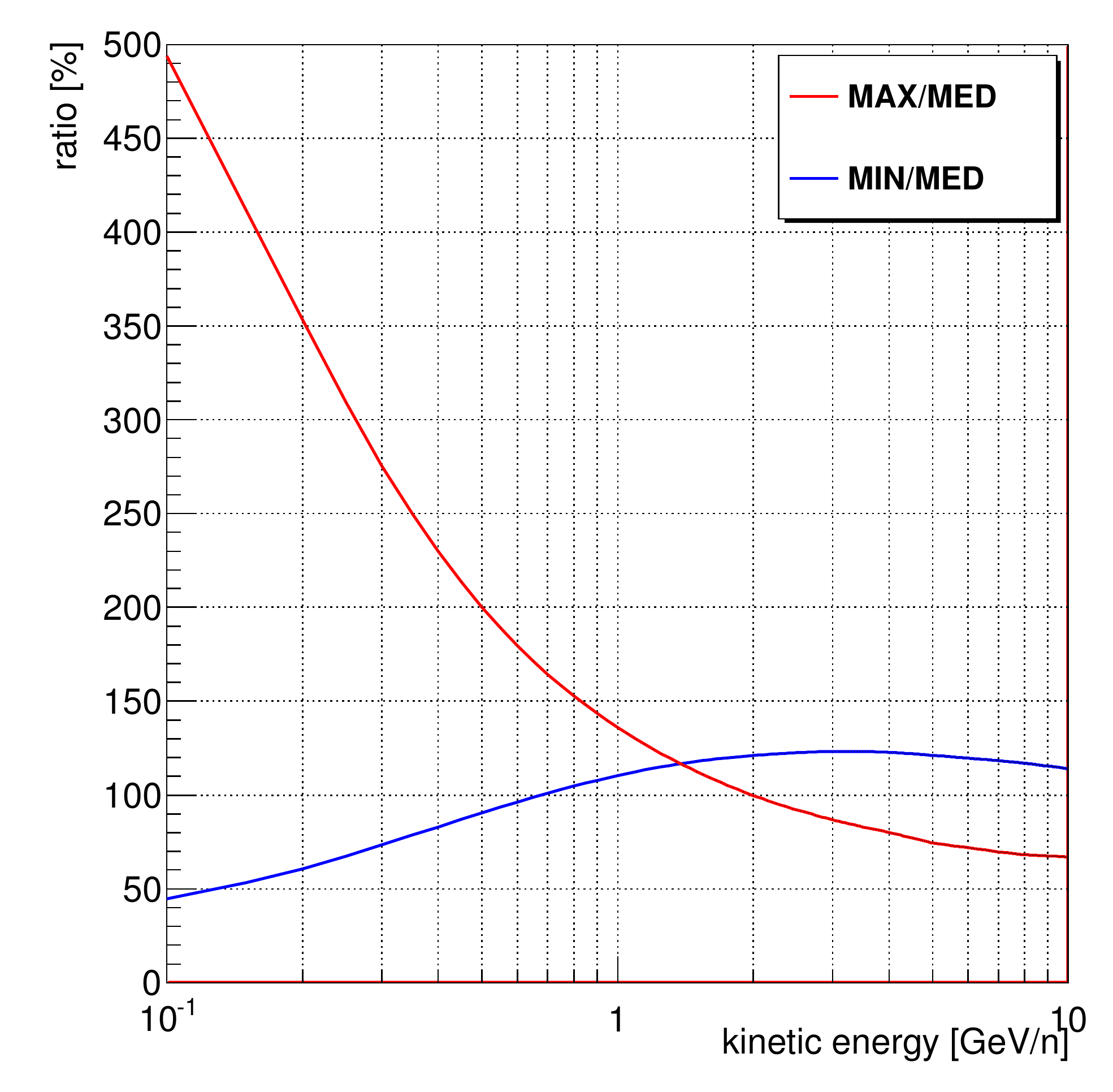}
\caption{Ratios of antideuteron fluxes from dark matter annihilation ($m_{\text{DM}}=100$\,GeV, $b\bar{b}$ channel) in the MAX and MIN propagation models with respect to the MED propagation model, based on \cite{Fornengo:2013osa}.}
\label{fig:antid_propleft}
\end{figure}

As benchmark scenarios, the three sets of parameters labeled as MIN, MED and MAX defined in \cite{Donato:2003xg} are adopted in this review. A large range of $L$ values can be found in the literature. The values $L=1$ and 15\,kpc are used in the MIN and MAX models, respectively. These values were proposed by~\cite{Donato:2003xg} to bracket the theoretical uncertainties on dark matter signal predictions, relying on the boron-to-carbon (B/C) analysis performed in~\cite{Maurin:2001sj}. Fig.~\ref{fig:antid_propleft} shows the ratios of antideuteron fluxes from dark matter annihilation ($m_{\text{DM}}=100$\,GeV, $b\bar{b}$ channel) in the MAX and MIN propagation models with respect to the MED propagation model. For low energies the relative difference between MIN and MAX covers about one order of magnitude.  Some correlation between the uncertainties for the propagation of astrophysical antideuterons and dark-matter induced antideuterons exists. However, this is not a dominant effect because the main uncertainty for the dark matter signal comes from the fact that it is produced in the halo and not, like the astrophysical background, in the Galactic disk. While the astrophysical background is nearly unchanged by the choice of halo size, a larger halo produces more dark-matter induced antideuterons.

\subsubsection{Recent experimental constraints on diffusion model parameters\label{subsec:twozone_constraints}}

The $K_0/L$ degeneracy can be broken by an observable that is sensitive to only one of the two parameters. In general, radioactive secondary-to-stable secondary ratios, such as $^{10}$Be/$^9$Be, have been used. The radioactive secondaries decay before they can reach the edge of the Galaxy and escape, and thus their modeling is independent of the Galactic diffusive halo size $L$. However, several studies have shown that this method is very sensitive to the modeling of the local ISM and strongly affected by the presence of a local under-density, known as a local bubble~\cite{1983ApJ...271L..59F}. In addition, the lack of precise measurements over a sufficiently large energy range does not permit a precise estimation of the halo size.

Indirect constraints on $L$ can be obtained from calculations of diffuse Galactic $\upgamma$-rays~\cite{2012ApJ...750....3A} or radio~\cite{2013MNRAS.436.2127O, 2012JCAP...01..049B} emissions, disfavoring very small (large) halo values $L \lesssim 2$\,kpc ($\gtrsim 10$\,kpc). Nevertheless, predictions of these observables rely on additional model parameters with non-negligible systematics and correlations, for example line-of-sight integrals depending on the astrophysical source, interstellar medium, and/or magnetic field distributions.

Recently, the authors of~\cite{2014PhRvD..90h1301L} demonstrated that low-energy secondary posi\-trons can directly constrain two-zone diffusion models with small haloes and large diffusion indices $\delta$, effectively excluding the MIN model. Secondary posi\-trons are produced by inelastic scattering of primary cosmic rays off the ISM. Due to energy losses, the mean free path of secondary posi\-trons is much smaller than that of nuclei, decreasing the dependence on $L$. For small $L$ models, boundary effects become important and the secondary positron flux scales roughly with $1/L$. This excludes small $L$ models which predict a higher than observed positron flux, and excludes high spectral indices through the spectral shape. Besides the availability of very precise positron data over a large energy range, this method is less sensitive to the modeling of the local interstellar medium compared to the standard approach.

\begin{figure}
\centering
\includegraphics[width = 0.48\textwidth]{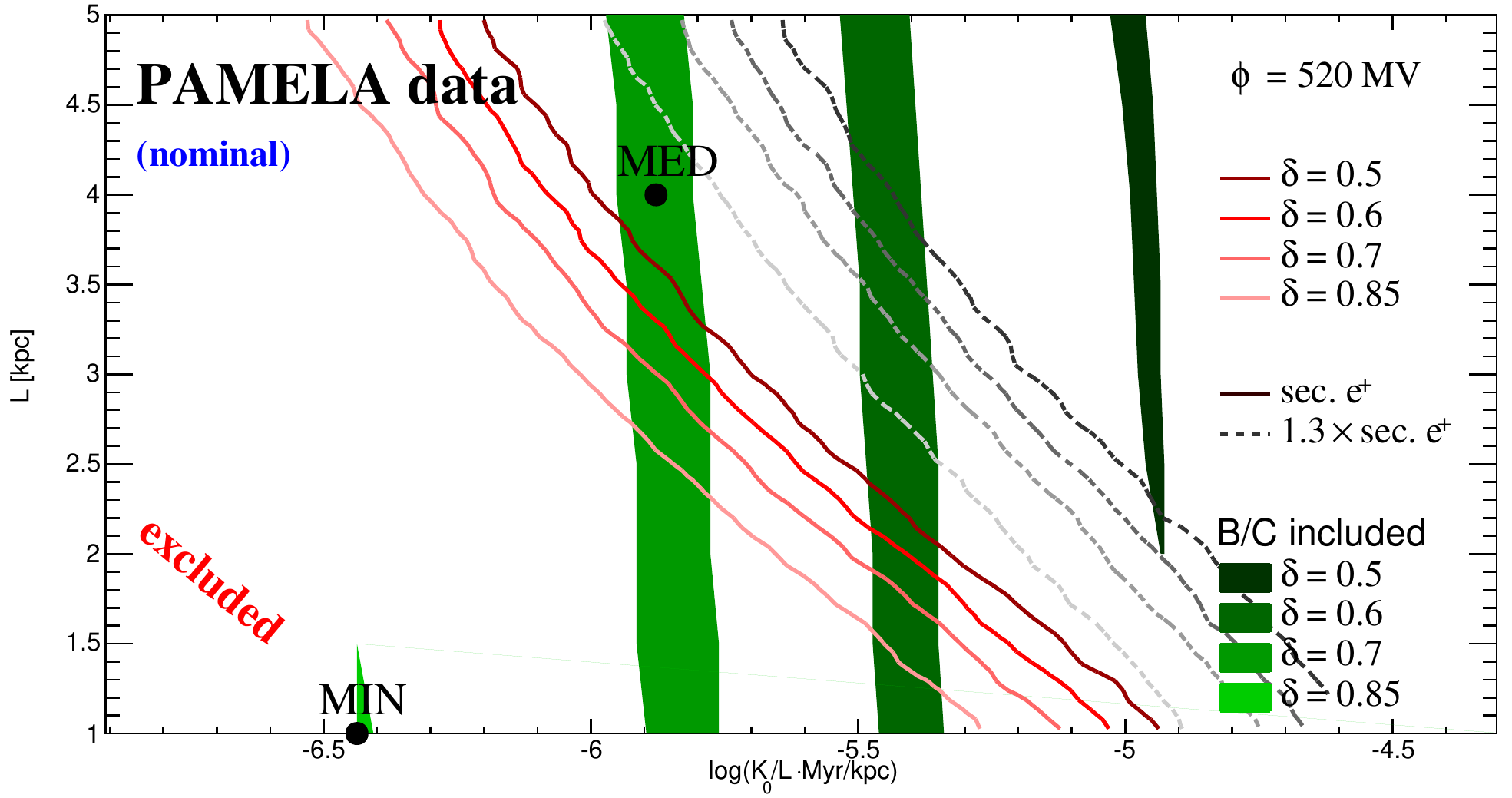}
\hfill
\includegraphics[width = 0.48\textwidth]{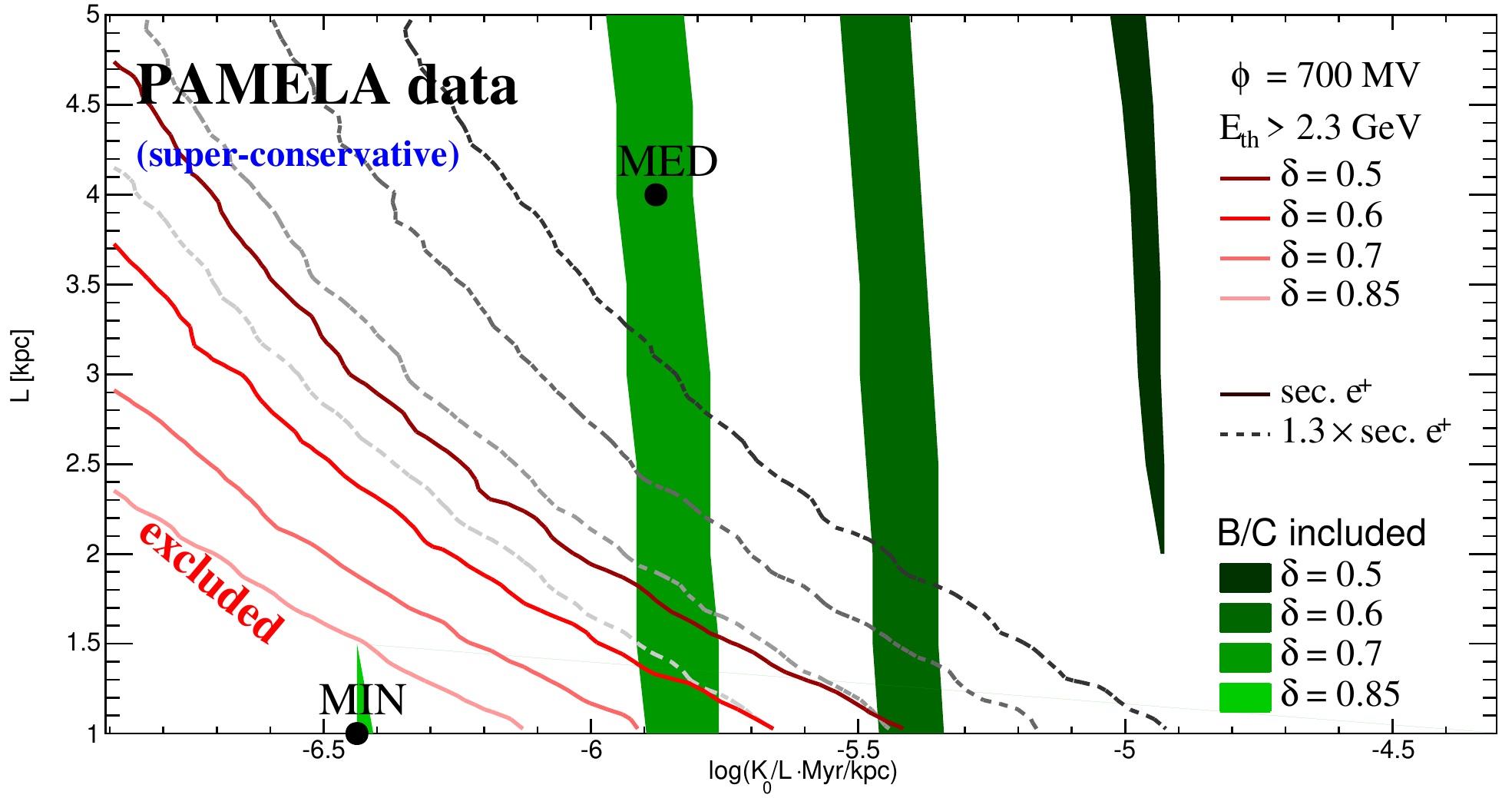}
\caption{Constraints on propagation parameters in the $\log(K_0/L \cdot 1\,\text{Myr}/ 1\,\text{kpc})$--$L$\,[kpc] plane. Lines are constraints from the positron flux (downward regions are excluded). Green filled contours are allowed by B/C data. The MIN and MED models of \cite{Donato:2003xg} are indicated by a filled black circle. \textit{Left}: Positron contours (from PAMELA data) for a realistic modulation level of $\phi = 520$\,MV. Dashed lines correspond to the limits if the secondary positron prediction is increased by 30\% to mimic a primary component. \textit{Right} (super conservative): Same, but with a solar modulation of 700\,MV and considering only data points above 2\,GeV. Figures are taken from \cite{2014PhRvD..90h1301L}.}
\label{fig:ExlusionMIN}
\end{figure}

The exclusion curves shown for different spectral indices $\delta$ (red solid lines) in Fig.~\ref{fig:ExlusionMIN} are obtained by comparing 500\,000 predicted secondary positron fluxes consistent with B/C constraints to the PAMELA data~\citep{2013PhRvL.111h1102A}. Models leading to fluxes in excess by more than three standard deviations with respect to the data are excluded. In the left panel, the nominal value of 520\,MV for the solar modulation (Sec.~\ref{subsec:transport_HELIO}) was adopted. In  the right panel, conservative results are shown by (i) considering only the data above $2.38$\,GeV and (ii) increasing the solar modulation potential to 700\,MV. In the same plots the B/C constraints of \cite{Maurin:2001sj} for the different slices of $\delta$ are represented by green bands. The obtained bounds are almost orthogonal to the B/C constraints, and thereby uncorrelated. In the nominal and conservative case, the MIN propagation model, and more generally large diffusion indices ($\delta \gtrsim 0.8$), are excluded. This is good news for cosmic-ray antideuteron searches, as the MED and MAX diffusion models predict at least an order of magnitude higher low-energy antideuteron fluxes, as seen in Fig.~\ref{fig:antid_propleft}. 

\subsubsection{Effect of dark matter density profiles} 
\label{subsec:DMdensity} 

Due to propagation effects, charged antideuterons in the relevant energy range cannot be traced back to the origin of dark matter annihilation or decay. Therefore, a proper determination of the effect induced by changing the dark matter profile cannot be separated from the size and properties of the diffusion zone. Nevertheless, a qualitative understanding can be obtained by realizing that the dark matter signals depend on the quantities:

\begin{align}
A &= \int_V \rho(r)^2 \text{d}V & &\text{for dark matter annihilation,}\\
D &= \int_V \rho(r) \text{d}V & & \text{for dark matter decay,}
\end{align}

\noindent with $V$ being a cylinder with radius and height from the propagation model under study and $\rho(r)$ the dark matter density. $D$ is quite insensitive to the halo profile, while $A$ instead has a more pronounced dependence. The reason is that for a cuspy profile most of the annihilations occur inside the diffusion zone, whereas for a cored profile, one can also have annihilations outside. In the latter case, some of the antimatter particles produced in the annihilation can escape into intergalactic space, and hence cause some differences in the fluxes at the Earth. This difference is not very significant, since the dark matter distribution is practically the same everywhere inside the diffusion zone except for the small region around the Galactic Center. Ref.~\cite{Donato:2003xg} explicitly investigated the effect of dark matter profile choice together with confinement zone and transport parameters for dark matter annihilating into antiprotons and found a typical difference of about 20\% or less for small confinement regions and a maximum difference of 70\% for very large confinement volumes. For dark matter decay the uncertainty from the choice of the halo profile is at the level of 10\% or less~\cite{2008JCAP...07..002I}. 

In addition to the more involved combination of dark matter profile and propagation, the local dark matter density $\rho_0$ just acts as a dark matter flux normalization parameter. A recent calculation~\cite{2015JCAP...12..001P} finds a local dark matter density value of $\rho_0=0.42$ with an error at the 10\%-level, which results in a dark matter profile uncertainty for dark matter decay of 10\% and dark matter annihilation of 20\%.

\subsubsection{Galactic transport of antihelium} 
\label{subsec:antiHe_transport} 

Galactic transport of antihelium nuclei can be modeled similarly to the antideuteron case, with suitable modifications concerning interactions with the ISM. The most relevant difference is in the cross sections for annihilating collisions between the antihelium nucleus and the hydrogen and helium nuclei of the ISM. These cross sections can be modeled with the parameterizations provided in Table 4.5 of \cite{DuperrayThesis}. Similar to the antideuteron flux, the antihelium interstellar flux can be expressed in terms of the propagation function $R$ as in Eq. (\ref{eq:antid_flux}). A parameterization of this function for several choices of the dark matter density profile is provided in \cite{Cirelli:2014qia}.

\subsection{Transport in the solar environment}
\label{subsec:transport_HELIO}

An important effect for cosmic-ray measurements in the GeV energy range is the modulation by the magnetic field of the Sun that depends on the solar cycle. When the antinuclei reach the heliosphere, they experience diffusion due to inhomogeneities in the solar magnetic field (SMF), begin drifting along the field lines of the SMF, and suffer adiabatic energy loss in the solar wind. Several approaches, including the simple force-field approximation and numerical simulations, have been discussed in the literature during the last decades.

\subsubsection{Transport equation}

Similar to the galactic environment, the propagation in the heliospheric environment is described by a transport equation~\cite{1965P&SS...13....9P}:

\begin{equation}
\frac{\partial f}{\partial t} = -(\vec{V}_{\rm sw}+\vec{v}_d)\cdot \nabla f + \nabla\cdot ({\rm K}\cdot\nabla f) + \frac{p}{3}\,(\nabla\cdot\vec{V}_{\rm sw})\,\frac{\partial f}{\partial p}\;,
\label{eq:solartransport}
\end{equation} 

where $f$ is the antideuteron phase space density averaged over momentum directions, $p$ is the antideuteron momentum, $K$ is the (symmetrized) diffusion tensor, and the vectors $\vec{V}_{\rm sw}$ and $\vec{v}_d$ are the velocity of the solar wind and the velocity associated to antideuteron drifts, respectively.

\subsubsection{Force-field approximation}

\begin{figure}
\centering
\includegraphics[width=0.65\linewidth]{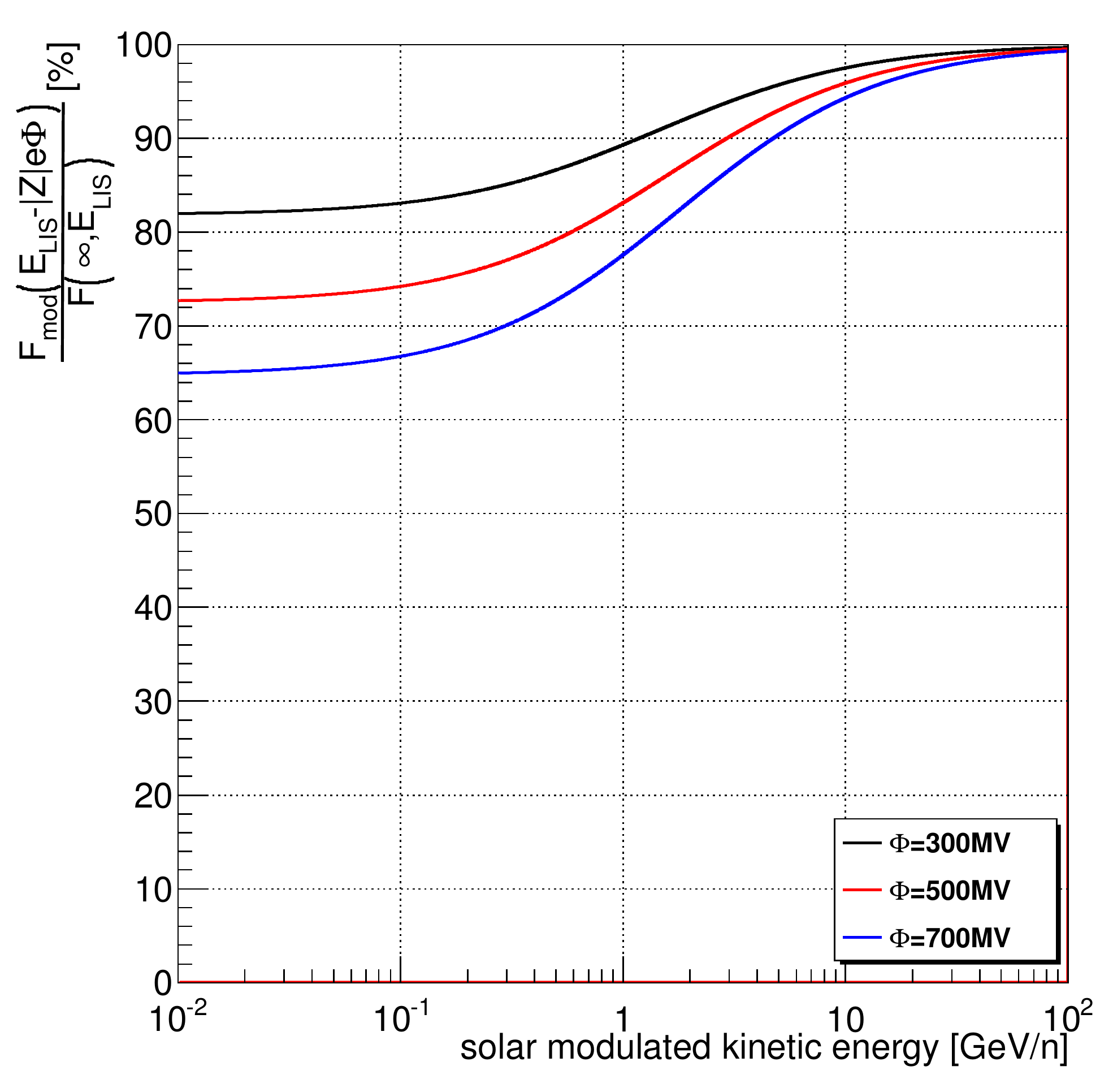}
\caption{Effect of solar modulation for antideuterons as a function of solar modulated kinetic energy (notation as in Eq.~\ref{e-forcef}).}
\label{f-sol}
\end{figure}

A simple solution to the propagation equation (\ref{eq:solartransport}) is the force-field model approximation:
\begin{equation} 
F_\text{mod}(E=E_\text{LIS}-|Z|e\Phi)=F(r,E_\text{LIS}-|Z|e\Phi)=F(\infty,E_\text{LIS})\cdot\frac{E^2-m_0^2}{E_\text{LIS}^2-m_0^2}\,,
\label{e-forcef}
\end{equation}
where $F(\infty,E_\text{LIS})$ is the flux in the local interstellar medium, $E_\text{LIS}$ is the total energy of the particle with mass $m_0$ and charge $Ze$ and the effective solar modulation parameter $\Phi$ for all particle species, which is tuned to experimental data. The relative effect of solar modulation using the force-field approximation for the antideuteron flux is shown in Fig.~\ref{f-sol} for three different choices of $\Phi$, representing different periods of solar activity. For larger values of $\Phi$ the flux measured in the solar system experiences a stronger modulation. The solar modulation potential $\Phi$ correlates with the sun activity, and thus low-energy antideuteron measurements are ideally carried out during solar minima. However, the effect of varying solar modulation parameters within a typical range (from 300 to 700\,MV at a kinetic energy of 100\,MeV/$n$) is only at the 25\% level. It is important to note that the parameter $\Phi$ is known with about 10\% accuracy from the measurements of other cosmic-ray species like protons~\cite{2011JGRA..116.2104U}. Starting from about 20\,GeV/$n$, solar modulation becomes only a small effect.

\subsubsection{Numerical simulation}

An alternative approach to solving Eq.~(\ref{eq:solartransport}), such as recently implemented in the {\tt HelioProp} numerical code~\cite{Maccione:2012cu}, consists in adopting a Monte Carlo treatment of heliospheric propagation, in which a sample of particles is injected at the modulation boundaries and traced to the position of the Earth. Here, the SMF is modeled as a large rotating spiral. Its geometry is characterized by the heliospheric current sheet, which is a surface separating magnetic field lines according to their polarity, with an exact shape depending on a parameter known as the tilt angle. The mean-free-path of antideuterons in the direction parallel to the SMF can be expressed as~\cite{Fornengo:2013osa}: 

 \begin{equation}
 \lambda_{\|} = \lambda_{0}\,\left(\frac{\mathcal{R}}{1~{\rm GV}}\right)^{\gamma}\,\left(\frac{B}{B_{\bigoplus}}\right)^{-1},
 \end{equation}
 
where $B$ is the intensity of the SMF ($B_{\bigoplus}=5~{\rm nT}$ is the value of the magnetic field at the Earth position~\cite{2011ApJ...735...83S,2012Ap&SS.339..223S}) and the normalization $\lambda_{0}$ as well as the exponent $\gamma$ are assumed to be constant. In contrast to the force-field approximation, this model includes the important effect of $\vec{v}_d$. As described in \cite{Fornengo:2013osa}, the extent to which the solar modulation affects the antideuteron fluxes depends on four parameters: the polarity of the SMF, the tilt angle $\alpha$, $\lambda_{0}$, and $\gamma$. Ref.~\cite{Fornengo:2013osa} showed that the difference between the more detailed approach and the force field approximation is small ($\approx20$\%) compared to the uncertainties in the antideuteron coalescence model and Galactic propagation. However, a big advantage of the Monte Carlo approach is its ability to predict event-by-event fluctuations and to take into account spectral features of dark matter annihilation and background flux. This allows to use the solar activity level in the modeling when evaluating a particular dark matter model.

\subsection{Geomagnetic deflection}
\label{subsec:geo_deflection} 

\begin{figure}
\centering
\includegraphics[width=0.65\linewidth]{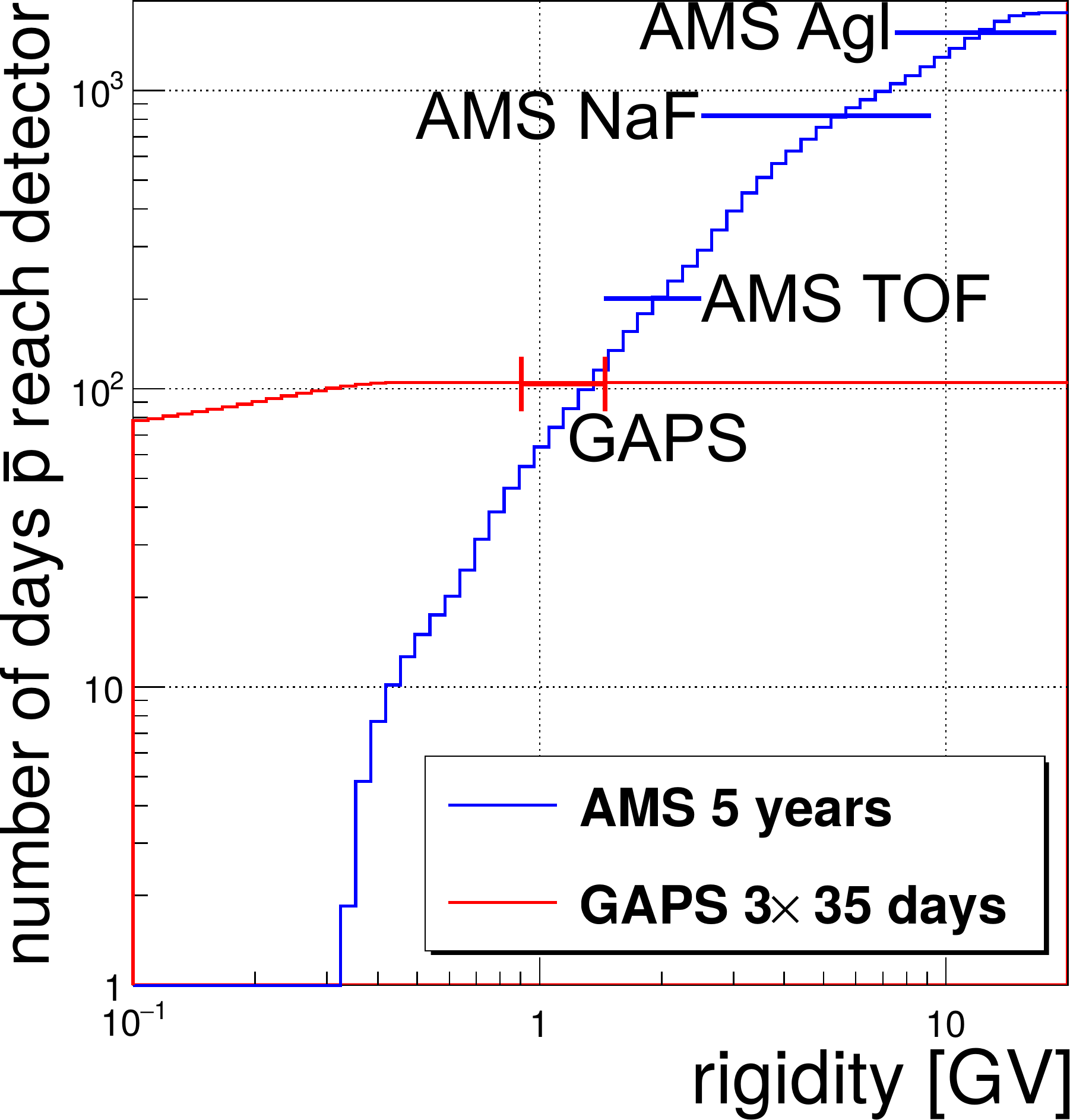}
\caption{Simulated number of days as a function of rigidity antiprotons are able to reach AMS-02 and GAPS through the geo\-magnetic field (horizontal lines indicate the means for different rigidity regions).}
\label{fig_geocutoff}
\end{figure}

The geomagnetic field plays a crucial role in the detection of charged cosmic rays. The geomagnetic field is roughly described by a tilted dipole field, which provides the strongest charged particle shielding at the equator and the weakest at the poles. Utilizing the Geant4-based {\tt PLANETOCOSMICS} \cite{planeto,2009PhDT........77V} framework for backtracing charged particles through the Earth's main magnetic field using the International Geomagnetic Reference Field \cite{GJI:GJI4804} and the Tsyganenko 2004 for the external magnetosphere~\cite{JGRA:JGRA17702}, Fig.~\ref{fig_geocutoff} compares the number of days that antiprotons can reach AMS-02 (over five years of orbit) and GAPS (over three 35-day Antarctic flights) as a function of particle rigidity. About 10\% of the measurement time AMS-02 is exposed to antideuterons of cosmic origin in the low-rigidity region (Sec.~\ref{sec:AMS}), which is most interesting for dark matter searches. The geomagnetic deflection is much lower above Antarctica and GAPS is sensitive 100\% of the time. For AMS-02 the relative exposures increase in the higher rigidity regions to about 45\% and 85\%, respectively. These geomagnetic comparisons do not account for  detector acceptances. It has also to be noted that solar magnetic disturbances influence the magnetic field in the vicinity of the Earth and can lead to considerable changes of the geomagnetic cutoff value~\cite{2002cosp...34E2424S,Shea2006209}, which have to be taken into account for the AMS-02 data analysis.

\subsection{Atmospheric influence}\label{s-atmo}

\begin{figure}
\centering
\includegraphics[width=0.65\linewidth]{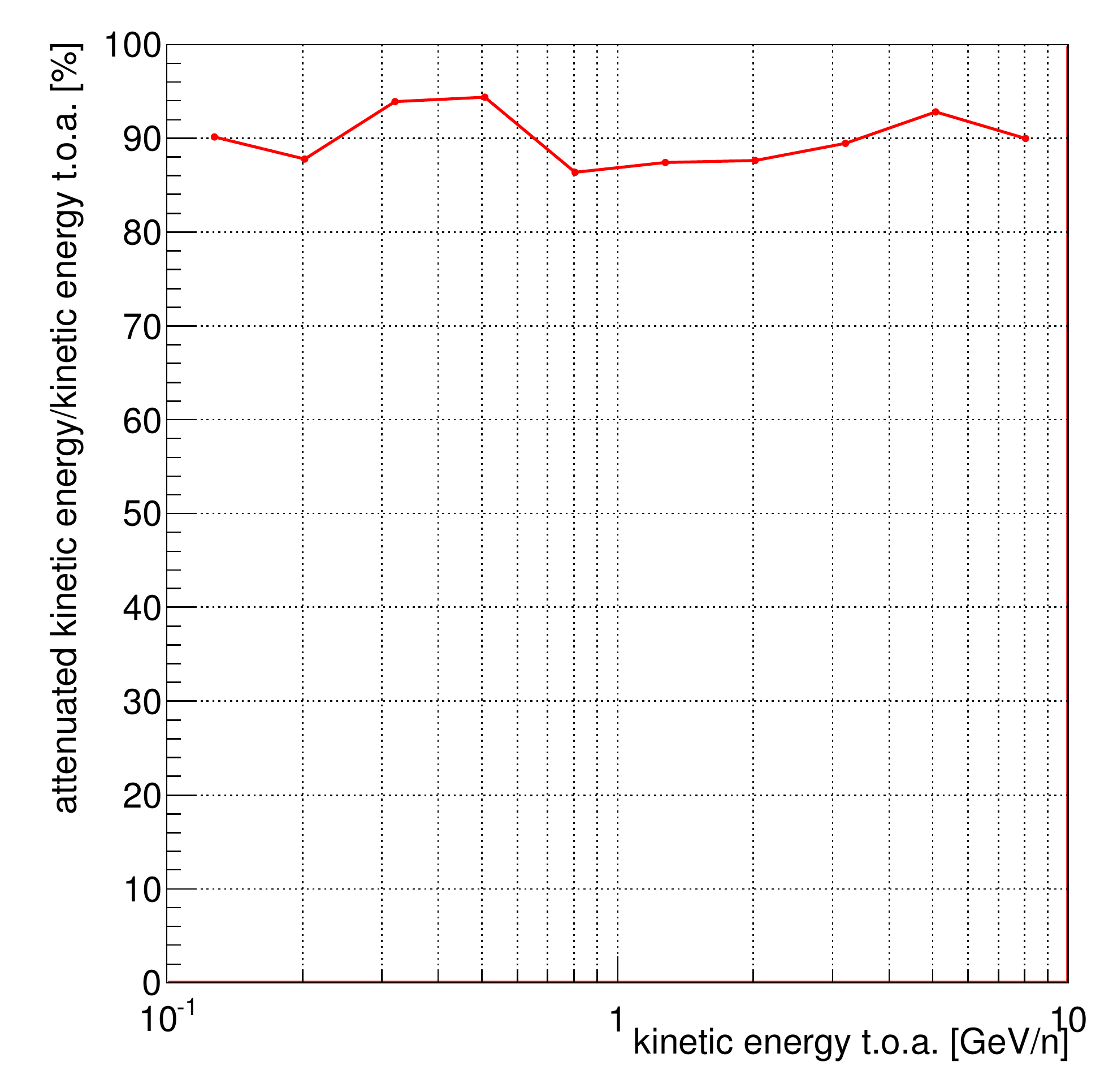}
\caption{Attenuation of antiprotons in Earth's atmosphere at 37\,km altitude for an Antarctic balloon flight trajectory.}
\label{f-atmo}
\end{figure}

One more important effect is the influence of the atmosphere. This is directly apparent when comparing the average amount of matter traversed by cosmic rays in the interstellar medium (6--10\,g/cm$^2$) to the overburden atmosphere ($\approx$6\,g/cm$^2$). Fig.~\ref{f-atmo} shows results of simulations that have been carried out in the same Geant4-based approach as the geomagnetic simulations discussed above~\cite{ucla12,Doetinchem2013pGAPS}. These studies have two components: ({\emph i}) the attenuation of cosmic-ray particles in the atmosphere and ({\emph ii}) the atmospheric background production. Fig.~\ref{f-atmo} compares the top-of-the-atmosphere kinetic energy of antiprotons to the average kinetic energy after reaching a typical balloon altitude of 37\,km above Antarctica. The energy loss in the atmosphere is at the 10\%-level, and thus the measured kinetic energy at balloon altitudes can be corrected to top-of-the-atmosphere kinetic energies without introducing large systematic uncertainties. Antideuterons at the same kinetic energy per nucleon lose about the same amount of energy before reaching a balloon-borne experiment like GAPS. However, the atmospheric production is also important for the space-based AMS-02 experiment. Atmospherically produced antideuterons can upscatter into space and will create low-energy antideuterons at high geomagnetic cutoff locations. These effects have been studied semi-analytically in \cite{Duperray:2005si}, but atmospheric antideuteron production and transport through the atmosphere and geomagnetic field to the ISS should be studied more in the future.

\subsection{Antideuteron interactions in particle physics detectors}\label{s-geant}

One of the standard tools for studying particle interactions with particle physics detectors is the Geant4 simulation suite \cite{Agostinelli2003250,2006ITNS...53..270A}. As mentioned in the last two sections, it was also used for the studies of the geomagnetic and atmospheric influence. Until recently Geant4 did not allow for the study of light antinucleus-nucleus interactions like antideuterons or antihelium with detector material. The authors of \cite{Galoyan:2012bh} added light antinuclei capabilities to Geant4 using the Glauber approach for the cross sections, the quark-gluon string model for annihilations and meson production, and the binary cascade for secondary interactions of low-energy mesons. This model was validated between 0.1\,GeV/$n$ and 1\,TeV/$n$. However, it has to be kept in mind that only very little data on the interactions of antideuterons with targets exist. It is key to apply the Geant4 extended capabilities to geomagnetic, atmospheric, and detector simulations in the future and to study the uncertainties in the cross section systematically.

\subsection{Extragalactic transport of antideuterons}\label{s-extrag}

While the focus of the section on propagation was so far on antideuterons originating in our Galaxy, one might also wonder if it were possible to search for a small component of antideuterons that originate in other galaxies, which could then be employed as a test of universal baryon symmetry.  The short answer is no, intergalactic magnetic fields provide an insurmountable barrier to the transport of antideuterons from cosmological distances at typical cosmic-ray energies and limits on diffuse $\upgamma$-ray emission provide strong constraints on large domains of primordial antimatter anywhere within the observable universe.   

The arguments follow those that have been used in the past to argue against using cosmic-ray antihelium as a probe of primordial antimatter \cite{Adams:1997ym,Cohen:1997ac}. Galactic magnetic fields are typically a few $\upmu$G.  At the intergalactic scale of about 1\,Mpc, one would naively expect that ejected galactic magnetic fields would contribute to intergalactic fields at the $\mathcal{O}(1\text{\,nG})$ level.  The leakage of the observed large dipolar magnetic fields in galaxies would contribute to the intergalactic field at least at the $\mathcal{O}(1\text{\,pG})$ level.  If disordered, these fields would prevent diffusion of $\mathcal{O}(1\text{\,GeV})$ cosmic rays from even the closest galaxies.  The only way around this is if cosmic rays were to follow ordered intergalactic fields stretched between galaxies.  Under these optimized conditions, Ref.~\cite{Adams:1997ym} argued that cosmic rays would have a diffusion distance of at most 32\,Mpc.  Additionally it was shown that galactic magnetic fields will result in an accessibility problem. If cosmic rays have a large accessibility to galaxies, they will enter them and be destroyed at each galactic encounter.  If extragalactic cosmic rays have a small galactic accessibility, they could in principle avoid destruction by galaxies en-route, but then they could not get into our galaxy to be observed at any significant level.

Both \cite{Adams:1997ym} and \cite{Cohen:1997ac} have shown that the observed diffuse $\upgamma$-ray background is inconsistent with the existence of significant antimatter domains within the Hubble volume.  Together, the extragalactic cosmic-ray transport argument and the $\upgamma$-ray argument explain why extragalactic antinuclei heavier than hydrogen have not been, nor will ever be observed in the cosmic rays.  Any antideuterons seen in the cosmic rays at typical GeV to TeV energies would have to originate within our Galaxy.

\section{Experiments for the detection of cosmic ray antideuterons\label{s-5}}

As the search for cosmic-ray antideuterons is a rare-event search, particle identification and background rejection are keys. 
BESS-Polar II, GAPS, and AMS-02 exploit different experimental designs to search for antideuterons. Both BESS-Polar II and AMS-02 utilize magnetic spectrometers to distinguish antiparticles, while GAPS uses a novel exotic atom capture and decay technique. As with all rare signal searches, these different detection techniques are essential to validate any possible detection.

The BESS Antarctic flight program, discussed in Sec.~\ref{sec:BESS}, has already provided the current best upper limit on the antideuteron flux, and ongoing analysis of BESS-Polar II data will soon update this limit. AMS-02, installed on the International Space Station, has already collected over four years of cosmic-ray data, and it will continue collecting data in the years to come. 
The AMS-02 collaboration is currently preparing its first antideuteron results, as detailed in Sec.~\ref{sec:AMS}. GAPS, described in Sec.~\ref{sec:GAPS}, completed a successful prototype balloon flight in June 2012, and is currently proposing for a first Antarctic balloon campaign. In the coming years, AMS-02 and GAPS should provide either the first detection of a cosmic-ray antideuteron or upper limits yielding harsh constraints on viable dark matter models. 

\subsection{BESS-Polar II: The Balloon-borne Experiment with Superconducting Spectrometer}
\label{sec:BESS}

The BESS-Polar program exploits particle tracking in a solenoidal magnetic field to identify antimatter. The original BESS-Polar experiment flew over Antarctica in late 2004, providing the current best antideuteron flux upper limits. The BESS-Polar II experiment collected about 30\,days of Antarctic flight data from December 2007 to January 2008. No antideuterons were observed, and progress towards an improved flux upper limit is reported here. 

\subsubsection{BESS-Polar II instrument design}

\begin{figure}
  \centering
  \includegraphics[width=1\linewidth]{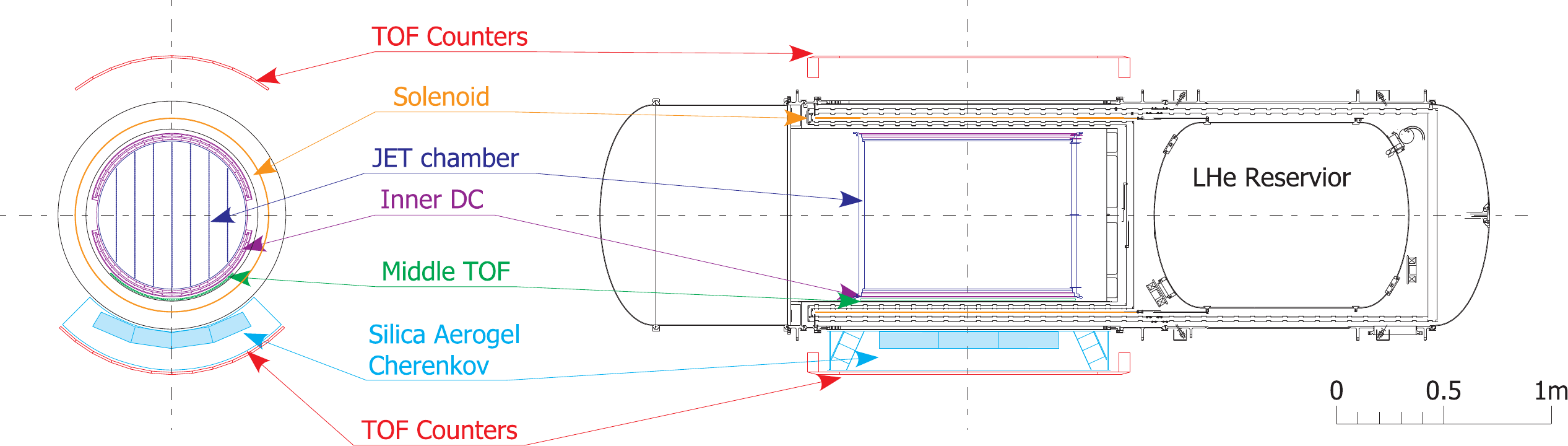} 
  \caption{Cross sectional view of the BESS-Polar II instrument. }
  \label{fig:bess-polar}
\end{figure}

BESS-Polar II consists of a large solenoidal magnet, filled by inner drift chambers (IDC) and a jet-type drift tracking chamber (JET), and surrounded by an aerogel Cherenkov counter (ACC) and a time-of-flight system composed of scintillation counter hodoscopes. 
These components are arranged in a coaxial cylindrical geometry, providing a large geometric acceptance of 0.23\,m$^{2}$\,sr.  

Fig.~\ref{fig:bess-polar} shows a cross sectional view of the BESS-Polar II instrument. A thin superconducting solenoid provides a uniform field of 0.8\,T. Tracking is performed by fitting up to 48 hit points in the JET and 4 hit points in the IDC, resulting in a magnetic-rigidity resolution of $0.4\%$ at 1\,GV and a maximum detectable rigidity of 240\,GV. The upper and lower scintillator hodoscopes provide time-of-flight (TOF) and $\text d E/\text d x$ measurements as well as trigger signals. The timing resolution of each hodoscope is 120\,ps, resulting in a $\beta^{-1}$ resolution of 2.5\%. The threshold-type Cherenkov counter, using a silica aerogel radiator with refractive index $n = 1.03$, can reject electron and muon backgrounds by a factor of 12\,000. The threshold rigidities for antiproton ($p$) and antideuteron ($d$) are 3.8\,GV and 7.6\,GV, respectively. In addition, a thin scintillator middle-TOF with timing resolution of 320\,ps is installed between the central tracker and the solenoid, in order to detect low-energy particles that cannot penetrate the magnet wall. 

\subsubsection{BESS-Polar II antideuteron identification}

The event selection criteria used to identify antideuterons follow closely that used for the antiproton analysis described in~\cite{besspbar}. The major difference is the background processes that must be considered. In the antiproton analysis, relativistic electrons, muons, and pions are the main background sources; for antideuterons, the main background are antiprotons. 

A clear detection of an antideuteron over the plentiful antiproton and $e/\mu/\pi$ background events requires reliable measurements of particle rigidity, $\beta$, and $\text d E/\text d x$. Special care must be taken to reduce tails of distributions, which are caused by scattering, interaction, and misidentified tracks in the detector. In the study outlined below, samples of protons and deuterons, as well as antiprotons with a mask applied to the antideuteron region, were used to optimize selections and to determine efficiencies.

\begin{figure}
\centering
	\includegraphics[height = 0.41\textwidth]{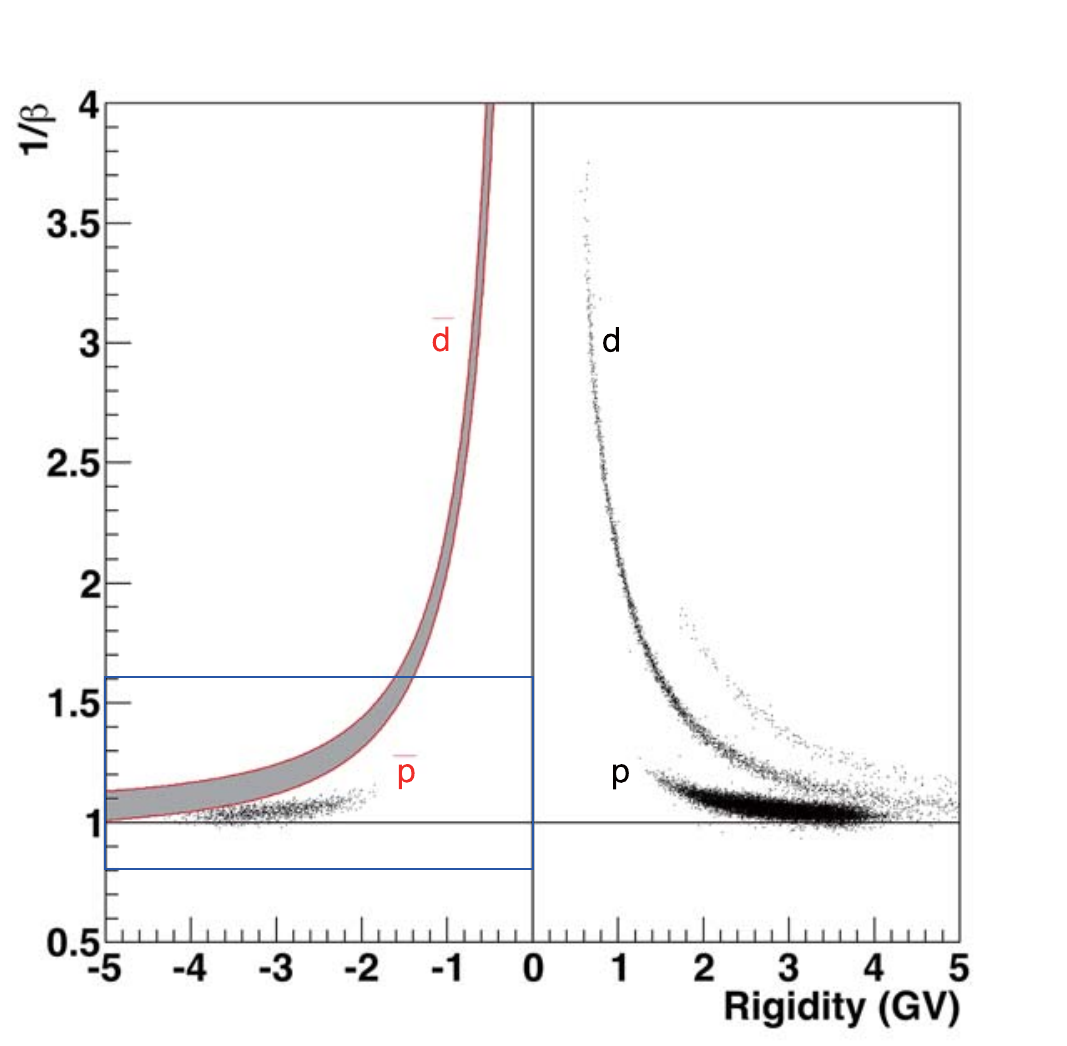}
	\hfill
	\includegraphics[height = 0.41\textwidth]{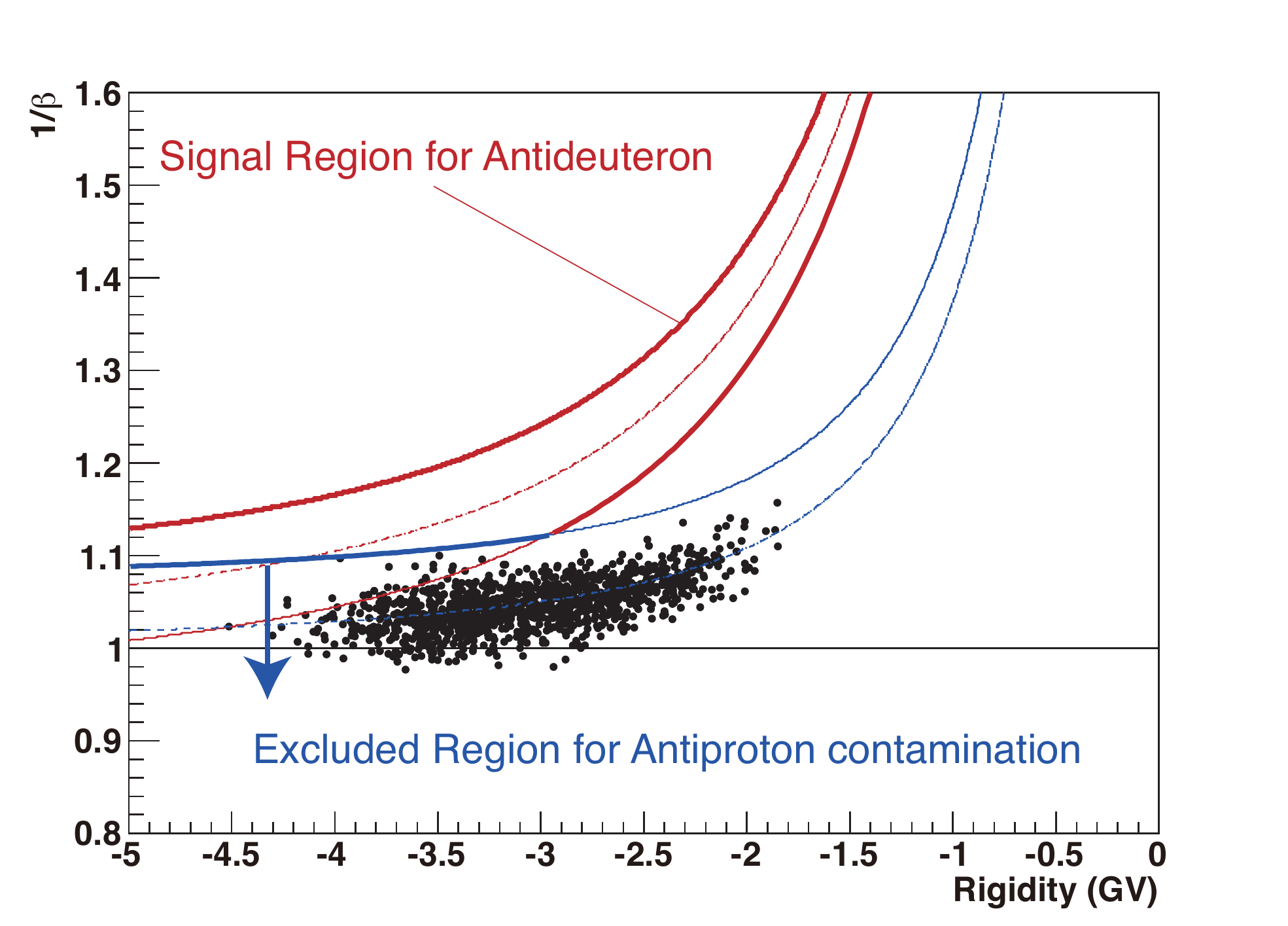}
	\caption{\textit{Left}: Plot of $\beta^{-1}$ vs. rigidity for events remaining after the $\text dE/\text dx$ and the ACC requirements. \textit{Right}: View of the region indicated by the blue box to the left. The red and blue curves show signal and excluded region, respectively. No antideuteron candidate was found in the signal region.}
 \label{fig:idplot}
\end{figure}

Below are the requirements used in the preliminary BESS-Polar II analysis to isolate antideuteron events:

\begin{description}
\item[Clean single track requirement:] First, well-reconstructed, non-interacting single track events that pass through the fiducial region are selected. 
Cuts are then applied to the $\chi^2$ of the track fit, the consistency between track and TOF hit information, and the number of $z$-hits in the IDC to ensure the quality of the track and correct timing information. 
These cuts also eliminate hard scattering events. Events with a large number of hits in the tracker that are not associated with a track, which may cause mismeasurement of the TOF timing and $\text d E/\text d x$, are also rejected.

\item[\boldmath$\text d E/\text d x$ requirement:] Next, bands of $\text d E/\text d x$ vs. rigidity space are identified  to distinguish (anti-)protons, (anti-)deuterons, helium, and $e/\mu/\pi$ events. Both positive and negative-curvature events are used, since various efficiencies for antideuterons can be estimated using positive-curvature deuteron events. $\text d E/\text d x$ vs. rigidity selections are applied to the upper TOF, lower TOF, and JET chambers. The JET $\text d E/\text d x$ has better resolution than the TOF $\text d E/\text d x$, providing good separation of antiprotons from antideuterons (and protons from deuterons) up to about 2\,GV.

\item[ACC requirement:] The ACC veto, based on Cherenkov counter measurements and designed originally for antiproton identification, are applied to reject relativistic $e/\mu/\pi$ samples. In addition, this cut eliminates antiprotons and protons with rigidities higher than the threshold $R>3.8$\,GV and reduces the number of events with interactions outside the tracker.

\item[Mass requirement:] Fig.~\ref{fig:idplot} shows the scatter plot of $\beta^{-1}$ vs. rigidity for events remaining after the $\text d E/\text d x$ and ACC requirements. Clear band structures are visible, each of which corresponds to particle mass as $m= \mathcal{R} \sqrt{1/\beta^2 -1}$. The antideuteron band width defines a three standard deviation detection region. The antideuteron region is initially masked for blind analysis. Antiproton events that survived the $\text d E/\text d x$ and ACC cuts are clearly observed. These remnants are carefully studied, and the selection criteria are iteratively optimized to remove this tail.
\end{description}

To select the final antideuteron events, an antiproton excluded region is defined as indicated in Fig.~\ref{fig:idplot}. This region is defined such that less than 0.1 event is expected to contaminate the antideuteron signal region, assuming that the antiproton $\beta^{-1}$ distribution is Gaussian. No antideuteron candidate is observed in the signal region.

\subsubsection{BESS-Polar II current status}

The BESS-Polar II flight was successfully carried out in December 2007 through January 2008, near solar minimum period. $4.7\cdot10^9$ cosmic-ray events were collected at altitudes of 34--38\,km (average residual air of 5.8\,g/cm$^2$). During the 24.5 days of observation, all detectors exhibited the expected performance, except for the central tracker which showed an instability due to high-voltage fluctuation. However, through the development of a time-dependent tracker calibration, more than 90\% of the data has been successfully calibrated. The data has been analyzed for antideuteron signals, as described above, and no candidate events were observed. A publication of this result, including a new current best upper-limit on the cosmic antideuteron flux, is in preparation.

\subsection{AMS-02: The Alpha Magnetic Spectrometer}
\label{sec:AMS}

The Alpha Magnetic Spectrometer (AMS-02) is a general purpose high-energy particle physics detector, installed on the International Space Station (ISS) since 19 May 2011~\cite{ams02first}. A detector prototype, AMS-01, was successfully flown aboard U.S.\ Space Shuttle Discovery in June 1998~\cite{ams01res}. Although its mass and volume are optimized to fit the Space Shuttle payload bay, the layout of the AMS-02 detector is typical of larger particle physics experiments, with particle identification performed using the tracking of charged particles in a magnetic field, momentum measurement in a time-of-flight system, and energy measurements performed in Cherenkov and electromagnetic calorimeter systems.

\subsubsection{AMS-02 instrument design}

\begin{figure}
\centering
\includegraphics[width=0.65\linewidth]{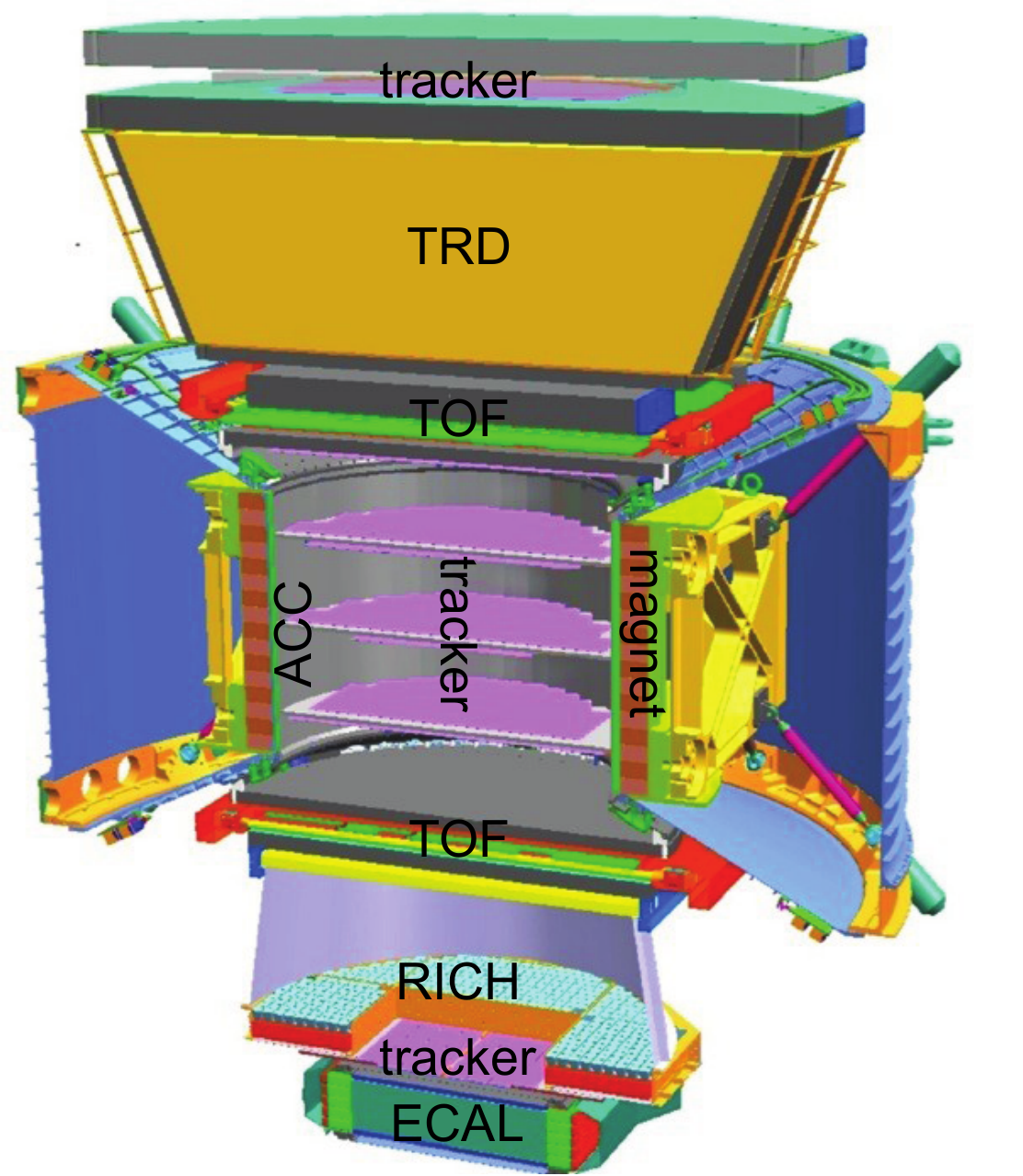}
\caption{\label{f-ams}Cross sectional view of the AMS-02 experiment.}
\label{fig:ams}
\end{figure}

The AMS-02 instrument~\cite{amsdet-kounine, amsdet-ting} is shown in Fig.~\ref{fig:ams}. It consists of: a permanent magnet ($B=0.14$\,T) and a 9-layer precision silicon tracker, enabling the measurement of rigidities up to about $2$\,TV and a position resolution of 10\,$\upmu$m for single charged particles~\cite{ams02trkicrc}; a time-of-flight (TOF) detector consisting of four detection planes, providing a time resolution of 120\,ps for $Z=1$; a transition radiation detector (TRD) with 20 layers, with the primary goal of $e/p$ separation at energies up to hundreds of GeV~\cite{ams02trdicrc}; a ring imaging Cherenkov detector (RICH) with a dual radiator, consisting of silica aerogel ($n=1.05$) and sodium fluoride (NaF, $n=1.33$), for accurate velocity measurement at $\beta > 0.95$ for aerogel and $\beta > 0.75$ for NaF~\cite{ams02richicrc}; an electromagnetic calorimeter (ECAL) composed of nine superlayers with a total thickness of 17 radiation lengths, for $e^+/e^-$ and photon measurement~\cite{ams02ecalicrc}; and anti-coincidence counters (ACC)~\cite{ams02acc}. The absolute value of particle charge is measured independently by the tracker, TOF, TRD, and RICH, while its sign is given by the combination of curvature information from the tracker and directional information from the TOF. The acceptance of AMS-02 (without requiring ECAL crossing) is approximately 0.5\,m${^2}$\,sr~\cite{ams02acceptanceicrc}.

\subsubsection{AMS-02 antideuteron identification}

A key element in antideuteron identification and background exclusion is the determination of particle mass. The mass is calculated from the particle's charge $Z$, its rigidity $\mathcal{R}$ and its velocity $\beta$: 
\begin{equation}
m = Z \mathcal{R} \sqrt{\frac{1}{\beta^2} - 1}\,.
\end{equation}
Three kinetic energy regions can be considered for potential antideuteron identification.
The lowest energy region, $E_{\text{kin,TOF}} \approx 0.2$--0.6\,GeV/$n$, is the most relevant for dark matter studies. In this region, the velocity measurement is provided by the TOF. Two higher-energy regions use the RICH detectors, and are thus limited by the Cherenkov thresholds of its radiators: $E_{\text{kin,NaF}} \approx 0.6$--3.8\,GeV/$n$ for NaF and $E_{\text{kin,Agl}} \approx 2.9$--8.5\,GeV/$n$ for aerogel. The NaF radiator covers about $10\%$ of the AMS-02 acceptance~\cite{ams02richdet}], but, compared to the TOF region, this is partly offset by the lower geomagnetic deflection (Fig.~\ref{fig_geocutoff}).

Nearly all major contributions to the cosmic-ray flux must be considered as potential backgrounds to the antideuteron search. The most challenging backgrounds are those from particles that have the same mass, charge, or mass/charge ratio as the antideuteron, or that may easily produce such a particle through interaction with the detector.

AMS-02 data contain many valuable variables for exclusion of these backgrounds. Selecting clean tracks without in-flight interactions is essential. A large number of hits in the ACC or off the main track is a typical signature of a noisy event or pile up of coincident events. Additionally, a match of TRD and tracker tracks and a minimum number of TOF hits may be required. In the lowest energy region, which uses TOF velocity for mass reconstruction, the fact that antideuterons are significantly slower than minimum ionizing particles leads to a higher average energy deposition in both ECAL and TOF detectors in comparison with lighter particles in the same energy range.

\begin{figure}
\centering
    \includegraphics[width=0.65\linewidth]{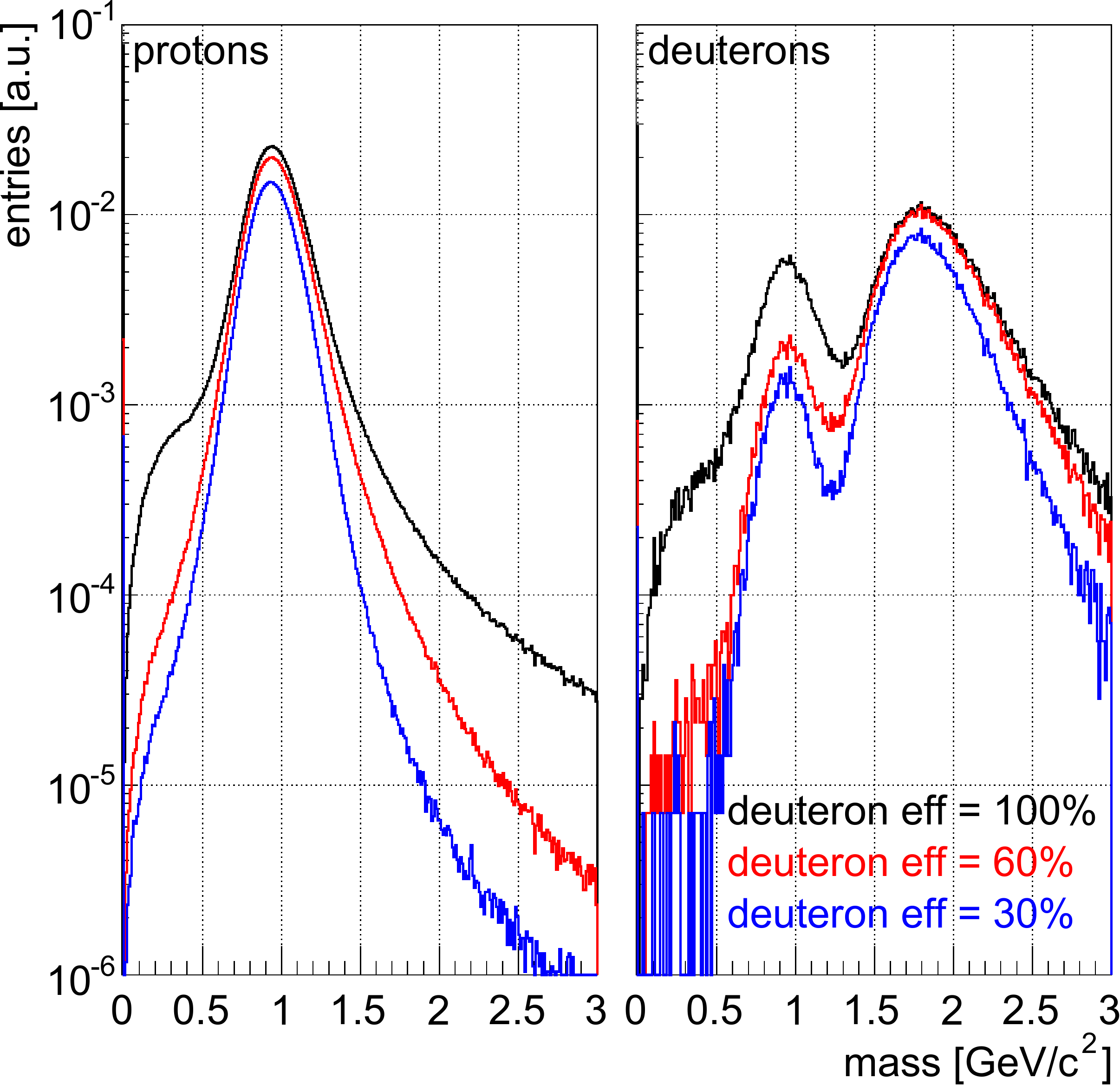}
\caption{Distribution of reconstructed mass in AMS-02 for simulated proton and deuteron events for a reconstructed kinetic energy range of 1--2\,GeV. In each case, the top distribution includes all events with a single reconstructed particle, while subsequent distributions show the result of applying quality cuts leaving 60\% and 30\% of the original well-reconstructed deuteron sample (defined as having $m_{\text{rec}}$ within $\pm 20\%$ of $m_d$).}
\label{fig_dpmassres}
\end{figure}

Specific features of different particle backgrounds are presented below.

\begin{description}

\item[Antiprotons] may present the greatest challenge, since they can only be separated from antideuterons by accurate mass measurement. As mentioned above, particle mass is derived from the combination of measured rigidity, velocity and charge. The AMS-02 collaboration is currently addressing the closely related problem of $d/p$ separation (Fig.~\ref{fig_dpmassres}), and this analysis will serve as a guide for the $\bar{d}/\bar{p}$ analysis. A rejection factor of about $10^4$--$10^6$ will be required. 

\item[Electrons] are the most abundant cosmic rays with negative charge. However, they can be discriminated from antideuterons using their much lighter mass and non-hadronic interaction signatures. Virtually all electrons measured by AMS-02 have $\beta \approx 1$, meaning that only those with a poor velocity reconstruction ($\beta \ll 1$) will pose a problem. Additional discrimination may be obtained using the energy deposition pattern in the TRD and the shower shape and total energy deposition in the ECAL. A rejection factor of about $10^6$--$10^8$ will be required.

\item[Positrons,] being less abundant than electrons and with an opposite charge, should be excluded efficiently using the same procedures applied to electrons. A rejection factor of about $10^5$--$10^7$ will be required.

\item[Protons and deuterons] have positive charge, meaning that any backgrounds due to these particles must be caused by charge signed misidentification. This may happen in AMS-02 if the rigidity is reconstructed in such a way that it has the wrong curvature. It could also occur if an upward traveling particle is misreconstructed as downward traveling due to an incorrect TOF measurement, or if an interaction occurs with a primary particle generating an upward traveling secondary. Although deuterons have the same mass as antideuterons, it is very unlikely for a deuteron that is misreconstructed in rigidity, velocity, or both to be reconstructed at the right mass. The masses of events with misreconstructed charge are expected to follow a continuous mass distribution rather than a peak at the (anti)deuteron mass. A rejection factor of about $10^8$--$10^{10}$ will be required.

\item[Helium] is by far the most abundant cosmic-ray component with $Z>1$. Helium nuclei frequently interact with the AMS-02 detector, producing secondary protons and deuterons that may be misidentified as antideuterons. The exclusion of such fragmentation events is therefore the main concern for this case. The identification of such events typically relies on a search for events containing a large number of particles crossing the detector, inconsistencies between charge measurements at different points of the detector, or an unexpected trajectory change. However, helium-induced background should be taken care of when suppressing particles with a positive charge (as discussed above). A rejection factor of about $10^7$--$10^9$ will be required.
\end{description}

\subsubsection{AMS-02 current status}

AMS-02 has been acquiring data on the International Space Station at an average rate of approximately $600$\,Hz~\cite{ams02first}. 
The total number of events detected as of 19 February 2015 is over $6.1 \cdot10{^{10}}$, with data collection expected to continue for a total period of about $20$~years~\cite{ams02first}.

The sensitivity of AMS-02 to antideuterons has been addressed in the past~\cite{ams02dbaricrc2007}.
However, this analysis was conducted in 2007, during the development phase of the experiment, and was based on the assumption of a
superconducting magnet configuration with $B = 0.86$\,T instead of the permanent magnet's $0.14$\,T and a substantially different tracker plane layout spanning about $1$\,m instead of about $2.5$\,m. The projected antideuteron sensitivity of AMS-02 shown in Fig.~\ref{f-dmdbar} is based on the superconducting-magnet configuration. An updated study based on the final configuration is therefore needed, and will profit from knowledge gained in recent years from improved simulations and detector operation.

Studies of particle identification are currently being performed using both AMS-02 data and large Monte Carlo samples produced using a detailed simulation of the AMS-02 detector based on the Geant4 toolkit~\cite{Agostinelli2003250, 2006ITNS...53..270A}. This is a continuous process where ISS data inform the detector construction and physics lists used in the simulations. For example, the high statistics of AMS-02 data mean that nuclear cross sections may become the dominant source of uncertainty for cosmic-ray predictions, which in turn are assessed using cosmic-ray flux ratios such as $d/^3$He~\cite{coste2012}. Other ratios like $d/^4$He may provide further constraints on propagation parameters already tuned to B/C data, allowing a better estimate of sensitivities for antiproton and antideuteron dark matter searches~\cite{tomassetti2014}.

Antiprotons will be the most challenging background to antideuteron searches. Fig.~\ref{fig_dpmassres} demonstrates the narrowing of the proton and deuteron mass peaks for the reconstructed kinetic energy range of 1--2\,GeV, shown after various stages of quality cuts (number of ACC hits, track fit quality, number of TOF hits, etc.) for simulated data. The selections reduce significantly the proton high mass tail while maintaining a high deuteron efficiency. In addition, a second peak visible in the deuteron case is due to protons produced in detector interactions. Thus a solid understanding of deuteron interactions with the detector will be required, which can be studied with experiments like NA61/SHINE (Sec.~\ref{s-na61}). The final antideuteron selection cut will ultimately be in a range around the deuteron mass peak. In simulated events, a mass resolution $\sigma_m/m \approx 12\%$ has been attained for protons at kinetic energies $E \approx 0.5$\,GeV. The AMS-02 deuteron and antideuteron analyses are currently ongoing.

The AMS-02 collaboration has already published results with unprecedented accuracy for fluxes and/or ratios on several of the most abundant particle species ($p$~\cite{PhysRevLett.114.171103}, He~\cite{ams02hefluxicrc}, $e^+/e^-$~\cite{ams02first,ams4,ams02elposflux2014}, B/C~\cite{ams02bcratioicrc}). Future publications are expected to include data on antiprotons, deuterons, and heavier nuclei.

\subsection{GAPS: The General Antiparticle Spectrometer}
\label{sec:GAPS}

The General Antiparticle Spectrometer (GAPS) experiment is a balloon-borne instrument optimized specifically for antideuteron and antiproton detection in an unprecedented low-energy range, providing essential complementarity to the AMS-02 antideuteron search. Traditionally, detectors have relied on magnetic spectrometers to discriminate between matter and antimatter particles. 
GAPS, however, utilizes a complementary technique that relies on the production and decay of short-lived exotic atom~\cite{Hailey2013290}. In GAPS, a time-of-flight detector measures the momentum of an incident low-energy particle, which then loses energy passing through layers of semiconducting Si targets/detectors and ultimately stops in the target material. A negatively charged antiparticle will be trapped by the nucleus of the Si detectors or Al frame and form an exotic atom in an excited state. The exotic atom will then emit X-rays of characteristic energy as the antiparticle quickly de-excites, before annihilating on the target nucleus, producing a shower of pions and baryons.

The X-ray energies, annihilation product multiplicity, $\text dE/\text dx$ energy loss, and stopping depth provide discrimination between antideuterons and its main background, antiprotons. An antiproton with the same velocity will penetrate less deeply, produce X-rays of different characteristic energies, and generate fewer annihilation products. The combination of these multiple signatures gives GAPS enormous rejection power against cosmic-ray protons, electrons, or other non-antimatter species. GAPS will also provide the first precision measurement of the cosmic antiproton spectrum below 0.25\,GeV/$n$, which will help both constrain propagation parameters and probe low-mass dark matter and primordial black hole models~\cite{Aramaki:2014oda}.  

The GAPS technique has been validated in antiproton beam tests~\cite{Aramaki2013}, and a prototype balloon instrument (pGAPS) was successfully flown from Taiki, Japan in June 2012~\cite{2014NIMPA.735...24M,Doetinchem2013pGAPS}. Since a magnet is not needed and the GAPS instrument is relatively simple, a balloon or satellite payload of significantly larger geometric acceptance is possible within a realistic mass budget. In addition, the GAPS trajectory will be close to the lowest-possible geomagnetic cutoff, dramatically increasing the number of detectable low-energy antiparticles (Fig.~\ref{fig_geocutoff}). GAPS is foreseen to execute several Long Duration Balloon (LDB) campaigns from Antarctica.

\subsubsection{GAPS instrument design}

\begin{figure}
\centering
\includegraphics[width=0.65\linewidth]{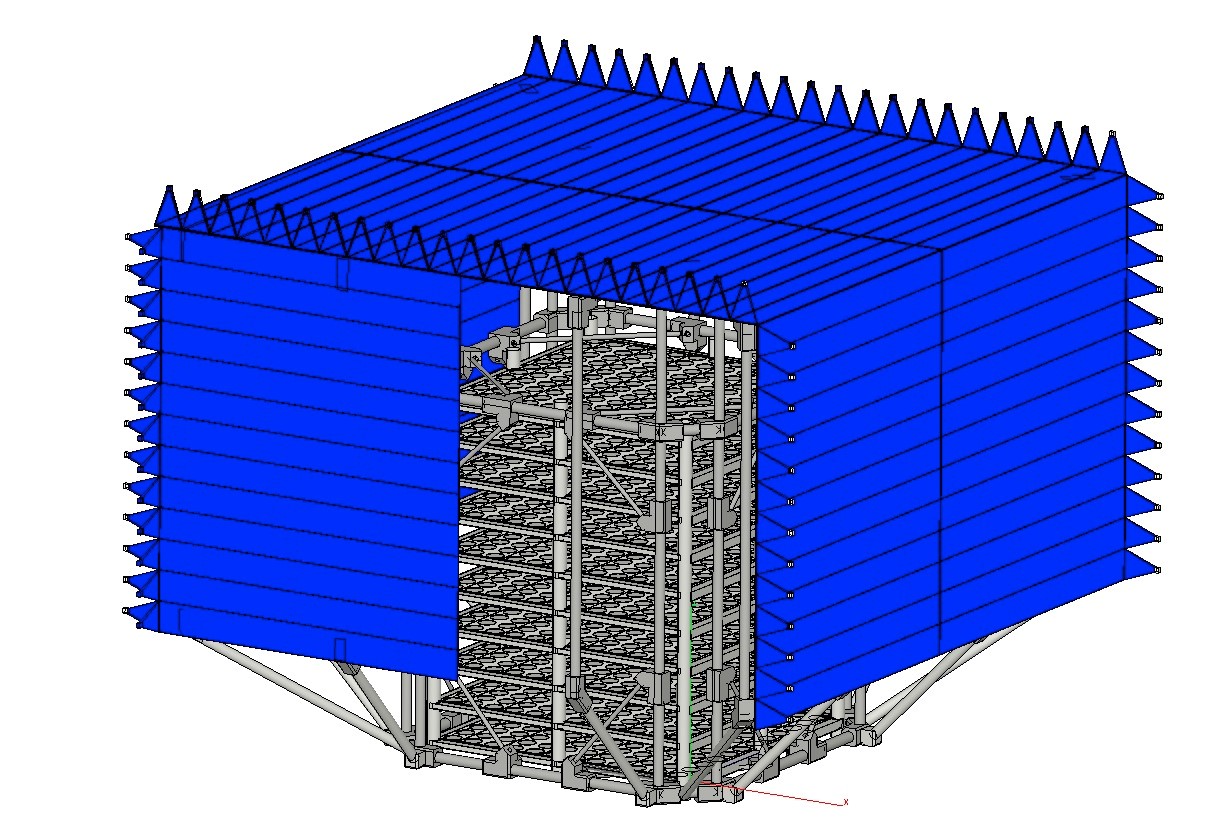}
\caption{\label{f-gaps}Mechanical drawing of the GAPS balloon instrument, with ten layers of Si(Li) detectors (gray) surrounded by a plastic scintillator TOF (blue).}
\label{fig_GAPS}
\end{figure}

GAPS consists of a thin plastic scintillator TOF surrounding a 1.6\,m $\times$ 1.6\,m $\times$ 2\,m cube of ten planes of lithium-drifted silicon (Si(Li)) detectors, as shown in Fig.~\ref{fig_GAPS}. The scintillator paddles will be 0.5\,cm thick and 160--180\,cm long. Each paddle will be read out on both ends, providing 500\,ps timing resolution. The Si(Li) detectors will be 2.5\,mm thick with a diameter of either 5 or 10\,cm, providing a total of 10.5\,m$^2$ of active area. The Si(Li) detectors serve as degrader; exotic atom target; tracker for the incident charged particle and annihilation products; and X-ray energy spectrometers. A strip size of 20\,cm$^2$ provides sufficient spatial resolution to differentiate between separate X-rays, annihilation products, and incident background particles. A readout with both a low-gain and high-gain channel will provide both sensitivity to heavy annihilation products and the 4\,keV X-ray energy resolution necessary to distinguish antiprotonic from antideuteronic exotic atoms.

GAPS will have a geometric acceptance of about $18$\,m$^2$\,sr in the 0.05--$0.25$\,GeV/$n$ energy range. As shown in Fig.~\ref{f-dmdbar}, the payload will offer a sensitivity of about $2\cdot10^{-6}$\,m$^{-2}$\,s$^{-1}$\,sr$^{-1}$\,(GeV/$n$)$^{-1}$ over the above energy range for 105 days of flight, achievable with three Antarctic LDB flights of modest duration. With this integration time and energy range, the antideuteron measurement is essentially background-free, with  about 0.01 expected antiprotons misidentified as antideuterons and a cosmic secondary/tertiary rate several orders of magnitude below the predicted sensitivity \cite{Aramaki2015}. The GAPS sensitivity to antideuterons is an improvement of more than two orders of magnitude over the current limits reported by the BESS collaboration. 

\subsubsection{GAPS antideuteron identification}

To estimate the sensitivity of GAPS to an antideuteron signal, a two-part Monte Carlo simulation based on Geant4~\cite{Agostinelli2003250, 2006ITNS...53..270A} was conducted~\cite{Aramaki2015}. The first simulation estimated the grasp and flux of stopped antideuterons, including energy loss and in-flight annihilation in the atmosphere and the instrument. The second simulation estimated the energy spectrum in the detector due to the decay of the exotic atoms, including all instrument interactions.

The grasp ($\Gamma$) is defined as the product of the geometrical acceptance ($G$) and the stopping efficiency. The grasp was estimated from simulations of antideuteron interactions as implemented in Geant4. This simulation predicts $\int\Gamma_{\bar p}\text d E = 0.15 (0.45)$\,m$^2$\,sr\,GeV/$n$ for antiprotons in the energy range 50-110 (110-250)\,MeV/$n$ at the top of the atmosphere, and $\int\Gamma_{\bar d}\text d E  = 0.16 (0.41)$\,m$^2$\,sr\,GeV/$n$ for antideuterons in the energy range 50-110 (110-250)~MeV/$n$.

The second simulation predicts the TOF and $\text dE/\text dx$ of the incident antiparticle, as well as the X-ray and nuclear annihilation product energy spectra. Since only low-energy antiprotons will produce a signature of an incoming particle with X-rays and decay products originating from a vertex, they are the only background considered here. The antiproton flux is more than four orders of magnitude larger than the predicted dark matter antideuteron flux. Thus an antiproton rejection factor of more than $10^5$ is required to obtain a 99\% confidence level for the detection of two antideuterons. Although, protons are the most abundant cosmic-ray species they are not able to fake the distinct exotic atom signature of antideuterons and antiprotons.

The following variables can be used to distinguish antideuterons from antiprotons and is described in detail in~\cite{Aramaki2015}:

\begin{description}

\item[X-ray energies:] An antideuteron exotic atom formed in a Si target will produce X-rays with energies 30\,keV, 44\,keV and 67\,keV; an antiproton will produce X-rays with energies 35\,keV, 58\,keV, and 107\,keV. The X-ray yields were estimated with the simple cascade model discussed in \cite{Aramaki2013}. The antiproton rejection factor due to one or more antideuteronic atomic X-ray detections is about 40,  while the antideuteron efficiency is $\approx$10\%.

\item[Nuclear annihilation product multiplicity:] The intranuclear cascade framework~\cite{Cugnon1989,Sudov1993} can be applied to the annihilation of the antideuteron with the Si nucleus, resulting in two different models. The first model assumes that the two antinucleons of the antideuteron interact with the Si nucleons simultaneously, while the second
model assumes that the antinucleons interact with nucleons separately. Requiring a pion multiplicity of five or more provides an antiproton rejection factor of about 15 with an average antideuteron efficiency of about 60\% for the two models. The intranuclear cascade model also predicts proton and neutron production. For antiprotons incident on Si, the average multiplicity of protons with energy larger than 60\,MeV, chosen to assure good track reconstruction from passage through three or more Si(Li) layers, is $\langle M_p \rangle= 0.37$. For antideuterons, the average multiplicity is $\langle M_p  \rangle = 2.06$ for the average of the two models. A proton multiplicity requirement $\langle M_p  \rangle \geq 3$ yields an antiproton rejection factor of 150 and an average antideuteron efficiency of about 35\% for the two models.

\item[Stopping depth and $\text dE/\text dx$ energy loss: ] The stopping range for anti\-deute\-rons is approximately twice as large as that for antiprotons with the same TOF. Therefore, antideuterons stopped in the same layer (same stopping range) can have a longer time-of-flight. For instance, antiprotons with the incoming angle of $(0\pm5)^{\circ}$ ($(45\pm5)^{\circ}$) and stopped in layer 3 will be rejected by a factor of more than a 1000 with an antideuteron selection cut, $t_{\text TOF}\geq13.3$\,ns (17.6\,ns), while the antideuteron efficiency is at about 80\%. Note that the stopped position of the incoming antiparticle can be determined by tracking the annihilation products back to a common vertex. Moreover, $\text dE/\text dX$ energy deposits in the Si(Li) detector can be also used as a powerful discriminator for any possible signal event since antideuterons at the same velocity as antiprotons have roughly twice as much kinetic energy, and deposit more energy in the Si(Li) detector before stopping. Also the different change of $\text d E/\text dx$ from layer-to-layer for antideuterons and antiprotons of the same velocity provides additional rejection power. In the future, this will be evaluated further with more detailed simulations and analyses, such as an event-by-event analysis and likelihood analysis.
\end{description} 

\subsubsection{GAPS current status}

All major mission challenges for the GAPS design have been solved. Optimization of the payload design is now being refined through detailed simulations. Si(Li) detectors with the requisite performance have been produced using in-house fabrication techniques. ISAS/JAXA has also demonstrated a cooling system using a passive Oscillating Heat Pipe (OHP) technique~\cite{iwata,6836265,Okazaki2014}, which provides the cooling necessary for the Si(Li) system at a reduced mass and power requirement compared to standard active pump systems.

\begin{figure}
\centering
\includegraphics[width=0.8\linewidth]{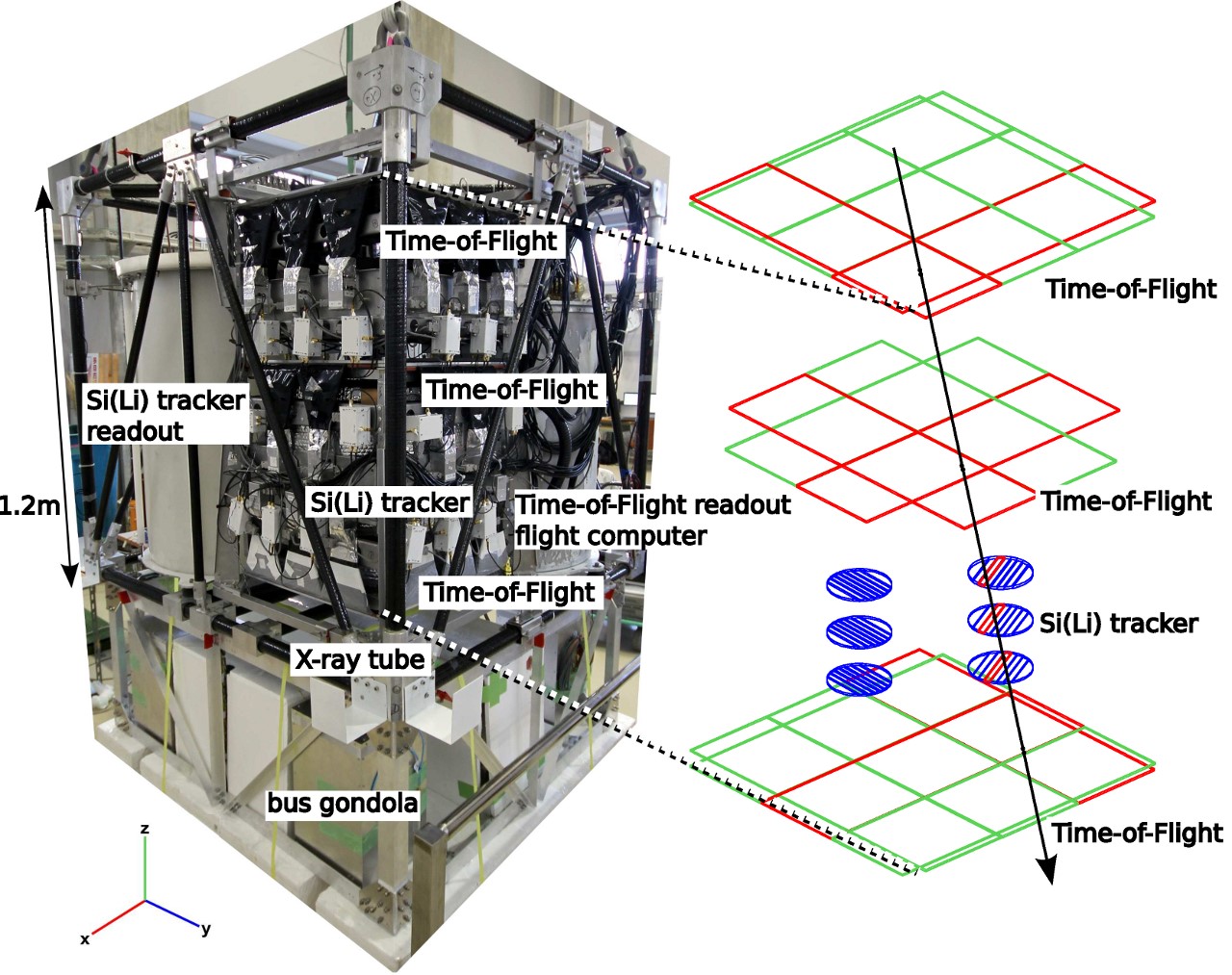}
\caption{\label{p-gaps}Photograph of the pGAPS instrument (\textit{left}) and schematic view of a reconstructed cosmic-ray event (\textit{right}).}
\label{pGAPS}
\end{figure}

To validate the GAPS instrument concept, a prototype GAPS payload was constructed and flown from the Taiki Aerospace Research Field in Japan on 3 June, 2012~\cite{2014NIMPA.735...24M,2014AdSpR..53.1432F,Doetinchem2013pGAPS}. The payload, shown in Fig.~\ref{pGAPS}, consisted of six commercially-acquired 10\,cm-diameter Si(Li) detectors arranged in three planes, with two layers of plastic-scintillator TOF above and one layer below. Because of the limited geometric acceptance, stopping power, and flight duration, pGAPS did not have any sensitivity to cosmic-ray antiparticles. The goals of the flight were instead to verify operation of the Si(Li) detectors in a balloon environment, measure the velocity of cosmic rays using the prototype TOF, verify X-ray and cosmic-ray backgrounds at flight altitude, and validate both the thermal model of the active pump cooling system and the prototype OHP system.

The total flight duration was about $5$ hours, with more than 3 hours at float altitude. For about $245$\,minutes, the payload operated in TOF trigger mode, with an additional about $50$\,minutes devoted to repeated illumination with an X-ray calibration tube and a total of about $29$\,minutes using a Si(Li) trigger mode for the study of incoherent X-ray backgrounds. 

All engineering and science goals of the flight were satisfied. Over 600\,000 cosmic-ray triggers were recorded. The thermal model was fully validated, and the OHP test was also successful. Both the TOF and the Si(Li) detectors performed in a very stable fashion, with performance as expected according to pre-flight measurements. Finally, the incoherent X-ray and cosmic-ray backgrounds were measured and present no issues to the operation of the full GAPS science mission.

\subsection{Experimental path forward}

The use of multiple, complementary experiments has been successfully employed by the direct dark matter detection community. Separate experimental designs yield different backgrounds and approaches to suppressing these backgrounds, allowing for independent confirmation of any observed signal. As a rare event search, the hunt for cosmic-ray antideuterons requires a similar approach. If AMS-02 sees one or more antideuteron events a confirming experiment is absolutely needed. GAPS and AMS-02 have complementary energy ranges, but also some overlap at low energy, allowing the study of a large energy range for confirming signals and the best chance for controlling systematic effects. The key virtue comes from the different antideuteron identification techniques. AMS-02 and BESS rely on magnetic spectrometers, and thus face different backgrounds than the GAPS exotic atom approach. A magnetic spectrometer does not only have to deliver the isotopic separation between antideuterons and antiprotons, but also between the very abundant protons and antideuterons. Limited magnetic field strength and tracking resolution can cause protons to be misreconstructed with negative charge and antideuteron mass. In contrast, in the GAPS experiment low-energy protons are not able to fake the exotic-atom annihilation signature because protons cannot replace a shell electron nor annihilate with the nucleus. A critical benchmark to reduce systematic uncertainties of AMS-02 and GAPS will be the comparison of low-energy (anti)proton fluxes as well as antiproton-to-proton ratios.

AMS-02 orbits at relatively high geomagnetic cutoffs, which reduces the number of detectable low-energy charged particles. By flying in Antarctica, the proposed GAPS trajectory is specifically tailored to low-energy particles, and thus GAPS will face a smaller geomagnetic cutoff correction. In addition, AMS-02 was launched at the beginning of the solar activity maximum. Although solar cycle 24 exhibits a relatively weak maximum compared to former cycles, a first GAPS flight would presumably happen during the next solar activity minimum~\cite{solar}, easing the uncertainties associated with low-energy measurements. 

Therefore, the combination of AMS-02 and GAPS antideuteron searches is highly desirable. Unless GAPS is begun soon, confirmation of AMS-02 results would be a long while forthcoming. Similarly, a non-detection by AMS-02 may simply mean that the high background in the AMS-02 orbit and the difficulty of rejecting backgrounds are indicative not of the absence of antideuterons, but of the presence of high background.

\section{Conclusion}

This article reviewed how antideuterons may be generated in dark matter annihilations or decays, offering a potential breakthrough in unexplored phase space for dark matter. The unique strength of a search for low-energy antideuterons lies in the ultra-low astrophysical background for this channel. Many dark matter models are capable of producing an antideuteron flux that is within the reach of the space-based AMS-02 and balloon-borne GAPS. The combination of these two experiments allows for independent experimental confirmation, which is critical for a rare event search such as the hunt for cosmic-ray antideuterons. After reviewing different classes of models, it is evident that cosmic-ray antideuterons are sensitive to a wide range of theoretical models, probing dark matter masses from $\mathcal{O}(1\text{\,GeV})$ to $\mathcal{O}(1\text{\,TeV})$. 

The important uncertainty of the antideuteron production was also extensively discussed. The standard approach is to describe the merging of the antiproton and antineutron into an antideuteron in coalescence models. These coalescence models are tuned to collider measurements of deuteron and antideuteron production. However, it is currently an open question if the antideuteron production depends on the exact underlying process and on the available center-of-mass energy or if  Monte Carlo generators need further refinement. The predicted cosmic antideuteron flux arriving at the Earth also relies on the modeling of charged-particle transport both in the Galactic medium and in the heliosphere. The modeling of Galactic transport is the main source of theoretical uncertainty in the low-energy antideuteron range, while solar modulation is more relevant to shaping the low-energy tail of the predicted flux. Galactic propagation is an important uncertainty in predicting antideuteron fluxes at Earth. Recent positron data exclude the MIN Galactic propagation model, which predicts the lowest antideuteron flux levels at Earth, supporting higher antideuteron flux predictions that could be detected by GAPS or AMS-02. After arriving in the solar system, antideuterons can be deflected away from balloon-borne or space-based detectors by the geomagnetic field. After traveling through the geomagnetic field antideuterons can interact with Earth's atmosphere. The latter two processes can be understood in the same framework and show that the choice of geomagnetic location is crucial, but that the influence of the atmosphere is a well-controlled systematic effect if a balloon experiment's flight altitude is high enough.

Cosmic antideuterons have not yet been detected, with the current best flux upper limits provided by the BESS experiment. More sensitivity will be provided by the AMS-02 experiment, which is currently taking data onboard of the International Space Station, and by the General Antiparticle Spectrometer (GAPS) experiment, which is planned for Antarctic balloon campaigns. The combination of AMS-02 and GAPS covers a large energy range with different detection methods and backgrounds.  Both BESS and AMS-02 utilize magnetic spectrometers to distinguish antiparticles, while GAPS uses a novel exotic atom capture and decay technique. The BESS collaboration will soon update their antideuteron limits. AMS-02 has already collected more than four years of cosmic-ray data, and it will continue collecting data until the end of the International Space Station. The AMS-02 collaboration is currently preparing its first antideuteron results. GAPS completed a successful prototype balloon flight in June 2012, and is currently proposing a first Antarctic balloon campaign. In the coming years, AMS-02 and GAPS should provide either the first detection of a cosmic-ray antideuteron or upper limits yielding harsh constraints on viable dark matter models. As with all rare signal searches, complementary experiments will be essential to validate any possible detection. Hence, searching for antideuterons with both AMS-02 and GAPS is very important.

In conclusion, with intensified activities in many related areas cosmic-ray antideuterons move away from being the most-unexplored indirect dark matter detection technique and are on the way to become an important additional channel with breakthrough potential for disentangling the nature of dark matter.

\section*{Acknowledgments}

The first dedicated cosmic-ray antideuteron workshop at UCLA in June 2014 was supported, in part, by the  University of California, Los Angeles. The organizers of the workshop, PvD and RO, would like to thank all participants for a successful and fruitful meeting. The work of MG was supported by the Forschungs- und Wissenschaftsstiftung Hamburg through the program ``Astroparticle Physics with Multiple Messengers'' and by the Marie Curie ITN ``INVISIBLES'' under grant number PITN-GA-2011-289442. The work of AI was partially supported by the DFG cluster of excellence "Origin and Structure of the Universe. KPs work was supported in part by the National Science Foundation under Award No.~1202958. SW was supported by the Studienstiftung des Deutschen Volkes (scholarship number 2008 0315) and by the TUM Graduate School.

\bibliographystyle{naturemag_noURL}
\bibliography{dbar}

\end{document}